\documentclass[twocolumn,groupedaddress,aps,nofootinbib,prd,superscriptaddress]{revtex4-1}
\usepackage{amsmath}
\usepackage{amssymb}
\usepackage{epsfig}
\usepackage{color} 
\usepackage{soul}
\usepackage{ulem}
\usepackage[dvipsnames]{xcolor}

\def\bi#1{\hbox{\boldmath{$#1$}}}

 \usepackage{multirow}
 \usepackage{enumitem}
 \usepackage{bbold}
 \usepackage[output-decimal-marker={.}]{siunitx}
\usepackage{hyperref}
\usepackage{tabularray}
\usepackage{tabularx}
\usepackage{graphicx}
\usepackage{slashed}
\usepackage[export]{adjustbox}
\usepackage{hhline}
 
\begin{document}

\title{Charmoniumlike channels $1^{+}$ with isospin $1$ from lattice and effective field theory}

\def\UL{Faculty of Mathematics and Physics, University of Ljubljana, Ljubljana, Slovenia}
\def\IJS{Jozef Stefan Institute, Ljubljana, Slovenia}
\def\UR{Institut für Theoretische Physik, Universität Regensburg, Regensburg, Germany}
\def\IMSc{The Institute of Mathematical Sciences, CIT Campus, Chennai, 600113, bulldogIndia}
\def\HBNI{Homi Bhabha National Institute, Training School Complex, Anushaktinagar, Mumbai 400094, India}
\def\HBUni{Department of Physics and Hebei Key Laboratory of Photophysics Research and Application, Hebei Normal University, Shijiazhuang, China}
\def\HBUniGEO{School of Management, Hebei GEO University, Shijiazhuang, China}
\author{Mitja Sadl}
  \email{mitja.sadl@gmail.com}
      \affiliation{\UL}
    
\author{Sara Collins}
    \affiliation{ \UR}

\author{Zhi-Hui Guo}
    \affiliation{\HBUni}

\author{M. Padmanath}
    \affiliation{ \IMSc}
    \affiliation{ \HBNI}

\author{Sasa Prelovsek}
  \email{sasa.prelovsek@ijs.si}
    \affiliation{\UL}
    \affiliation{\IJS}

\author{Lin-Wan Yan}
    \affiliation{\HBUni}
    \affiliation{\HBUniGEO}

\begin{abstract} 

Many exotic charmoniumlike mesons have already been discovered experimentally, of which the $Z_c$ mesons with isospin 1 are prominent examples. We investigate $J^{PC}=1^{+\pm}$ states with flavor $\bar cc \bar qq$ ($q=u,d$) in isospin 1 using lattice QCD. This is the first study of these mesons employing more than one volume and involving frames with nonzero total momentum. We utilize two $N_f=2+1$ CLS ensembles with $m_{\pi}\simeq\SI{280}{MeV}$. The simulations are performed with unphysical light quark masses at a single lattice spacing of $a\simeq \SI{0.086}{fm}$ and omit $\psi(2S)\pi$, $\psi(3770)\pi$ and three-particle decay channels, so our results provide only qualitative insights. Resulting eigenenergies are compatible or just slightly shifted down with respect to noninteracting energies, where the most significant shifts occur for certain $D\bar D^*$ states. Both channels $1^{+\pm}$ have a virtual pole slightly below the threshold if $D\bar D^*$ is assumed to be decoupled from other channels. In addition, we perform a coupled channel analysis of $J/\psi \pi$ and $D\bar D^*$ scattering with $J^{PC}=1^{+-}$ within an effective field theory framework. The $J/\psi \pi$ and $D\bar D^*$ invariant-mass distributions from \mbox{BESIII} and finite-volume energies from several lattice QCD simulations, including this work, are fitted simultaneously. All fits yield two poles relatively close to the $D\bar D^*$ threshold and reasonably reproduce the experimental $Z_c$ peaks. They also reproduce lattice energies up to slightly above the $D\bar{D}^*$ threshold, while reproduction at even higher energies is better for fits that put more weight on the lattice data. 
Our findings suggest that the employed effective field theory can reasonably reconcile the peaks in the experimental line shapes and the lattice energies, although those lie close to noninteracting energies. 
We also study $J/\psi \pi$ scattering in s wave and place upper bounds on the phase shift.

\end{abstract}

\maketitle
 
\section{Introduction}\label{sec:intro}

The first glimpse into the realm of unconventional mesons happened with Belle's 2003 discovery of the $\chi_{c1}(3872)$ \cite{X(3872)_exp_1}. While its quantum numbers $I(J^{PC})=0(1^{++})$ initially align with a conventional $\bar{c}c$ content, its mass and decay pattern suggest a more intricate composition related to the nearby $D\bar D^*$ threshold. In 2013, the BESIII and Belle collaborations unveiled the $Z_c^+(3900)$ with  explicitly exotic flavor composition $\bar cc\bar du$ since it was observed in $J/\psi\pi^+$ and $(D\bar{D}^*)_{I=1}$ final states  \cite{Zc(3900)_exp_1,Zc(3900)_exp_2,Zc(3900)_exp_5,Zc(3900)_exp_6}.\footnote{$D\bar D^*$ generically refers to a linear combination of $D\bar D^*$ and $\bar DD^*$ with appropriate charge parity $C=\pm$ throughout the text.} 
Recent determinations assert that the $Z_c(3900)$ is a $1(1^{+-})$ state with a mass of $M=3887.1(2.6)\,\mathrm{MeV}$ and decay width $\Gamma=28.4(2.6)\,\mathrm{MeV}$ \cite{PhysRevLett.119.072001,Workman:2022ynf}. These charmoniumlike states reside near open-charm threshold $D\bar D^*$, which is the energy region explored in this paper.

Various binding mechanisms have been proposed for the $Z_c(3900)$: it could be a hadronic molecule \cite{Wang:2013cya,Dong:2013iqa,Wilbring:2013cha,PhysRevD.90.016003,Gong:2016hlt,Chen:2023def,Liu:2024ziu,Liu:2024nac}, possess a compact tetraquark structure \cite{Maiani:2013nmn}, or arise from a kinematic effect (either as threshold cusp effect \cite{Liu:2013vfa,Swanson:2014tra,Swanson:2015bsa} or triangle singularity \cite{Liu:2015taa,Szczepaniak:2015eza,Chen:2023def,Liu:2024ziu}). Although many articles propose it to be a molecular state, there are no model-independent conclusions yet. Several binding mechanisms reasonably reproduce $Z_c$ line shapes according to the analysis by JPAC \cite{Pilloni:2016obd}.
Functional methods suggest it has significant $J/\psi\pi$ and $D\bar{D}^*$ Fock components \cite{Hoffer:2024alv}. A more extended discussion of phenomenological studies is given in Sec.~\ref{subsec:pheno1+-}.

Several lattice simulations of the $Z_c(3900)$ have been performed recently. Two studies employing the HAL QCD approach \cite{PhysRevLett.117.242001,Ikeda_2018} highlighted the significance of cross-channel interaction, aligning with the conclusions of Ref. \cite{Ortega2019}. However, studies employing the Lüscher formalism could not confirm a (narrow) resonancelike peak near the threshold. This includes \cite{Zc_Sasa_1,Zc_Sasa_2,Zc_CLQCD,Lee:2014uta,Cheung2017}, as well as the coupled channel analysis of \cite{CLQCD:2019npr}. Notably, no additional finite-volume eigenstates were found and the energy shifts relative to the noninteracting levels were insignificant. Direct comparisons between methods are challenging as the HAL QCD approach does not provide information on the energy shifts.

While experimental observations have revealed charmoniumlike states with $I(J^{PC})=1(1^{+-})$, no $\bar{c}c\bar{q}q$ states with $1(1^{++})$ have been detected.\footnote{In the following, $\bar{q}q \in\{\bar ud, \bar uu -\bar dd, \bar du\}$ since $I=1$ is considered.} Such a state would be an isospin partner of $\chi_{c1}(3872)$.
An isospin-1 partner was predicted by diquark-antidiquark models \cite{Maiani:2004vq,Anwar:2018sol}, and there are arguments for its existence also in the molecular scenario  \cite{Ji:2022uie,Zhang:2024fxy}, where reasons for nonobservation in experiment are provided \cite{Zhang:2024fxy}.  
Two lattice QCD studies \cite{TOP_1_5,PRD_2015}, which found the state $\chi_{c1}(3872)$ slightly below the $D\bar{D}^*$ threshold, also did not find significant interaction in the isospin-1 system. However, the scattering amplitude for the $1(1^{++})$ channel has not yet been extracted from the lattice, and this is one of the aims of the present paper.

This article presents the results of a lattice study of channels with the flavor composition $\bar cc \bar qq$ and the quantum numbers $I(J^{PC})=1(1^{+\pm})$.\footnote{In practice, we implement the charge-neutral ($I_z=0$) combination $ \bar uu -\bar dd$.} Our central result is the extraction of the discrete spectrum of QCD eigenstates within a finite volume. To realize this, we compute matrices of two-point correlation functions using a large basis of interpolators. We construct the interpolators from pairs of meson operators (each meson operator with definite momentum) that are projected onto two different total momenta, $|{\bi P}|L/(2\pi)=0,1$. The matrices of correlation functions are evaluated on two ensembles with different spatial extents and numerous excited finite-volume energies are extracted. In the $1^{++}$ channel, the $J/\psi\rho$ threshold is positioned below the $D\bar{D}^*$ threshold. The $1^{+-}$ channel features two thresholds, $J/\psi\pi$ and $\eta_c\rho$, below $D\bar{D}^*$. These lower-lying thresholds result in inelastic $D\bar{D}^*$ scattering and introduce significant complexity to extracting the pertinent $D\bar{D}^*$ eigenenergies and scattering matrices.

Resonances and near-threshold (virtual) bound states manifest as peaks in hadron scattering amplitudes. These amplitudes are computed on the lattice (and also extracted from experimental measurements) at real values of the scattering energy. A comprehensive understanding of these amplitudes comes from considering their singularity structure in the complex energy plane. Scattering amplitudes can exhibit pole singularities that may be identified with bound states and virtual states (situated on the real axis below the lowest threshold\footnote{Scattering particles have positive (negative) imaginary momenta at the energy of the bound (virtual) state.}) or resonances (positioned off the real axis). 

After obtaining finite-volume energies from two distinct volumes and two frames, we attempt to determine scattering amplitudes. Extracting the energy dependence of these amplitudes is intricate. In this work, we perform the scattering analysis following two different approaches, as detailed in the following two paragraphs.

In the first approach, we try to address the question of whether a charmonium\-like, isospin 1 four-quark state with $1^{++}$ or $1^{+-}$ is present near the $D\bar{D}^*$ threshold with our lattice setup. We aim to discern whether this state exists, particularly in the scenario where the interaction between $D\bar{D}^*$ and other channels has a negligible impact. Therefore, scattering amplitudes are fitted from our eigenenergies using the Lüscher method assuming one-channel $D\bar{D}^*$ scattering. The energy dependence of scattering amplitude is parametrized using an effective range expansion.

The second approach is applied only to $1^{+-}$: the $J/\psi \pi$-$D\bar{D}^*$ coupled-channel system within a covariant framework of a contact effective field theory (EFT) is studied. More precisely, contact interactions $D\bar{D}^*$-$D\bar{D}^*$ and $D\bar{D}^*$-$J/\psi \pi$ are considered. We perform a joint fit to the experimental $J/\psi \pi$ and $D\bar{D}^*$ invariant-mass distributions from BESIII \cite{PhysRevLett.119.072001,PhysRevD.92.092006}, our and other \cite{Cheung2017,CLQCD:2019npr} lattice finite-volume energy levels. The resulting coupled-channel scattering amplitude features resonance poles near $D\bar D^*$ threshold. A similar investigation was carried out in Ref. \cite{Yan:2023bwt}, where the lattice results of this work were not included.

\vspace{0.2cm}

This paper is structured as follows. The next three sections discuss the lattice setup, the creation/annihilation operators and the extraction of finite-volume energy levels. These levels are presented in Sec.~\ref{sec:results}. Scattering amplitudes in the one-channel approximation are extracted in Sec.~\ref{sec:scattering}. Finite-volume energies are fitted together with the experimental data within a covariant framework of a coupled-channel system in Sec.~\ref{sec:comparison}. Other theoretical studies are discussed in Sec.~\ref{sec:pheno}. The practically vanishing $J/\psi \pi$ energy shifts indicate a negligible interaction between $J/\psi$ and $\pi$, and we put the bound on the $J/\psi \pi$ scattering length in Sec.~\ref{sec:jpsipi}. This is followed by the conclusions.

\section{Lattice details}\label{sec:details}

We employ two ensembles of gauge field configurations generated with $N_{\textrm{f}}=2+1$ nonperturbatively $\mathcal{O}(a)$ improved Wilson dynamical fermions, provided by the Coordinated Lattice Simulations (CLS) consortium \cite{Bruno2015,PhysRevD.94.074501}. The quark masses $m_{u/d,s}$ are chosen to lie on a trajectory that approaches the physical point holding the flavor average quark mass, $2m_{u/d}+m_s$, constant. The ensembles have volumes $N_T\times N_L^3=128\times 24^3$ (denoted U101) and $96\times32^3$ (labeled H105). We utilize $255$ and $492$ configurations on the smaller and larger volumes, respectively \cite{PhysRevD.95.074504}. Both ensembles have an inverse gauge coupling $\beta=6/g^2=\SI{3.4}{}$, lattice spacing $a=0.08636(98)(40)\,$fm and pion mass $m_\pi = \SI{280(3)}{MeV}$.
Open boundary conditions in time are imposed \cite{Luscher:2012av} and the sources and sink time slices of the correlation functions are located in the bulk away from the boundary. 

The study is performed with a slightly larger than physical charm quark mass \cite{sasa_charmonia_vector}. The masses of the pion, $D$, $D^*$ mesons and the spin-averaged 1S-charmonium mass determined on the larger ensemble are shown in Table~\ref{tab:mH105}. The hadron masses on both ensembles and the corresponding experimental values are given in Appendix \ref{sec:app:masses}. These masses indicate that the $D^*$ meson is stable on our lattices.

\begin{table}[ht!]
\caption{Masses of the pion, $D$, $D^*$ mesons and $M_{\textrm{av}}=\left(m_{\eta_c}+3m_{J/\psi}\right)/4$ measured on the larger lattice with $N_L=32$.}
\label{tab:mH105}
\begin{tabularx}{0.48\textwidth}
{>{\centering\arraybackslash}X
>{\centering\arraybackslash}X
>{\centering\arraybackslash}X
>{\centering\arraybackslash}X}
 \hline  \hline
$m_{\pi}$ [MeV] & $m_D$ [MeV] & $m_{D^*}$ [MeV] & $M_{\textrm{av}}$ [MeV] \\ \hline
$280(3)$ & $1927(1)$ & $2049(2)$ & $3104.3(4)$ \\ \hline \hline 
\end{tabularx}
\end{table}
\begin{table*}[ht!]
\caption{Our setup for the $\bar cc \bar qq$ system with total momenta (${\bi P}$), spatial lattice symmetry groups (LG), lattice irreps ($\Lambda^{(P)}$), total angular momentum--parities ($J^P$) and partial waves ($\ell$) that contribute to each $\Lambda^{(P)}$. For every combination of $\Lambda^{(P)}$, charge parity ($C$) and the number of lattice sites in spatial direction ($N_L$), the total number of interpolators used is presented. We list only $J\leq 3$ and $\ell\leq 2$. We do not calculate the correlation function for the case $\Lambda=A_2$ with $C=-$ and larger $N_L=32$ since it carries the most populated spectrum, and extraction of its energy levels would be too noisy. We denote the interpolator types whose lowest energy levels are above the $D\bar{D}^*$ threshold as \textit{less important} and put them into parentheses.}
\label{tab:irreps}
\centering
\begin{tblr}{
  width = \linewidth,
  colspec = {Q[90]Q[78]Q[54]Q[86]Q[95]Q[58]Q[280]Q[68]Q[118]},
  cell{2}{1} = {r=4}{},
  cell{2}{2} = {r=4}{},
  cell{2}{3} = {r=4}{},
  cell{2}{4} = {r=4}{},
  cell{2}{5} = {r=4}{},
  cell{2}{6} = {r=2}{},
  cell{2}{7} = {r=2}{},
  cell{4}{6} = {r=2}{},
  cell{4}{7} = {r=2}{},
  cell{6}{1} = {r=4}{},
  cell{6}{2} = {r=4}{},
  cell{6}{3} = {r=4}{},
  cell{6}{5} = {r=4}{},
  cell{6}{6} = {r=2}{},
  cell{6}{7} = {r=2}{},
  cell{8}{6} = {r=2}{},
  cell{8}{7} = {r=2}{},
  hline{1-2,10} = {-}{},
}
\hline
${\bi P}$    & LG               & $\Lambda^{(P)}$ & $J^P$      & $\ell$   & $C$  & (Less) important \mbox{interpolator} types                                     & $N_L$ & Total \# of \mbox{interpolators} \\
$(0,0,0)$         & O$_{\textrm{h}}$ & $T_1^+$         & $1^+,3^+$  & $0,2 $   & $- $ & $J/\psi \pi, \eta_c \rho, D\bar D^*, (D^*\bar D^*, h_c\pi)$                    & 24    & 15                        \\
                &                  &                 &            &          &      &                                                                  & 32    & 21                        \\
                &                  &                 &            &          & $+ $ & $J/\psi \rho, D\bar D^*, \chi_{c0}\pi, (\chi_{c1}\pi)$            & 24    & 5                         \\
                &                  &                 &            &          &      &                                                                  & 32    & 10                        \\
                \hline
$(0,0,1)\frac{2\pi}{L}$ & Dic$_4$          & $A_2 $          &            & $0,1,2 $ & $- $ & $J/\psi \pi, \eta_c \rho, D\bar D^*$                              & 24    & 21                        \\
                &                  &                 & $0^-,1^+,$ &          &      &                                                                  & 32    & /                         \\
                &                  &                 & $2^-,3^+$  &          & $+ $ & $J/\psi \rho, D\bar D^*, \chi_{c0}\pi, (\chi_{c1}\pi, \eta_ca_0)$ & 24    & 13                        \\
                &                  &                 &            &          &      &                                                                  & 32    & 17                        
\\ \hline                
\end{tblr}
\end{table*}

The Wick contractions are evaluated using the full \cite{HadronSpectrum:2009krc} and stochastic \cite{Morningstar:2011ka} distillation method with 90 and 100 Laplacian eigenvectors for $N_L=24$ and $N_L=32$, respectively. The setup for calculating the quark propagators is the same as that employed in Refs.~\cite{sasa_charmonia_vector,Prelovsek:2020eiw}, except for the number of eigenvectors utilized. The correlation matrices are averaged over all three polarizations (momentum directions) and up to nine source-time slices to increase the statistical precision. All uncertainties quoted are statistical only and are determined using the bootstrap method. We quote symmetric and asymmetric uncertainties, as defined in appendix A of Ref. \cite{Prelovsek:2020eiw}, using $N_{\textrm{b}}=1000$ bootstrap samples. Since this investigation utilizes lattice gauge ensembles featuring two distinct physical volumes, a single lattice spacing, and nonphysical quark masses, it is essential to note that our findings allow only a qualitative comparison with the experimental data.

\section{Two-hadron interpolating operators with $I=1$}\label{sec:interpolators}
 
The standard approach for determining finite-volume energies is employed, wherein we calculate two-point correlation functions of the form
\begin{equation}
C_{ij} (t) =
\langle 0|O_i(t)O_j^{\dagger}(0)|0\rangle\>.
\label{E0}
\end{equation}
Here, $O_i$ ($O_j^{\dagger}$) is an operator that annihilates (creates) a state with specific quantum numbers. These interpolating operators\footnote{We interchangeably use notions operator and interpolator for simplicity.} are constructed as a product of one or more quark and anti\-quark fields. We implement only meson-meson type inter\-polators
\begin{equation}
\bar c\Gamma_i c(|{\bi p}_i|^2)\ \bar q\Gamma_j q(|{\bi p}_j|^2)\quad \textrm{and} \quad \bar c\Gamma_i q(|{\bi p}_i|^2)\ \bar q\Gamma_j c(|{\bi p}_j|^2)\>.
\label{HLMM}
\end{equation}
One-meson interpolators $\sum_{ {\bi x}}e^{i{\bi p}\cdot{\bi x}}\bar\psi({\bi x})\Gamma\psi({\bi x})$ have an appropriate Dirac structure $\Gamma$ and are separately projected to a definite momenta, where the total momentum is ${\bi P}={\bi p}_i+{\bi p}_j$.\footnote{Throughout the article, ${\bi p}_{i,j}$ and $\left|{\bi p}_{i,j}\right|^2$, which represent the momentum of a single meson, will be given in units of $2\pi/L$ and $(2\pi/L)^2$, respectively, since ${\bi p}_{i,j}={\bi n}_{i,j}\frac{2\pi}{L}$, with ${\bi n}_{i,j} \in N_L^3$.} One-hadron interpolators of type $\bar{c}\Gamma c$ are absent in the $I=1$ sector. We omit local diquark-antidiquark inter\-polators since previous lattice studies~\cite{Cheung2017,PRD_2015,Zc_Sasa_2} suggest they have little influence on the rest of the spectrum. Some meson-meson channels represented by the left interpolator in \eqref{HLMM} include effects from light resonances, specifically the $\rho$ and $a_0$, which are unstable at our pion mass. Despite this, we only implement two-hadron interpolators; no three-hadron interpolators are included. The latter would be more suitable for capturing the physics of decays involving three hadrons in the final state, e.g., $Z_c(3900)\to\eta_c\rho\to\eta_c\pi\pi$. 

For $I=1$, the Wick contractions for the matrix of correlation functions (Eq.~\eqref{E0}) involve diagrams where the light quarks exclusively propagate from the source to the sink. Two types of quark line contractions arise for the charm quarks: one where the charm quarks propagate from the source to the sink time slice and another where they propagate from and to the same source/sink time slice. The latter annihilation diagrams can mix with channels that do not contain charm quarks. Similar to most previous works, we omit these diagrams as it is very challenging to include this mixing in the lattice calculation. We remark that experiments do not observe these decay channels in the region of interest. 

The setup for all the interpolators implemented in this study is provided in Table~\ref{tab:irreps}. Their free energies, in the center-of-momentum (cm) frame, mainly span the region below $\SI{4.2}{GeV}$. For more comprehensive lists, we refer to Tables~\ref{tab:Cm} and \ref{tab:Cp} in Appendix \ref{sec:app:interpolators}. To study s-wave $D\bar D^*$ scattering with spin-parity $1^+$, we focus on two finite-volume irreducible representations (irreps) of the spatial lattice symmetry groups $O_h$ ($|{\bi P}|=0$) and $\textrm{Dic}_4$ ($|{\bi P}|=1\cdot2\pi/L$), namely $\Lambda^P=T_1^+$ and $\Lambda=A_2$, respectively.

\begin{figure*}[ht!]
\begin{minipage}{0.45\linewidth}
\centerline{\includegraphics[width=1.2\textwidth]{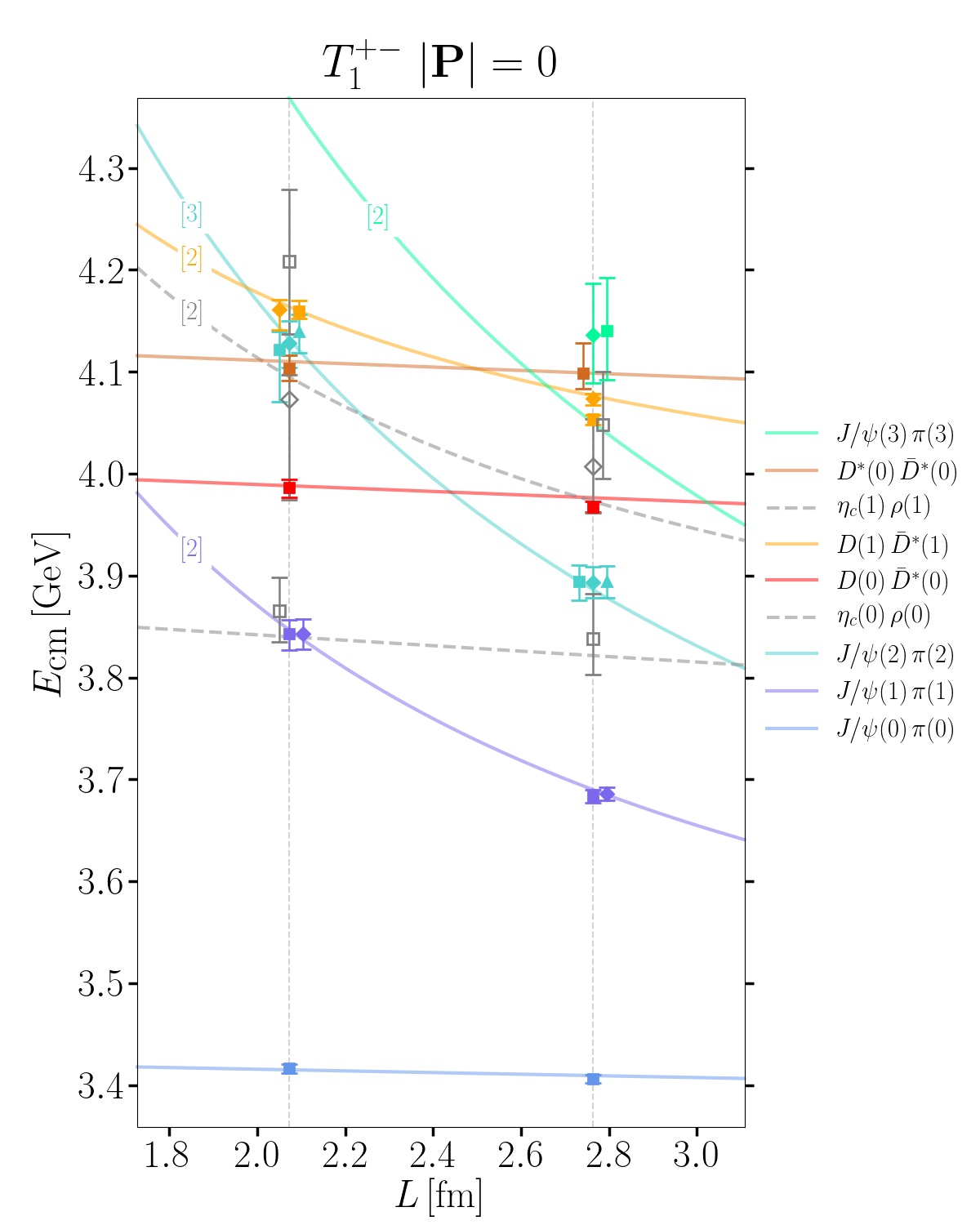}}
	\end{minipage}
\hspace*{1.55cm}
\begin{minipage}{0.45\linewidth}
\centerline{\includegraphics[width=1.2\textwidth]{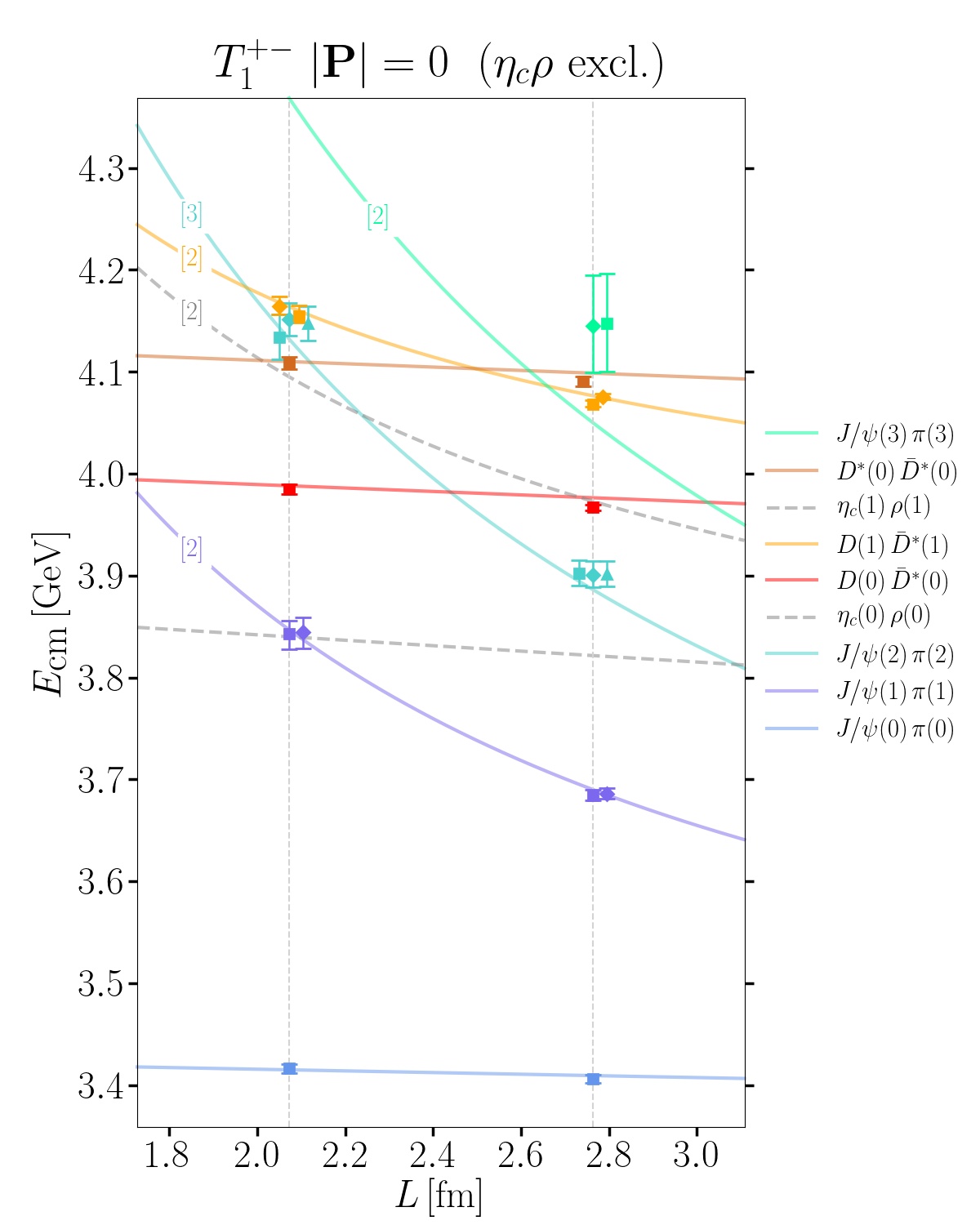}}
	\end{minipage}
\caption{Eigenenergies in the rest frame ($|{\bi P}|=0$) finite-volume irrep $T_1^{+-}$ relevant for the quantum numbers $I(J^{PC})=1(1^{+-})$. The left and right panes represent the energy levels (points, $E_{n,\textrm{cm}}$) and noninteracting energies (lines, $E_{\mathrm{ni,cm}}^{\mathrm{con}}$) for the case where all interpolators are included and the $\eta_c\rho$ interpolators are excluded, respectively. Empty gray symbols and dashed gray lines represent states with dominant overlap to $\eta_c \rho$ interpolators containing the $\rho$ resonance, which in our case is treated as stable under the strong interaction. The noninteracting energies and thresholds of the neglected channels (e.g., the three-particle channel $\eta_c\pi\pi$) and the noninteracting energies of the lowest neglected interpolators with higher internal momenta are shown in Fig.~\ref{fig:omit_1+-} of Appendix~\ref{sec:app:omitted_states}, together with the information on which levels are incorporated in various scattering analyses. Numbers within the square brackets refer to the multiplicity of certain noninteracting levels. If there is a possibility to relate an extracted energy level to a certain partial wave, then its symbol is a square, diamond or triangle for s wave ($J=1$), d wave ($J=1$) or d wave ($J=3$), respectively. Finite-volume eigenenergies are shown with $1\sigma$ statistical uncertainty, and some are slightly horizontally shifted for clarity. The main text explains that the noninteracting lines are drawn with each meson mass being a linear function of the lattice size, $m(L)$.}
\label{fig:T1pm}
\end{figure*}

\begin{figure*}[ht!]
\centering
\begin{minipage}{0.45\linewidth}
\centerline{\includegraphics[width=1.2\textwidth]{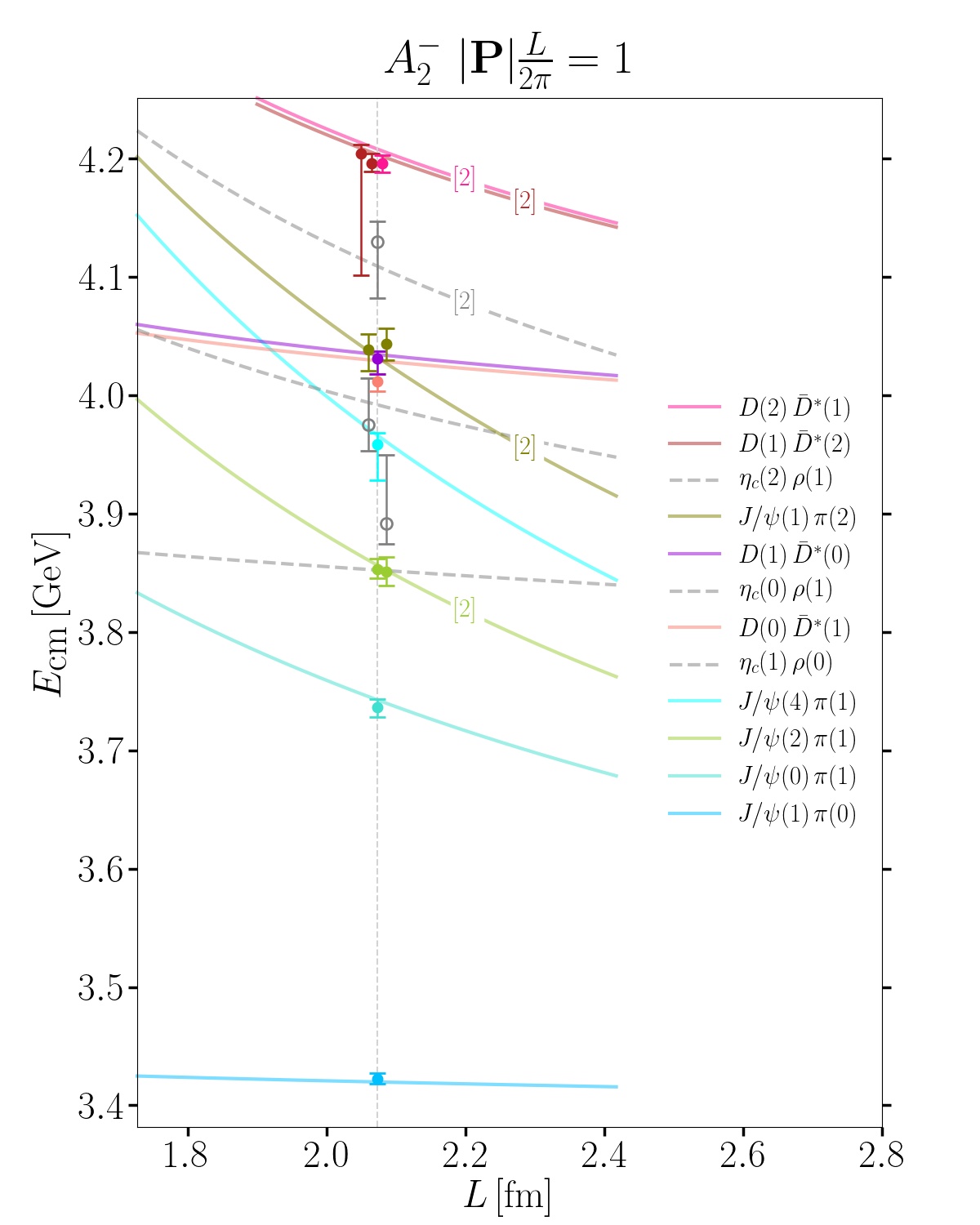}}
	\end{minipage}
\hspace*{1.2cm}
\begin{minipage}{0.45\linewidth}
\centerline{\includegraphics[width=1.2\textwidth]{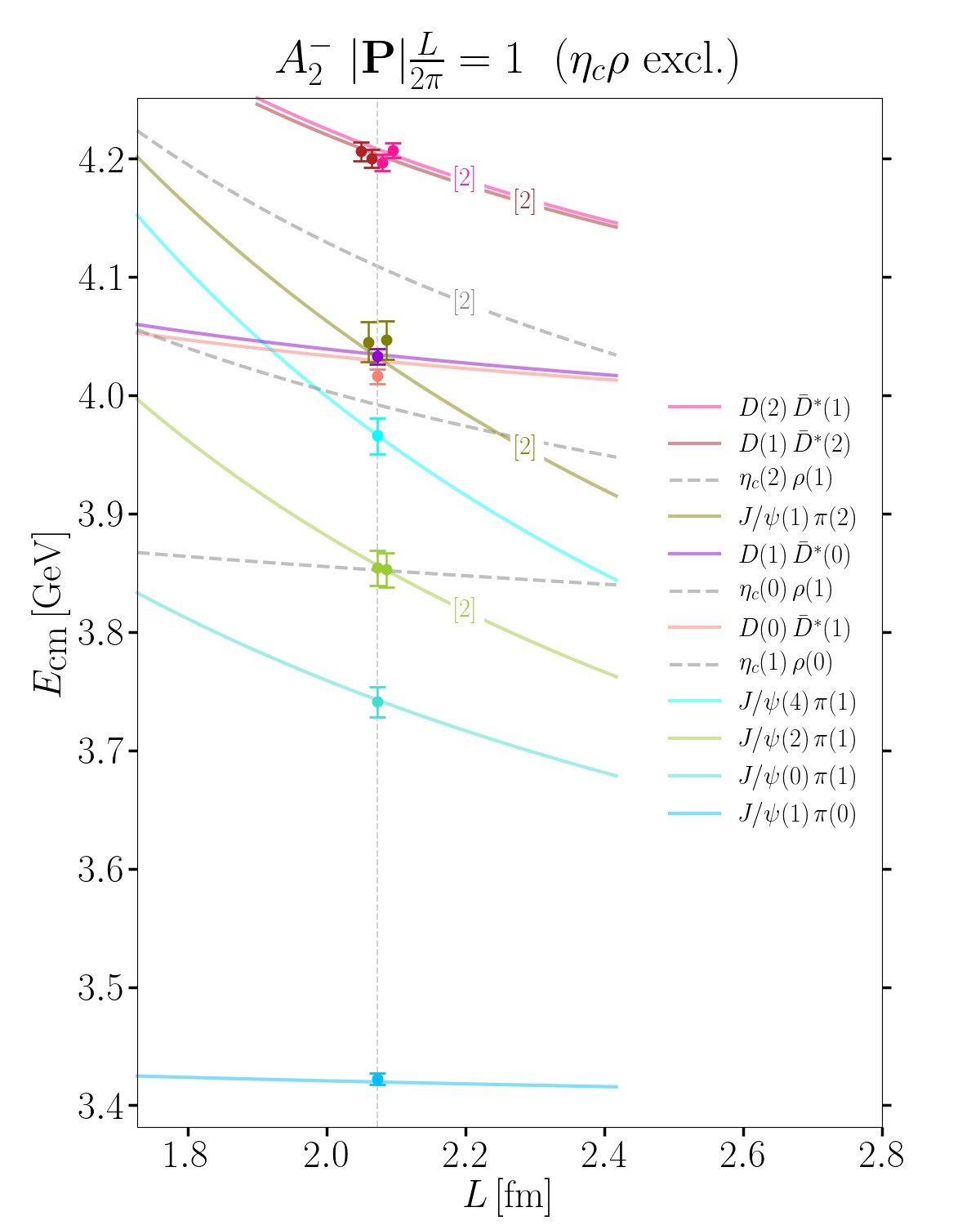}}
	\end{minipage}
\caption{Eigenenergies in the moving frame ($|{\bi P}|L/(2\pi)=1$) finite-volume irrep $A_2^{-}$ relevant for the quantum numbers $I(J^{PC})=1(1^{+-})$. The left and right panes represent the energy levels (points, $E_{n,\textrm{cm}}$) and noninteracting energies (lines, $E_{\textrm{ni,cm}}^{\textrm{con}}$) for the case where all interpolators are included and the $\eta_c\rho$ interpolators are excluded, respectively. Our calculations in this irrep are performed only on the smaller lattice (with $N_L=24$). For more information see the caption of Fig.~\ref{fig:T1pm}}
\label{fig:A2m}
\end{figure*}

\begin{figure*}[ht!]
\centering
\begin{minipage}{0.45\linewidth}
\centerline{\includegraphics[width=1.3\textwidth]{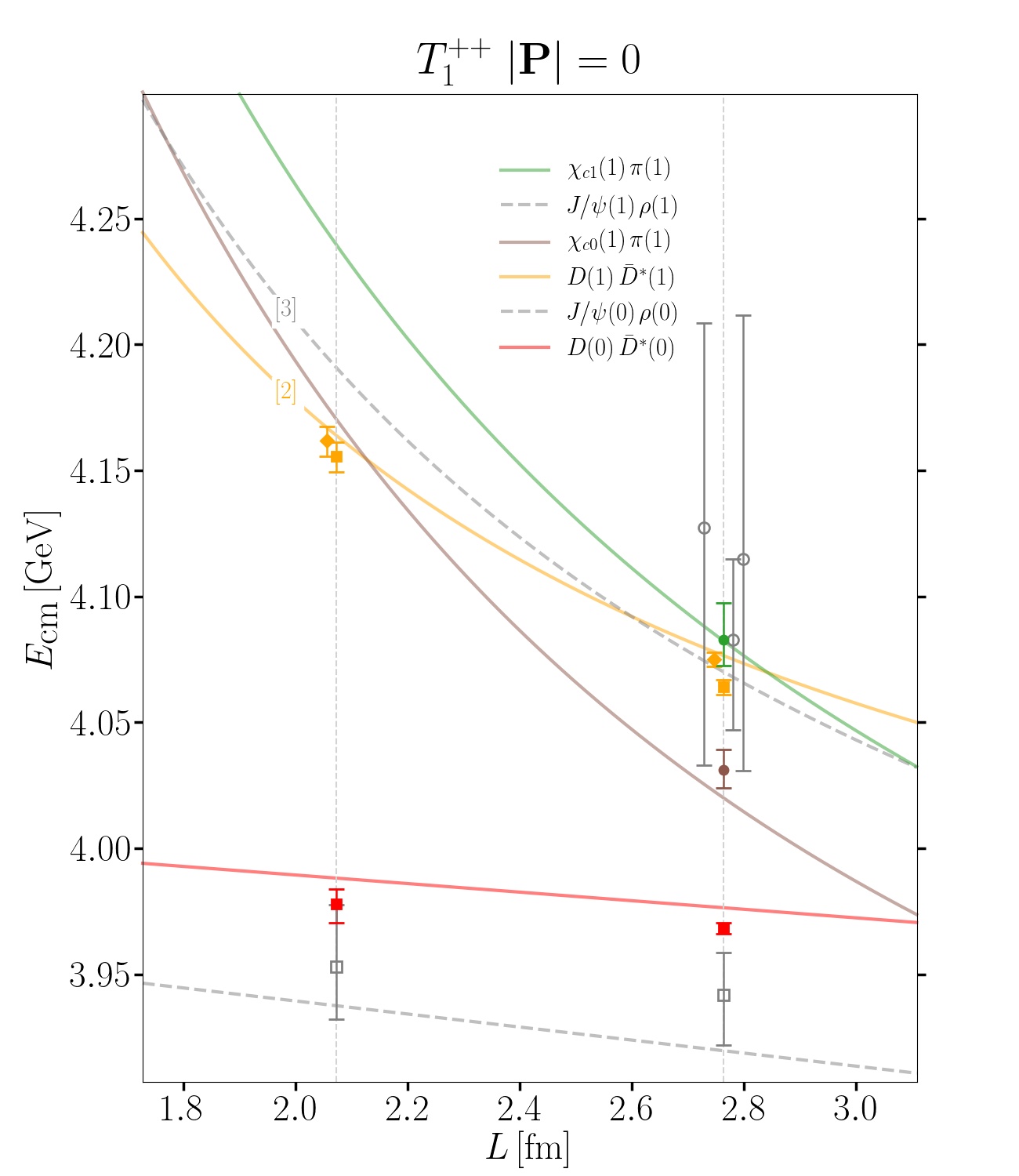}}
	\end{minipage}
\hspace*{1.5cm}
\begin{minipage}{0.45\linewidth}
\centerline{\includegraphics[width=1.3\textwidth]{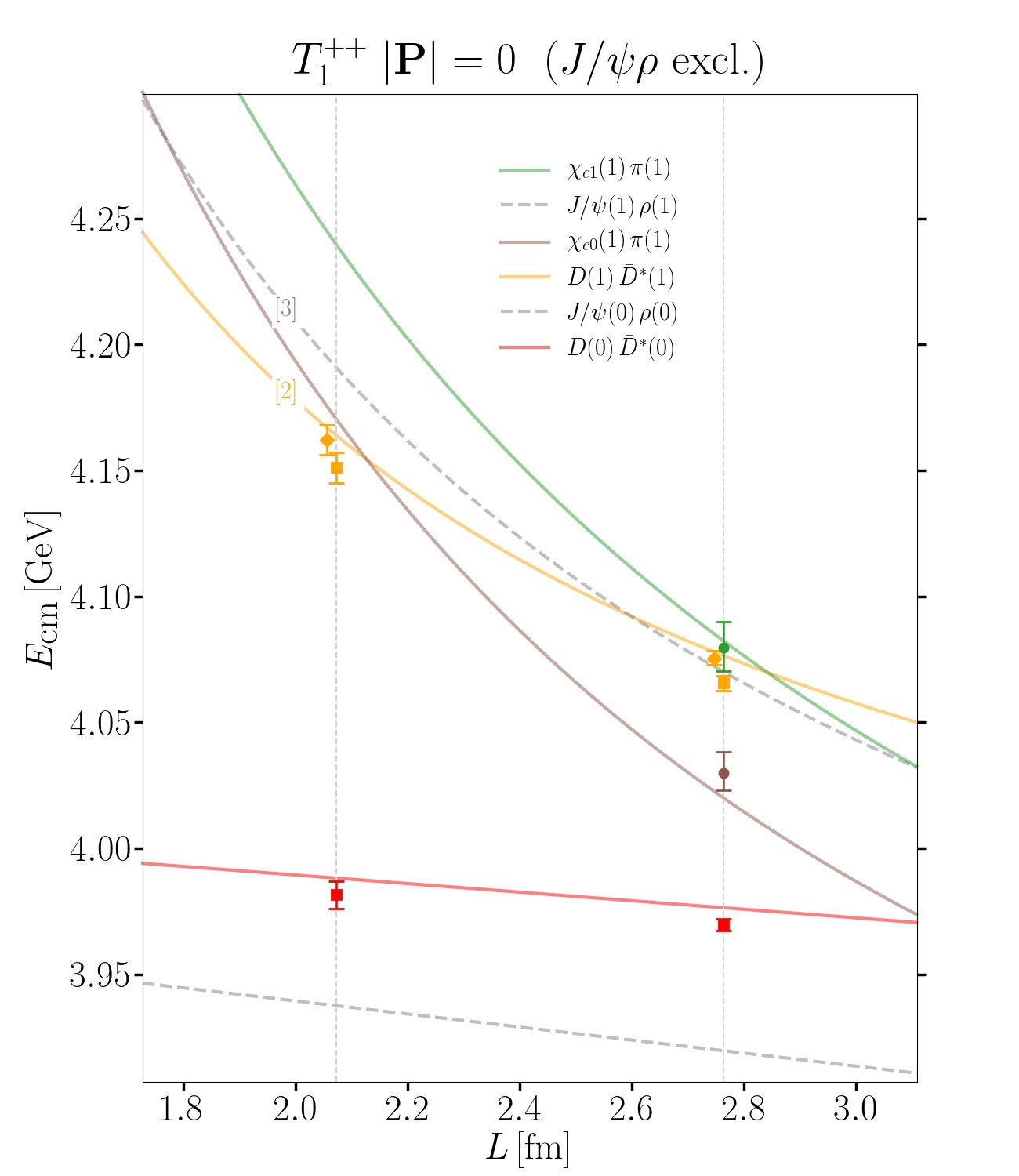}}
	\end{minipage}
\caption{Eigenenergies in the rest frame ($|{\bi P}|=0$) finite-volume irrep $T_1^{++}$ relevant for the quantum numbers $I(J^{PC})=1(1^{++})$. The left and right panes represent the energy levels (points, $E_{n,\textrm{cm}}$) and noninteracting energies (lines, $E_{\textrm{ni,cm}}^{\textrm{con}}$) for the case where all interpolators are included and the $J/\psi\rho$ interpolators are excluded, respectively. Empty gray symbols and dashed gray lines represent states coupled to $J/\psi \rho$ interpolators containing the $\rho$ resonance, which is treated as stable in our case. The noninteracting energies and thresholds of the neglected channels and the noninteracting energies of the lowest neglected interpolators with higher internal momenta are shown in Fig.~\ref{fig:omit_1++} of Appendix~\ref{sec:app:omitted_states}, together with the information on which levels are incorporated in the scattering analyses. The threshold of the three-particle channel $J/\psi\pi\pi$ that corresponds to $J/\psi\rho$ is lower than the lowest state shown here. 
For more information see caption of Fig.~\ref{fig:T1pm}.}
\label{fig:T1pp}
\end{figure*}

\begin{figure*}[ht!]
\centering
\begin{minipage}{0.45\linewidth}
\centerline{\includegraphics[width=1.2\textwidth]{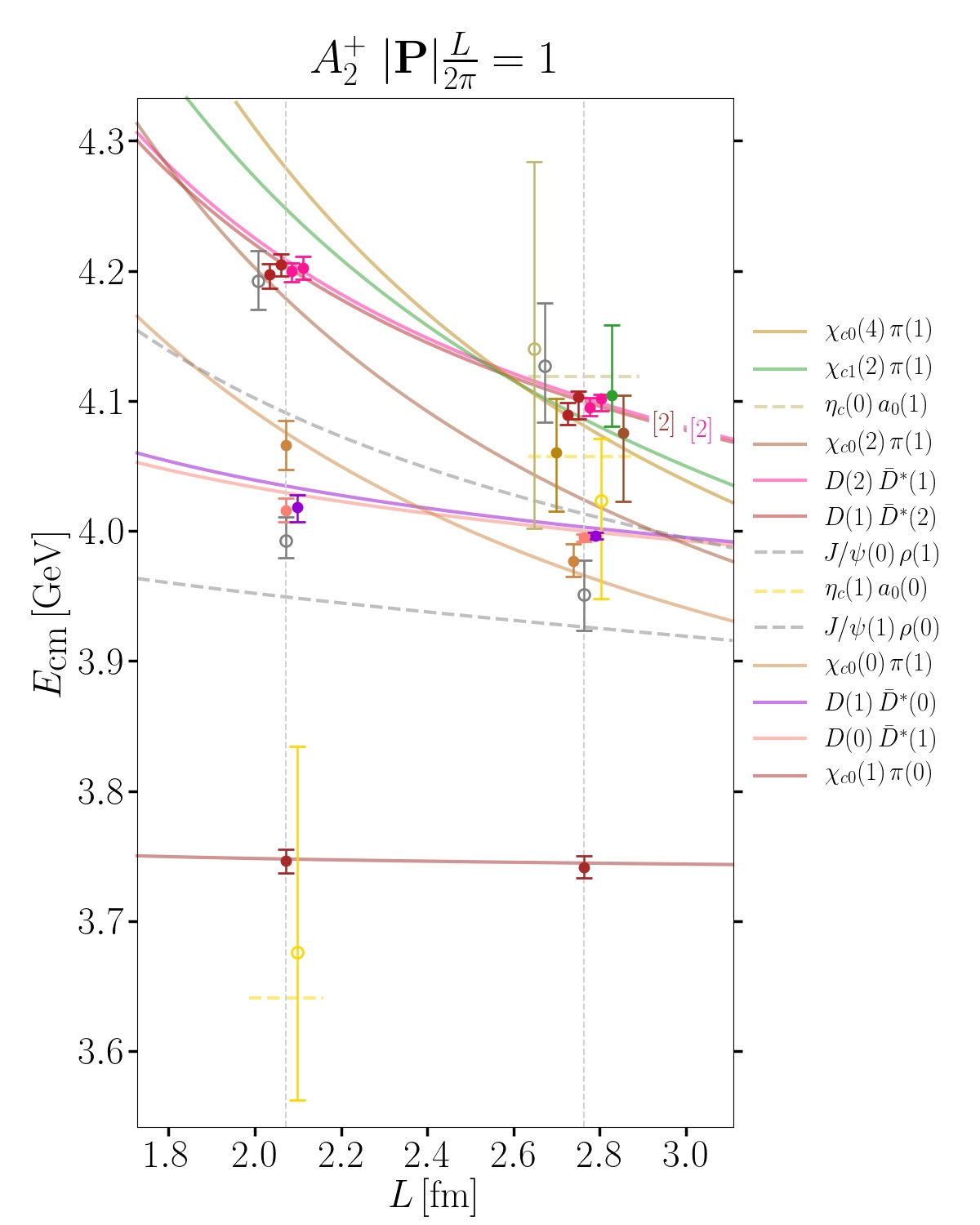}}
	\end{minipage}
\hspace*{1.2cm}
\begin{minipage}{0.45\linewidth}
\centerline{\includegraphics[width=1.2\textwidth]{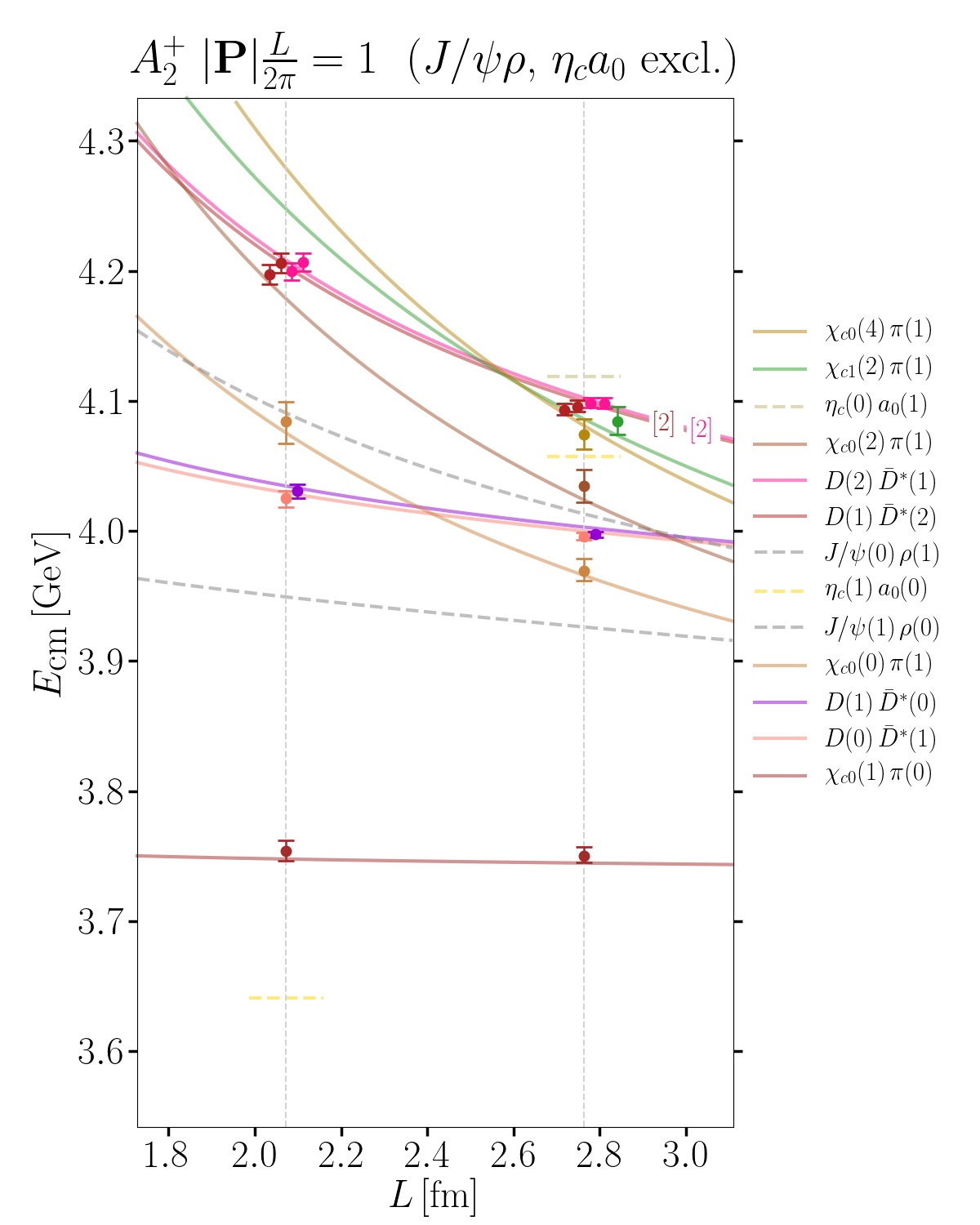}}
	\end{minipage}
\caption{Eigenenergies in the moving frame ($|{\bi P}|L/(2\pi)=1$) finite-volume irrep $A_2^{+}$ relevant for the quantum numbers $I(J^{PC})=1(1^{++})$. The left and right panes represent the energy levels (points, $E_{n,\textrm{cm}}$) and noninteracting energies (lines, $E_{\textrm{ni,cm}}^{\textrm{con}}$) for the case where all interpolators are included and the $J/\psi\rho$ and $\eta_c a_0$ interpolators are excluded, respectively. Empty symbols and dashed lines represent states coupled to $J/\psi \rho$ and $\eta_c a_0$ interpolators containing the $\rho$ and $a_0$ resonances, which in our case are treated as stable. As commented in the main text, we present the $\eta_c a_0$ noninteracting energies discontinuously since the ``masses'' of $a_0$ measured on both ensembles differ significantly. The noninteracting energies of the lowest neglected interpolators with higher internal momenta and the thresholds of the neglected channels (e.g., three-particle channel $J/\psi\pi\pi$) are shown in Fig.~\ref{fig:omit_1++} of Appendix~\ref{sec:app:omitted_states}.
For more information see the caption of Fig.~\ref{fig:T1pm}.}
\label{fig:A2p}
\end{figure*}

\begin{figure}[ht!]
\begin{center}
\includegraphics[width=.5\textwidth]{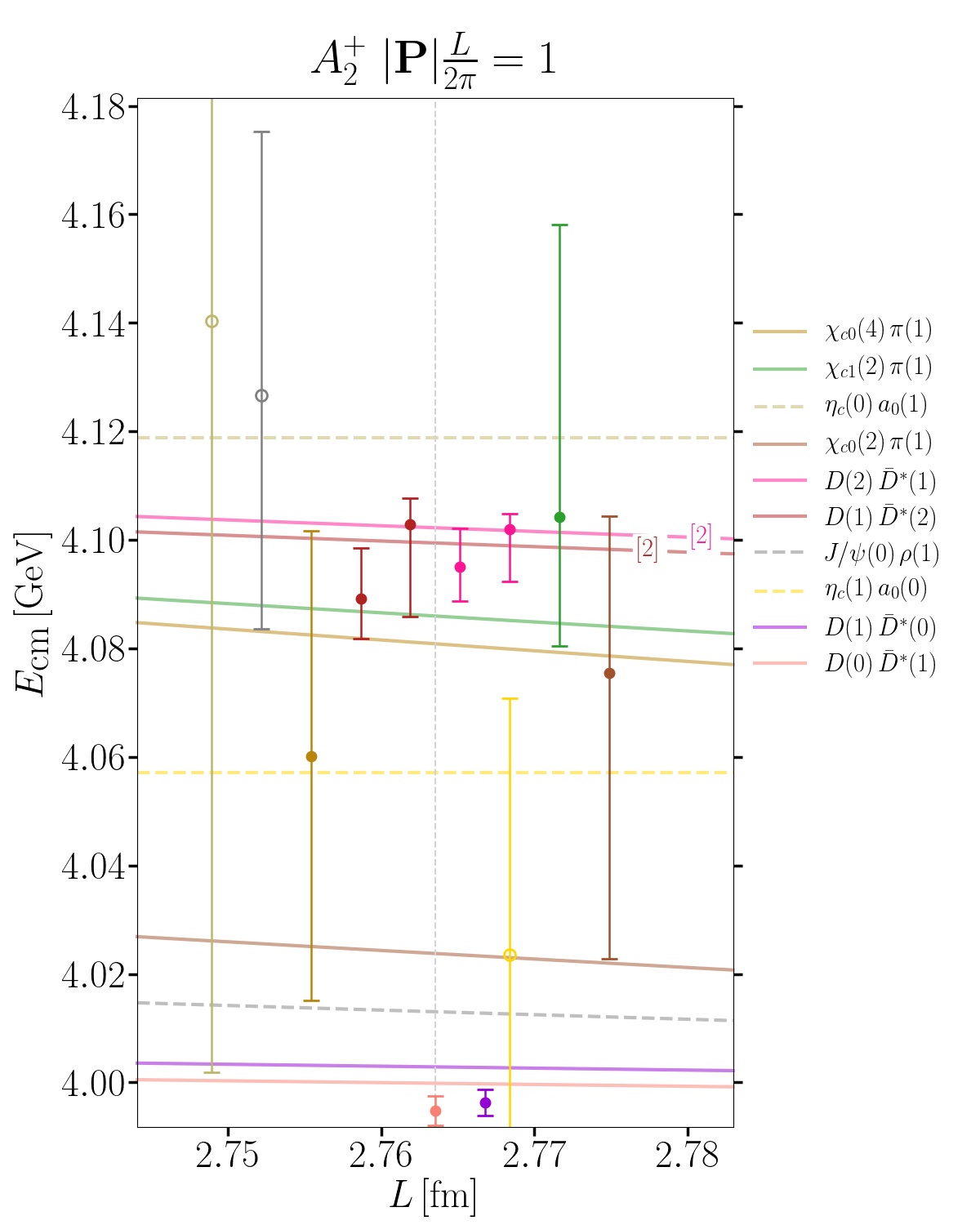}
\caption{A close-up of the densest region of the spectrum for irrep $A_2^+$ on the $N_L=32$ ensemble shown in the left-hand side plot in Fig.~\ref{fig:A2p}. Points are shifted horizontally for clarity.}
\label{fig:A2pCloseup}
\end{center}
\end{figure}

The partial-wave method and the projection method (see Ref. \cite{Prelovsek2017}) are employed for the construction of inter\-polators in the case of $|{\bi P}|=0$ and $|{\bi P}|=1\cdot2\pi/L$, respectively. 
Let us consider the noninteracting limit when the two hadrons do not interact.
In this case, two-particle states can exhibit energy degeneracy if at least one particle has a nonzero spin. This degeneracy depends on the spin and momentum of both particles. The multiplicity of noninteracting levels is determined by the number of possible linearly independent interpolators.
As an illustrative example, consider the case of $J/\psi\pi$ (vector-pseudoscalar) inter\-polators that transform according to a certain ``row'' of the 3D irrep $T_1^{+-}$. The total spin of a pseudoscalar-vector two-meson system is $S=1$. Considering $J^P=1^+,\ 3^+,\  \ldots\ $ that contribute to $T_1^{+-}$, one can have the partial waves $\ell=0,\ 2,\ \ldots\ $. When both particles are at rest, then $\ell=0$ (s wave) 
\begin{equation}
\begin{aligned}
O^{J/\psi(0)\pi(0)}_{S=1,\ell=0,J=1}&=\bar c\gamma_z c({\bi 0})\ \bar q\gamma_5 q({\bi 0})\>.
\label{EJpsipideg00}
\end{aligned}
\end{equation}
Considering $J/\psi(1)\pi(1)$, the $\ell=0$ as well as $\ell=2$ (d wave) appear and the partial-wave projected interpolators are \cite{Prelovsek2017}
\begin{align}
O^{J/\psi(1)\pi(1)}_{S=1,\ell=0,J=1}&=\bar c\gamma_z c(\hat{{\bi e}}_x)\,\bar q\gamma_5 q(-\hat{{\bi e}}_x)\,+\,\bar c\gamma_z c(-\hat{{\bi e}}_x)\,\bar q\gamma_5 q(\hat{{\bi e}}_x)\nonumber\\ 
&+\bar c\gamma_z c(\hat{{\bi e}}_y)\,\bar q\gamma_5 q(-\hat{{\bi e}}_y)\,+\,\bar c\gamma_z c(-\hat{{\bi e}}_y)\,\bar q\gamma_5 q(\hat{{\bi e}}_y)\nonumber\\
&+\bar c\gamma_z c(\hat{{\bi e}}_z)\,\bar q\gamma_5 q(-\hat{{\bi e}}_z)\,+\,\bar c\gamma_z c(-\hat{{\bi e}}_z)\,\bar q\gamma_5 q(\hat{{\bi e}}_z)\nonumber\\ 
O^{J/\psi(1)\pi(1)}_{S=1,\ell=2,J=1}&=\bar c\gamma_z c(\hat{{\bi e}}_x)\,\bar q\gamma_5 q(-\hat{{\bi e}}_x)\,+\,\bar c\gamma_z c(-\hat{{\bi e}}_x)\,\bar q\gamma_5 q(\hat{{\bi e}}_x)\nonumber\\ 
&+\bar c\gamma_z c(\hat{{\bi e}}_y)\,\bar q\gamma_5 q(-\hat{{\bi e}}_y)\,+\,\bar c\gamma_z c(-\hat{{\bi e}}_y)\,\bar q\gamma_5 q(\hat{{\bi e}}_y)\nonumber\\ 
&-2\,\bar c\gamma_z c(\hat{{\bi e}}_z)\,\bar q\gamma_5 q(-\hat{{\bi e}}_z)\nonumber\\
&-2\,\bar c\gamma_z c(-\hat{{\bi e}}_z)\,\bar q\gamma_5 q(\hat{{\bi e}}_z)\>.
\label{EJpsipideg}
\end{align}

Most of our interpolators consist of pseudo\-scalar (P) and vector (V) mesons. As shown in Eq.~\eqref{EJpsipideg}, the case $P(1)V(1)$ exhibits a multiplicity of two linearly independent interpolators, $O^{P(1)V(1)}_{S=1,\ell=0,J=1}$ and $O^{P(1)V(1)}_{S=1,\ell=2,J=1}$. A similar degeneracy appears when considering $P(3)V(3)$ interpolators, $O^{P(3)V(3)}_{S=1,\ell=0,J=1}$ and $O^{P(3)V(3)}_{S=1,\ell=2,J=1}$. Furthermore, there are three linearly independent $P(2)V(2)$ interpolators: $O^{P(2)V(2)}_{S=1,\ell=0,J=1}$, $O^{P(2)V(2)}_{S=1,\ell=2,J=1}$ and $O^{P(2)V(2)}_{S=1,\ell=2,J=3}$. The $P(1)V(2)$, $P(2)V(1)$ and $V(1)V(1)$ energy levels have degeneracies of 2, 2 and 3, respectively.

Our objective is to extract all these eigenstates, including those that are degenerate in the noninteracting limit. Therefore, for every given inter\-polator type, we implement all its linearly independent combinations.

\section{Extraction of finite-volume energy levels}\label{sec:technical}

On a periodic lattice with spatial size $L=N_La$, a two-hadron system exhibits discrete energies in noninteracting (ni) limit
\begin{equation}
E_{\textrm{ni}}^{\textrm{lat}}=E_{H_i({\bi p}_i)}^{\textrm{lat}}+E_{H_j({\bi p}_j)}^{\textrm{lat}}\>.
\label{EEnilat}
\end{equation}
In the continuum limit, the noninteracting energies follow the relativistic dispersion relation, and
\begin{equation}
E_{\textrm{ni}}^{\textrm{con}}=E_{H_i({\bi p}_i)}^{\textrm{con}}+E_{H_j({\bi p}_j)}^{\textrm{con}}\>,\ E_{H_i({\bi p}_i)}^{\textrm{con}}=\sqrt{m^2_{H_i}+\left|{\bi p}_i\right|^2}\>.
\label{EEnicon}
\end{equation}
Throughout this article, these energies are presented in the cm-frame,
\begin{equation}
E_{\textrm{ni,cm}}^{\textrm{con}}=\sqrt{\left(E_{\textrm{ni}}^{\textrm{con}}\right)^2-\left|{\bi P}\right|^2}\>.
\label{EEniconCM}
\end{equation}

In the interacting theory, finite-volume eigenenergies $E_n^{\textrm{lat}}$ are extracted from {\it ab-initio} lattice simulations. Specifically, these energies are obtained through single-exponential fits to the eigenvalues $\lambda^{(n)}(t,t_0)\propto e^{-E_n^{\textrm{lat}}(t-t_0)}$ of the generalized eigenvalue problem (GEVP)
\begin{equation}
C(t){\bi u}^{(n)}(t)=\lambda^{(n)}(t,t_0)C(t_0){\bi u}^{(n)}(t)\>,
\label{EGEVP}
\end{equation}
where $t_0/a\in\{2,3\}$ have been used \cite{Michael:1985ne}. Additional details about our specific implementation can be found in Appendixes \ref{sec:app:overlaps} and \ref{sec:app:En}. 

We emphasize that the finite-volume spectra are determined at a single lattice spacing. Consequently, we cannot quantify the uncertainty arising from lattice discretization effects, which may lead to the energy levels differing from continuum expectations. These lattice artifacts are more significant for charm quarks than light quarks since $am_c$ is not small for our lattices. The measured single-hadron energies $E_{H_i({\bi p}_i)}^{\textrm{lat}}$ deviate slightly from their continuum counterparts $E_{H_i({\bi p}_i)}^{\textrm{con}}$, especially for the smaller volume where the unit of lattice momentum is larger. To mitigate these effects, we first evaluate the energy shift of each interacting eigenstate with respect to the noninteracting state 
\begin{equation}
\Delta E_n=E_n^{\textrm{lat}}-E_{\textrm{ni}}^{\textrm{lat}}\>.
\label{EdeltaEn}
\end{equation}
Subsequently, we determine corrected finite-volume eigen\-energies using Eq.~\eqref{EEnicon}
 \begin{equation}
E_n=\Delta E_n+E_{\textrm{ni}}^{\textrm{con}}
\label{EEn}
\end{equation}
and present them in the cm-frame
\begin{equation}
E_{n,\textrm{cm}}=\sqrt{E_n^2-\left|{\bi P}\right|^2}\>.
\label{EEncm}
\end{equation}
Note that the energies $E_n$ are, by construction, equal to $E_n^{\textrm{lat}}$ in the continuum limit ($a\rightarrow 0$).

This procedure for extracting $E_n$ in Eq.~\eqref{EEn} requires a clear determination of the relevant noninteracting level for the eigenstate $n$.
A possible incorrect identification could modify $E_n$; however, the resulting differences typically remain within the presented uncertainties, as demonstrated in an example in Appendix \ref{sec:app:states}.

\section{Main results: finite-volume eigenenergies}\label{sec:results}

The interacting energy levels evaluated using Eq.~\eqref{EEncm} are depicted in Figs.~\ref{fig:T1pm}, \ref{fig:A2m} for $J^{PC}=1^{+-}$ and Figs.~\ref{fig:T1pp}, \ref{fig:A2p} for $1^{++}$. These plots also include the noninteracting energies determined using Eq.~\eqref{EEniconCM} presented as lines. Numerous eigenstates feature in our spectra, which become increasingly dense for nonzero total momentum and for the larger spatial volume. In particular, many energy levels reside in the region above the $D(1)\bar D^*(0)$/$D(0)\bar D^*(1)$ energies in the irrep $A_2^+$ on the $N_L=32$ lattice and, for clarity, the spectrum for this reduced energy region is presented in Fig.~\ref{fig:A2pCloseup}.

We use numbers inside square brackets in the figures to indicate the degeneracy of energy levels in the noninteracting limit. The origin of this degeneracy is discussed in Sec.~\ref{sec:interpolators}. Each linearly independent interpolator yields a corresponding finite-volume interacting eigenenergy.
Furthermore, the states we extract typically couple well to interpolators constructed using the partial-wave method.

Although the light mesons $\rho$ and $a_0$ are resonances, in our analysis, we treat them as stable under the strong inter\-action. Note that a reliable extraction of their eigenenergies would necessitate the inclusion of three-particle operators, which is beyond the scope of the present work. As a result, our analysis does not provide complete spectra, i.e., our $\eta_c\rho$ operator is not expected to generate all relevant $\eta_c\pi\pi$ eigenstates. 
However, our findings offer insights into whether the energies of these ``light mesons''  are modified in the presence of charmonium. In Figs.~\ref{fig:T1pm}--\ref{fig:A2p} (as indicated by the empty symbols), we observe significant uncertainties in the eigenenergies associated with interpolators containing $\rho$ and $a_0$.\footnote{We refrain from representing the noninteracting energies related to the $a_0$ meson as continuous lines. A significant difference exists between $m_{a_0}$ extracted on the $N_L=24$ and $N_L=32$ ensembles.
The unexpectedly low value for $m_{a_0}$ on the smaller volume likely represents the sum of the masses of the $\pi$ and $\eta$ mesons, which are allowed strong decay products of the $a_0$.}

The lines representing the noninteracting energies where both mesons are at rest would appear horizontal if the meson masses were independent of $L$. This is not valid exactly in our simulation; therefore, the lines for noninteracting energies are interpolated linearly between the $N_L=24$ and $32$ values in the figures.

In the $J^{PC}=1^{+-}$ sector, states coupled to $J/\psi \pi$ (apart from the highest levels) align with the noninteracting expectations, suggesting that interaction between $J/\psi $ and $\pi$ is very small. This topic is covered in Sec.~\ref{sec:jpsipi}, where an upper bound on the scattering length of elastic s-wave $J/\psi\pi$ scattering is estimated.

Our main objective is to find out whether there is any attraction\footnote{An attraction manifests as an energy level lying significantly below the free energy.} between $D$ and $\bar D^*$ for both the $1^{+-}$ and $1^{++}$ channels, specifically when employing the full correlation matrix (as shown in the left-hand side plots of Figs. \ref{fig:T1pm}--\ref{fig:A2p}). The states that couple predominantly to $D(0)\bar D^*(0)$, s-wave $D(1)\bar D^*(1)$, $D(0)\bar D^*(1)$ and $D(1)\bar D^*(0)$ exhibit slightly negative energy shifts. Exceptions are the states in the $1^{+-}$ channel on the $N_L=24$ ensemble in the $T_1^{+-}$ irrep that couple strongly to $D(0)\bar D^*(0)$ and s-wave $D(1)\bar D^*(1)$, as well as $A_2^-$ levels that couple to $D(1)\bar D^*(0)$.

To qualitatively check how strongly the $J/\psi \pi$ and $D \bar D^*$ channels couple to $\eta_c \rho$ for $1^{+-}$, we omit the latter set of interpolators and investigate the differences in the resulting spectrum. Energy levels are extracted from correlation matrices that are submatrices of the original ones and do not contain $\eta_c \rho$ inter\-polators. These levels are presented on the right-hand side in Figs.~\ref{fig:T1pm} and \ref{fig:A2m} and the corresponding $\eta_c \rho$ energy levels do not feature in these spectra. Apart from that, the spectra from the left- and right-hand side plots agree within uncertainties, except for the s-wave $D(1)\bar D^*(1)$ state (for the $N_L=32$ ensemble), which exhibits a greater degree of attraction when the $\eta_c \rho$ interpolators are included.

A similar procedure is carried out for $1^{++}$, where eigenenergies are extracted from submatrices that do not contain $J/\psi \rho$ and $\eta_c a_0$ inter\-polators, whereas the $D \bar D^*$ inter\-polators are still present. To ascertain the impact of the aforementioned inter\-polators, compare the left- and right-hand sides in Figs.~\ref{fig:T1pp} and \ref{fig:A2p}. The conclusion is similar to that for the $1^{+-}$ sector; the spectra agree within uncertainties.

The meson-meson operators used in this work are limited to those involving local single meson structures, which are insufficient to access the $\psi(2S)\pi$ and $\psi(3770)\pi$ channels for $1^{+-}$; therefore, our present study neglects these channels. In our previous study, including these two channels using nonlocal single-meson interpolators had an insignificant effect on the rest of the finite-volume spectrum \cite{Zc_Sasa_2}. Note that $Z_c$ has not been seen in the decay to either of these.

\section{Approximation of one-channel s-wave $D\bar{D}^*$ scattering}\label{sec:scattering}

We aim to calculate hadron scattering amplitudes using lattice QCD and examine their singularity content. 
The former can be done using the Lüscher finite-volume formalism and its extensions \cite{Luscher1986,Luscher:1990ux,Luscher1991,PhysRevD.89.074507} (see Ref. \cite{particle_scattering_review_1} for a review), which relate finite-volume energies with the infinite-volume scattering amplitudes.
In this section, we assume that the $D\bar{D}^*$ channel is decoupled and concentrate on single-channel $D\bar{D}^*$ scattering in partial wave $\ell=0$ near the threshold for both $J^{PC}=1^{+-}$ and $1^{++}$. Our goal is to check whether a charmoniumlike isospin 1 four-quark state with $J^P=1^{+}$ close to $D\bar{D}^*$ threshold exists if the coupling of $D\bar{D}^*$ to other channels is negligible. Note that this strategy cannot determine whether the cross-channel interactions substantially affect an exotic four-quark state. Possible effects from the left-hand cut (lhc) arising from one-pion exchange are omitted in our analysis, as detailed at the end of this section.

For partial wave $\ell$, the scattering amplitude $T_{\ell}$ is defined through the $S$ matrix, 
$S_{\ell}=e^{2i\delta_{\ell}}=1-i\frac{k}{4\pi E_{\mathrm{cm}}}T_{\ell}$, where
\begin{equation}
T_{\ell}=-\frac{8\pi E_{\mathrm{cm}}}{k\cot{(\delta_{\ell})}-i k}\>,
\label{Et}
\end{equation}
$\delta_{\ell}$ is the phase shift and $k$ is the magnitude of the spatial momentum in the cm-frame, derived from $E_{\textrm{cm}}=\sqrt{m_D^2+k^2}+\sqrt{m_{D^*}^2+k^2}$.
The total angular momentum is $J=1$, and the partial wave is $\ell=0$.\footnote{All energy levels significantly overlapping with d-wave $D\bar D^*$ operators align consistently with their corresponding noninteracting energies. Contributions from partial waves with $\ell>0$ near the $D\bar D^*$ threshold---our region of interest---are effectively suppressed by the phase space factor $k^{2\ell}$. Therefore, we assume $T_{\ell\ge 2}=0$ and consider the mixing between $\ell= 2$ and $\ell= 0$ in $J = 1$ negligible.}
Utilizing Lüscher relation 
\cite{PhysRevD.89.074507}, we evaluate
\begin{equation}
k\cot{\delta_{0}(k)}=\frac{2\mathcal{Z}^{{\bi d}}_{00}\left(1;(k\frac{L}{2\pi})^2\right)}{\pi^{1/2}\gamma L}\>,
\label{ELuscher}
\end{equation}
where $\mathcal{Z}^{{\bi d}}_{00}$ is the Lüscher zeta function \cite{Luscher:1990ux}, and ${\bi d}={\bi P}\frac{L}{2\pi}$ is the normalized total momentum.
Finally, the following parametrization is chosen for energies near the threshold:
\begin{equation}
k\cot{\delta_{0}}=\frac{1}{a_{0}}+\frac{r_{0}}{2}k^2\>,
\label{EkcotL0}
\end{equation}
which is an effective range expansion in $k^2$ up to the leading two terms. The parameters $a_{0}$ and $r_{0}$ are determined such that the relation Eq.~\eqref{EkcotL0} is optimized simultaneously for all the considered $D\bar D^*$ energy levels. One-channel scattering is analyzed using a correlated fit with the determinant residual method described in \cite{MORNINGSTAR2017477}, while additional details on the implementation can be found in Appendix C of Ref. \cite{Prelovsek:2020eiw}.

\begin{figure*}[ht!]
\centering
\begin{minipage}{0.45\linewidth}
\centerline{\includegraphics[width=1.2\textwidth]{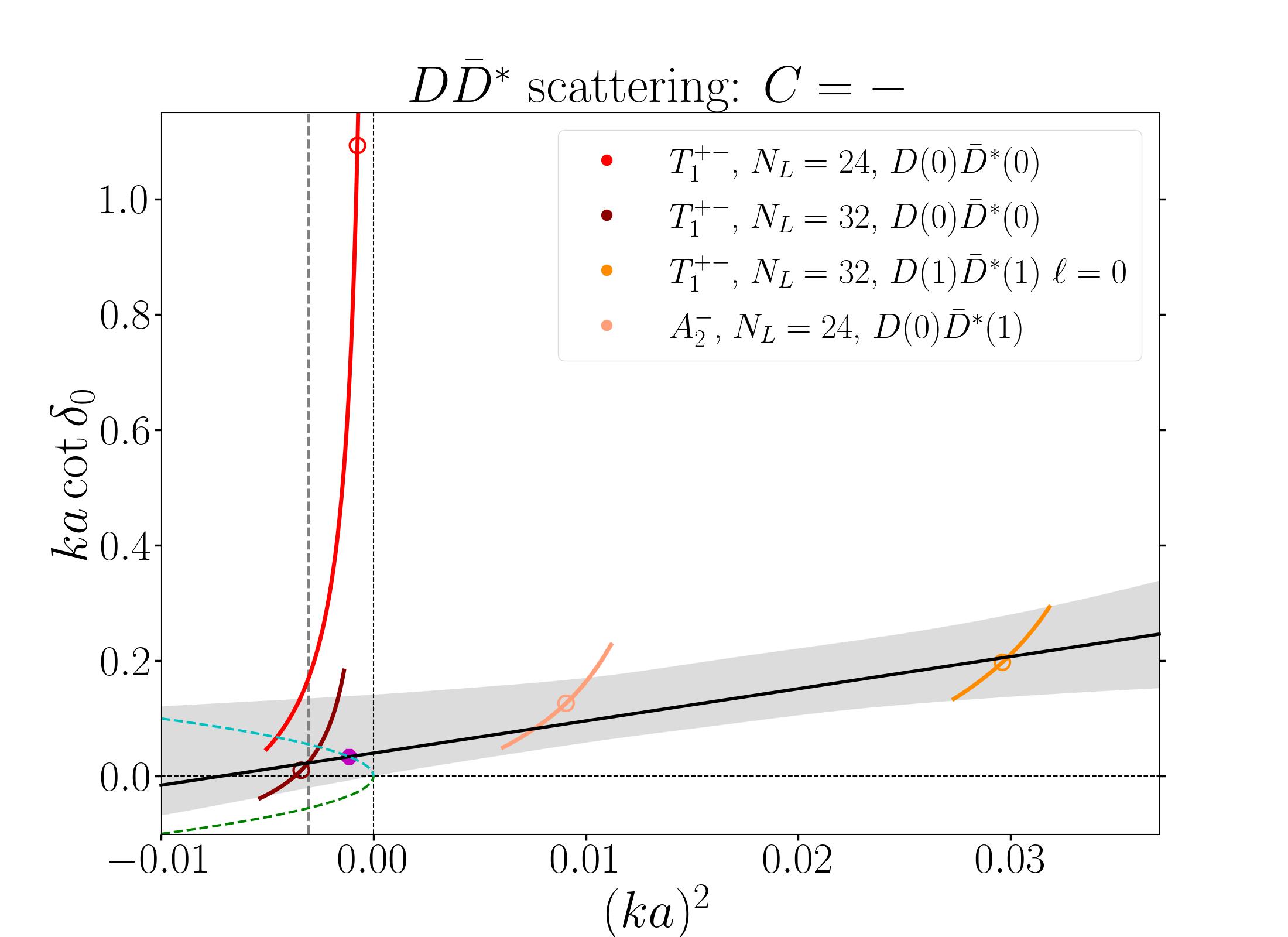}}
	\end{minipage}
\hspace*{1.cm}
\begin{minipage}{0.45\linewidth}
\centerline{\includegraphics[width=1.2\textwidth]{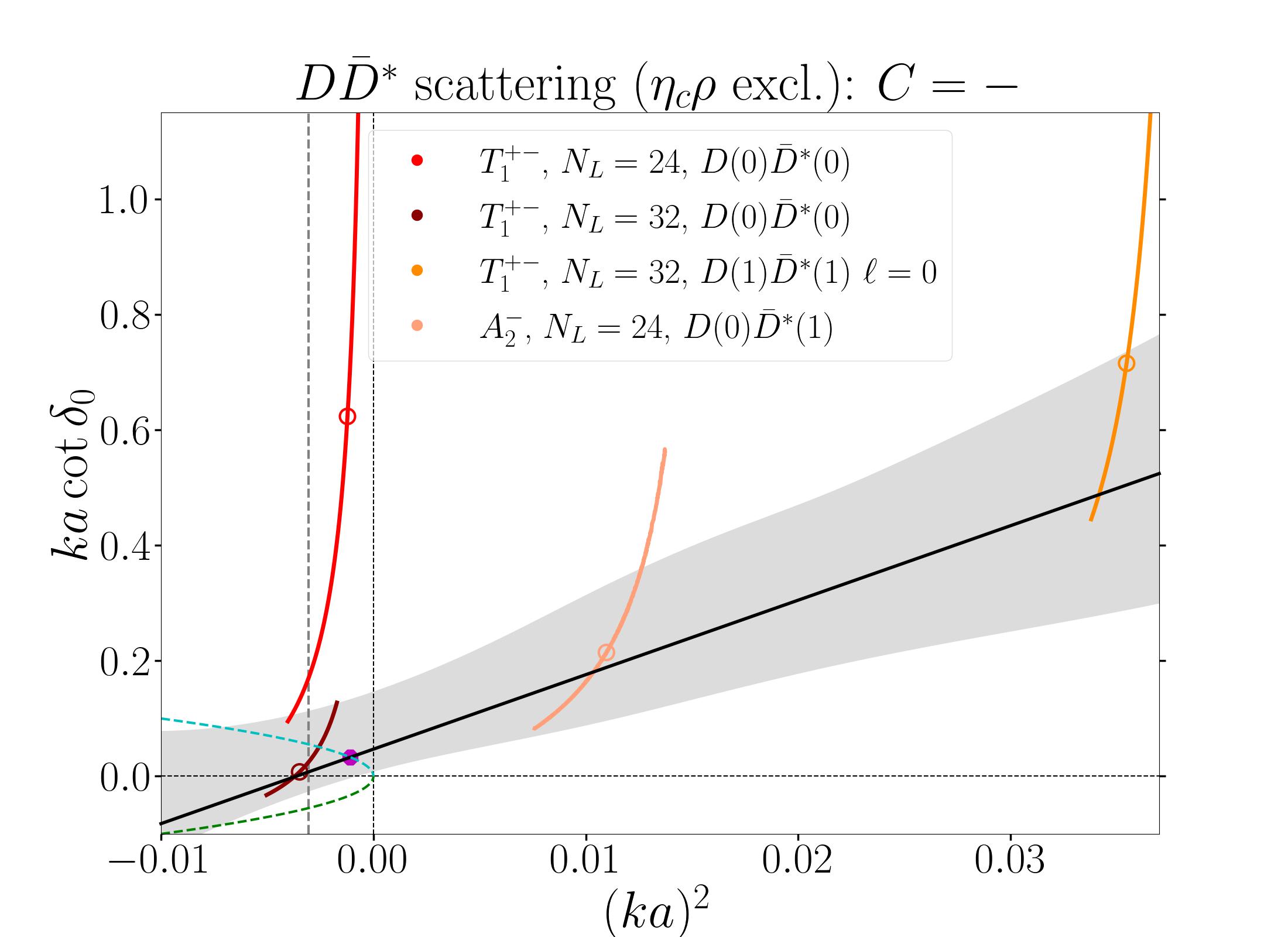}}
	\end{minipage}
\caption{$D\bar{D}^*$ scattering in partial wave $\ell = 0$ in the one-channel approximation with $J^{PC}=1^{+-}$. All interpolators were included in the variational analysis on the left, while the $\eta_c\rho$ interpolators were excluded on the right. Finite-volume energy levels incorporated into the left and right fit (see the legend) can be found in Figs.~\ref{fig:T1pm} and \ref{fig:A2m} in the left and right plots, respectively. The black solid line displays $ka\cot{\delta_{0}}$ [see Eq.~\eqref{EkcotL0}] evaluated using the best fit parameters. The magenta octagon at the intersection of the black and dashed cyan line (where $ik = +|k|$) represents the presence of a virtual state pole. The gray vertical dashed lines indicate the position of the branch point of the left-hand cut.}
\label{fig:fit1pm}
\end{figure*}

\begin{figure*}[ht!]
\centering
\begin{minipage}{0.45\linewidth}
\centerline{\includegraphics[width=1.2\textwidth]{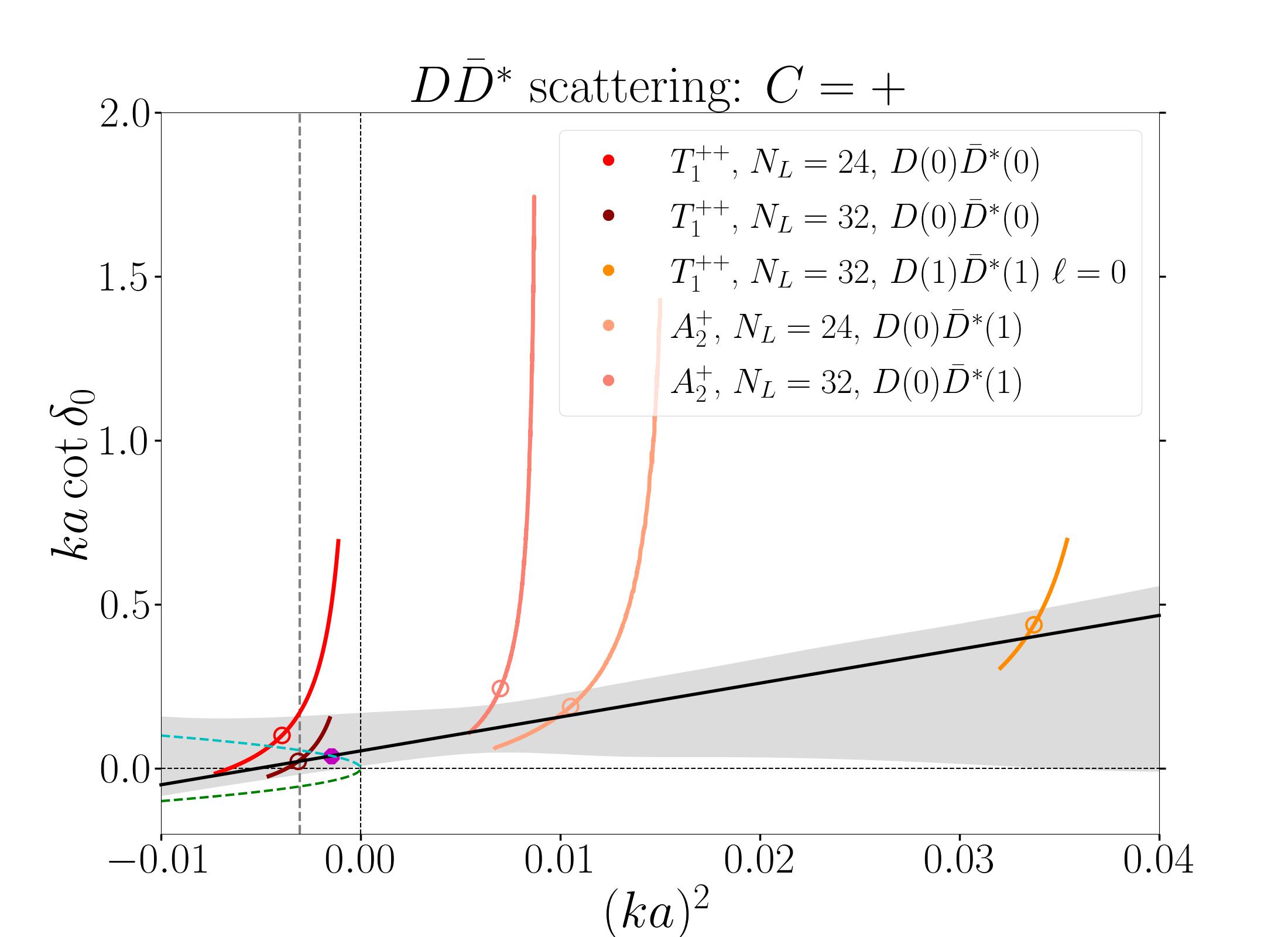}}
	\end{minipage}
\hspace*{1.0cm}
\begin{minipage}{0.45\linewidth}
\centerline{\includegraphics[width=1.2\textwidth]{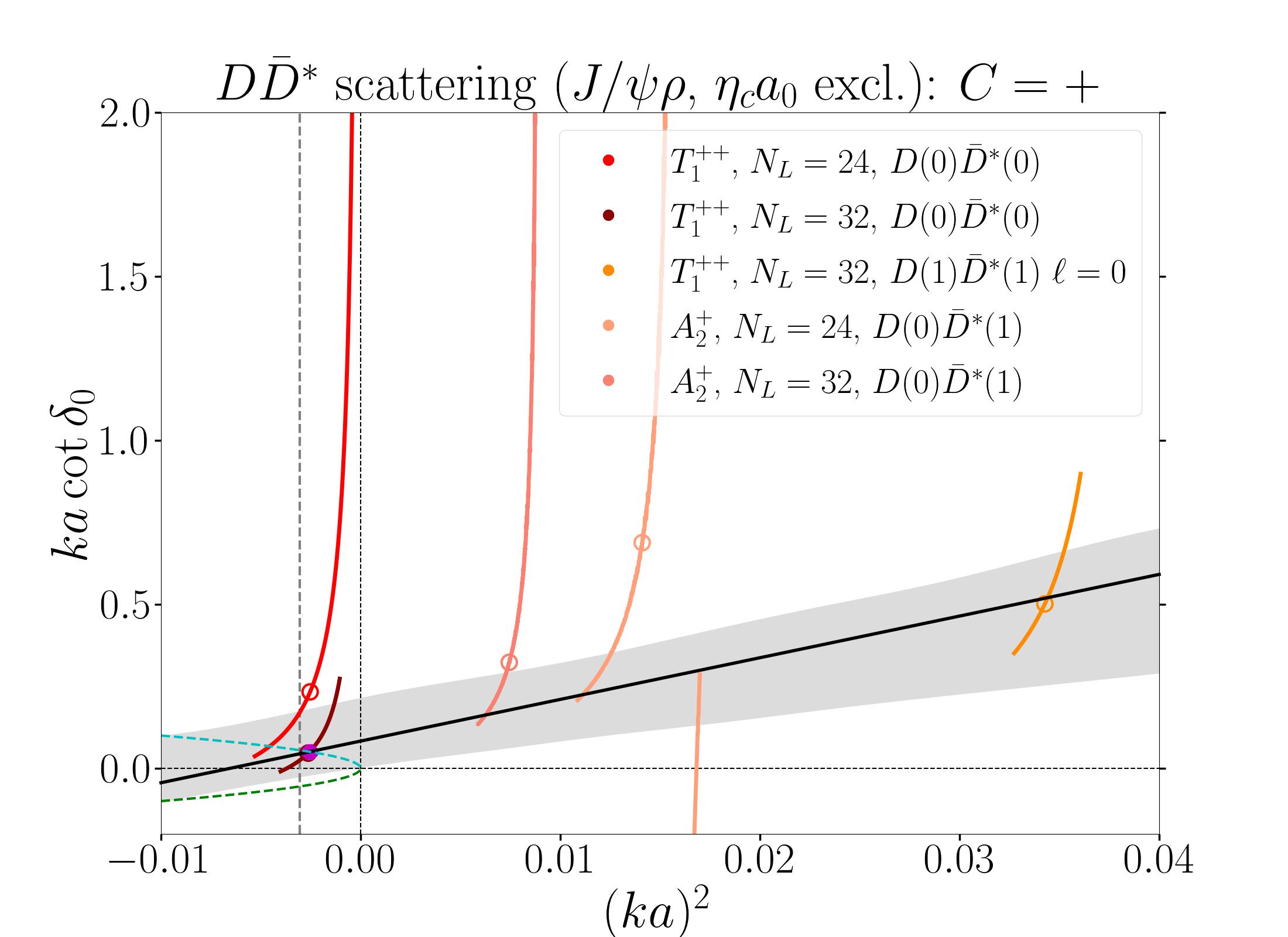}}
	\end{minipage}
\caption{$D\bar{D}^*$ scattering in partial wave $\ell = 0$ in the one-channel approximation with $J^{PC}=1^{++}$. All interpolators were included in the variational analysis on the left, while the $J/\psi\rho$ and $\eta_c a_0$ were excluded on the right. Finite-volume energy levels incorporated into the left and right fit (see the legend) can be found in Figs.~\ref{fig:T1pp} and \ref{fig:A2p} in the left and right plots, respectively. For more information see the caption of Fig.~\ref{fig:fit1pm}.}
\label{fig:fit1pp}
\end{figure*}

We present two cases for each $J^{PC}=$ $1^{+-}$ and $1^{++}$: one where the full correlation matrices are employed and the other where the meson-meson interpolators with $\rho$ and/or $a_0$ mesons are omitted from the basis. Therefore, four different fits are performed. The resulting scattering lengths $a_0$ and the effective ranges $r_0$ are positive, as presented in Table~\ref{tab:param:DDscatt} and Figs.~\ref{fig:fit1pm} and \ref{fig:fit1pp}. Four and five $D\bar D^*$ energy levels from different irreps are incorporated in the fits for $1^{+-}$ and $1^{++}$, respectively, as denoted in the legends of Figs. \ref{fig:fit1pm} and \ref{fig:fit1pp} and by black outlined symbols in (top) left panes of Figs.~\ref{fig:omit_1+-}--\ref{fig:omit_A2+}. We have verified that including/excluding additional energy levels has a negligible effect on the resulting fit parameters, which remain within the given uncertainty range. Our results are also robust under the change in parametrizations $k\cot{\delta_{0}}=1/a_{0}$, $k\cot{\delta_{0}}/E_{\textrm{cm}}=A+Bs$ and $k\cot{\delta_{0}}/E_{\textrm{cm}}=A+Bs+Cs^2$.

\begin{table}[ht!]
\caption{Table with parameters of four different fits to the $D\bar D^*$ scattering amplitude in the one-channel approximation; the scattering length, effective range and pole energy of the virtual state $\Delta m_{\textrm{V}} =E_{\textrm{cm}}^{\textrm{p}}-m_D-m_{D^*}$. Two fits for each $J^{PC}=1^{+-}$ and $1^{++}$ are performed, one from the analysis with full correlation matrices and the other one with certain interpolators excluded, as discussed in the text.}
\label{tab:param:DDscatt}
\bgroup
\def\arraystretch{1.3}
\begin{tabular}{cccccc} \hline \hline
$J^{PC}$ & Interpolators & \begin{tabular}[c]{@{}c@{}}$1/a_0\,[\textrm{fm}^{-1}]$\\\end{tabular} & $r_0\,[\textrm{fm}]$ & $\chi^2/N_{\textrm{dof}}$ & $\Delta m_{\textrm{V}}\,[\textrm{MeV}]$ \\ \hline
\multirow{2}{*}{$1^{+-}$} & all & $0.46^{+1.16}_{-0.45}$ & $0.96^{+0.43}_{-0.73}$ & $0.13$ & $-3.0^{+3.0}_{-31.1}$  \\
 & $\eta_c\rho\ \textrm{excl.}$ & $0.54^{+1.07}_{-0.44}$ & $2.23^{+0.95}_{-1.08}$ & $0.24$ & $-2.8^{+2.6}_{-17.1}$ \\\hline
\multirow{2}{*}{$1^{++}$} & all & $0.62^{+1.30}_{-0.51}$ & $1.78^{+0.25}_{-2.44}$ & $0.18$ & $-3.8^{+3.6}_{\footnote{Uncertainty is so large that it is unbounded from below.}}$ \\
 & $J/\psi\rho,\,\eta_c a_0\ \textrm{excl.}$ & $0.96^{+1.42}_{-0.91}$ & $2.19^{+0.36}_{-1.00}$ & $0.15$ & $-6.7^{+6.7}_{-19.5}$ \\ \hline\hline
\end{tabular}
\egroup
\end{table}

\begin{figure*}[ht!]
\centering
\begin{minipage}{0.45\linewidth}
\centerline{\includegraphics[width=1.2\textwidth]{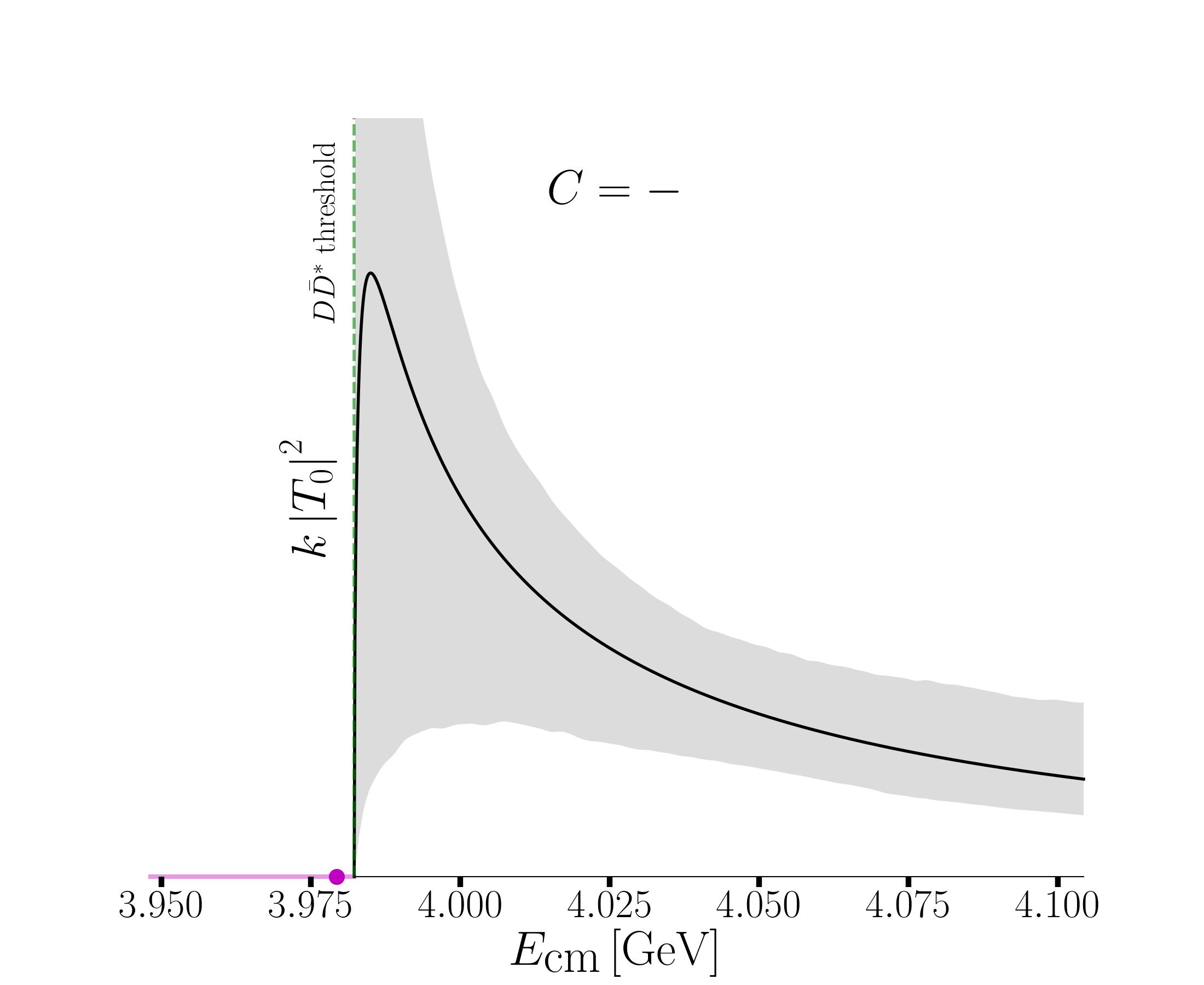}}
	\end{minipage}
\hspace*{1.0cm}
\begin{minipage}{0.45\linewidth}
\centerline{\includegraphics[width=1.2\textwidth]{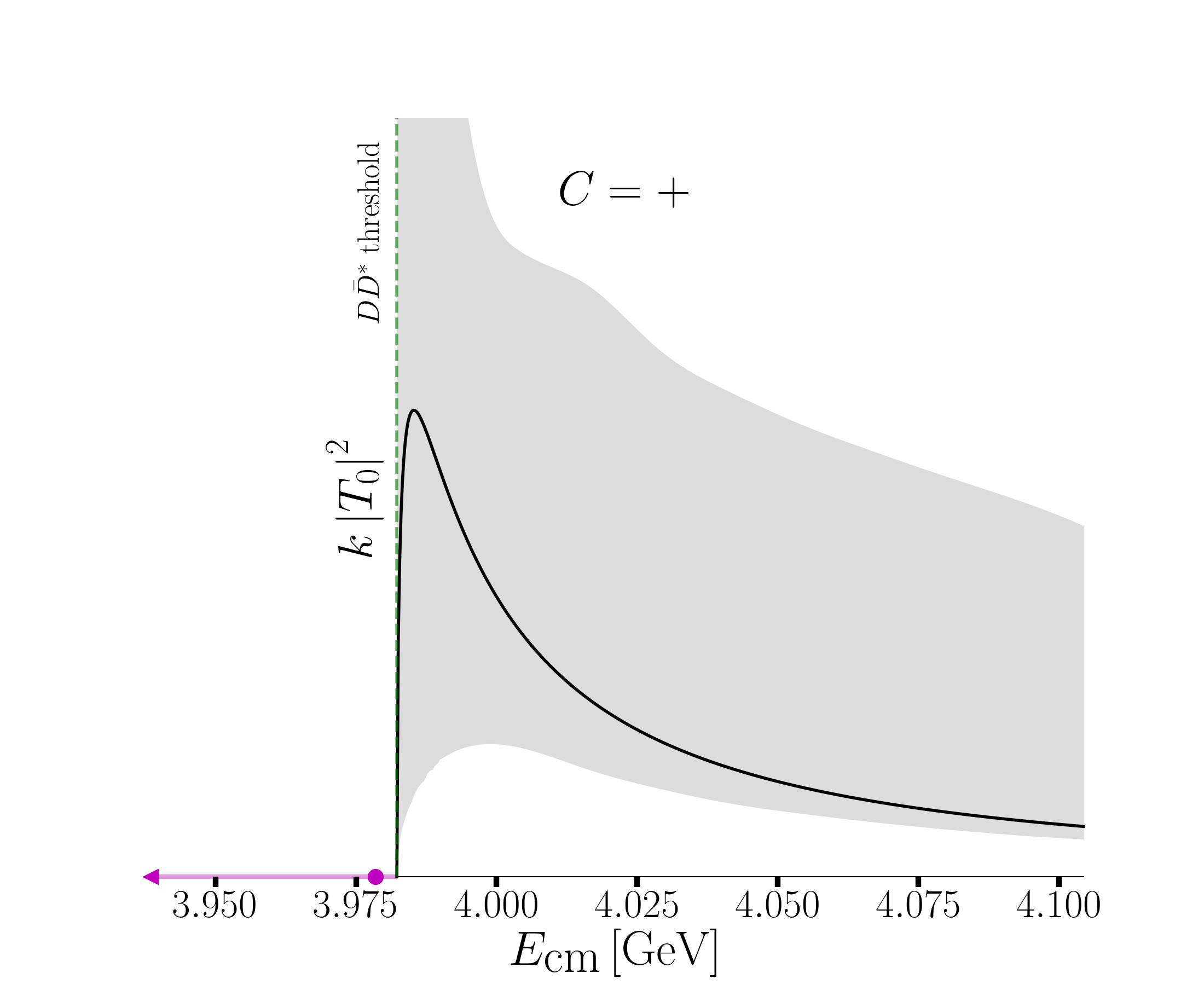}}
	\end{minipage}
\caption{Quantity $k\left|T_0\right|^2$ proportional to the rate for $D\bar{D}^*$ scattering in the one-channel approximation for $J^{PC}=1^{+-}$ (left) and $1^{++}$ (right). Here, all interpolators were included in the variational analysis. The black line represents the central value corresponding to the effective range parameters in Table~\ref{tab:param:DDscatt}. The gray band shows the significant 1$\sigma$ uncertainty arising from the uncertainty of the virtual pole position presented in violet.}
\label{fig:rate}
\end{figure*}

Specific energy levels exhibit a negative shift, indicating attraction. This results in a positive $k\cot{\delta_{0}}$ that increases with $k^2$. However, some levels are consistent with noninteracting energies, resulting in significant errors in the values of $k\cot{\delta_{0}}$ and consequently large errors in the fit parameters. Slightly below the threshold, the central values of these fits intersect with the dashed cyan line representing $ik = +|k|$ (see the magenta octagons). This gives us a pole in the scattering amplitude \eqref{Et} at a real energy below the $D\bar D^*$ threshold ($E_{\textrm{cm}}^{\textrm{p}}<m_D+m_{D^*}$). Specifically, the poles correspond to a virtual state, located at the scattering momentum $k_{\textrm{p}}$, where $k_{\textrm{p}}=-i|k_{\textrm{p}}|$.\footnote{The scattering amplitude below the virtual state pole is not constrained by our eigenenergies related to this channel, and we do not attribute physical significance to $T_{\ell=0}$} at those energies.

A pole below the threshold can enhance the scattering rate above the threshold, where the enhancement is more significant if the pole is closer to the threshold. The effect of our virtual poles on the s-wave $D\bar{D}^*$ scattering rate (which is proportional
to $k\left|T_0\right|^2$) is shown in Fig.~\ref{fig:rate}. Due to the large uncertainties in the scattering amplitude, the significance of the enhancement is similarly uncertain.

A virtual state pole slightly below the threshold was also found in a recent EFT study \cite{Zhang:2024fxy}, which argues why the molecular state with $1(1^{++})$ does not produce a sizable experimental signal and has not been observed so far. $Z_c$ has also been found as a virtual pole in a phenomenological study of $D\bar D^*$ scattering \cite{PhysRevD.88.054007}.

The standard Lüscher scattering formalism breaks down when applied to energies below the left-hand cut arising from the single-pion exchange. It predicts a real scattering amplitude in cases where it should be complex. We have not considered the effects stemming from the lhc, which is---in our scenario with an unphysically high pion mass---at $\left(ak_{\textrm{lhc}}\right)^2<-0.003$. This renders the extracted one-channel phase shifts below the lhc unreliable and an effective range expansion of the scattering amplitude is insufficient. Addressing these issues would require a modification of the standard Lüscher method \cite{Dawid:2023jrj,Du:2023hlu,Raposo:2023oru,Meng:2023bmz,Hansen:2024ffk}. 

The results obtained under the strong assumptions within this section should be taken as a first step to unveil the hadronic features hidden behind the densely populated finite-volume eigenenergies we extract. These results qualitatively suggest only a feeble interaction between $D$ and $\bar D^*$, which is not strong enough to provide a real bound state, while it is not possible to make further conclusions at this stage. An obvious next step is to consider extracting the energy dependence of the amplitude in the coupled channel system, including at least one of the channels in the calculation that is ignored in the above analysis. We discuss such an approach in the next section.

\section{Reconciling $Z_c(3900)$ data from lattice studies and experiment?}\label{sec:comparison}

A rigorous theory study of $Z_c$ requires the determination of the energy dependence of the scattering matrix $T_{ij}(E)$ for all relevant channels $i,j$. We attempted this and found that too many parameters are required to reliably constrain the energy dependence of this matrix for a wide energy range from the $J/\psi \pi$ threshold up to $\SI{4}{GeV}$, although we employed eigenenergies from two volumes and two total momenta.

In order to facilitate this challenging task, we follow a simplified innovative approach presented in Ref.~\cite{Yan:2023bwt}. The authors conducted a comprehensive analysis that combines both experimental $J/\psi\pi$, $D\bar{D}^*$ event distributions $\textrm{d}N/\textrm{d}E_{\textrm{cm}}$ \cite{PhysRevLett.119.072001,PhysRevD.92.092006} and the lattice finite-volume energy levels determined in Refs. \cite{Cheung2017,CLQCD:2019npr}. 
In this section, we utilize the same approach as in \cite{Yan:2023bwt}, including our main lattice results in the analysis along with the results of the other works. Compared to \cite{Cheung2017,CLQCD:2019npr}, our lattice results provide additional information, in particular, valuable energy levels from the system in the moving frame and higher energies in the rest frame. This section only considers the $J^{PC}=1^{+-}$ channel and addresses the $J/\psi \pi$-$D\bar{D}^*$ coupled system within a covariant effective field theory framework. The aim is to constrain the unknown parameters in the $J/\psi\pi$ and $D\bar{D}^*$ scattering amplitudes, thereby providing more definitive insights into the properties of $Z_c(3900)$. For a detailed derivation of the fitted experimental event distributions $\textrm{d}N/\textrm{d}E_{\textrm{cm}}$, the procedure to relate scattering amplitudes and finite-volume energies, the parameters used and other details, we refer the reader to Ref. \cite{Yan:2023bwt}. Here, we only briefly summarize the approach.

\subsection{The underlying EFT and the procedure behind the global fits}
\label{subsec:derivation}

We employ the effective Lagrangians within a rela\-tivistic framework, specifically emphasizing the energy region near the $D\bar D^*$ threshold. As argued later, no pion exchange terms are considered, so interactions are approximated using only local contact four-meson terms, guided by the symmetries inherent in the theory:

\begin{itemize}
\item The diagonal coupling $D \bar{D}^* \leftrightarrow D \bar{D}^*$ 
is obtained from the most general Lagrangian consistent with the heavy quark spin symmetry incorporated via $H=\left[D^*_{\mu}\gamma^{\mu}-D\gamma_5\right](1+\slashed{v})/2$ \cite{ALFIKY2006238,Mehen:2011yh}
\begin{equation}
\begin{aligned}
\mathcal{L}_{D \bar{D}^* D \bar{D}^*}= & -C_{0 a}\left(D^{\dagger} D+D_\mu^{* \dagger} D^{* \mu}\right)\left(\bar{D} \bar{D}^{\dagger}+\bar{D}_\mu^* \bar{D}^{* \mu \dagger}\right) \\
& +C_{0 b}\left(D^{\dagger} D_\mu^*+D_\mu^{* \dagger} D\right)\left(\bar{D}^{* \mu} \bar{D}^{\dagger}+\bar{D} \bar{D}^{* \mu \dagger}\right)\>.
\label{ELDDDD}
\end{aligned}
\end{equation}
Only terms at the leading-order $\mathcal{O}(p^0)$ in the chiral expansion are retained, while next-to-leading-order terms $\mathcal{O}(p^2)$ are neglected. The latter have a minor contribution since the heavy meson momenta are small. For the $1^{+-}$ channel, only one combination of $C_{0a}$ and $C_{0b}$ contributes \cite{PhysRevD.88.054007,ALFIKY2006238,PhysRevD.85.114037} and we refer to it as $\hat{\lambda}_1$.

Pion exchange interaction is not included explicitly. One reason is that the short-range pion exchange can be absorbed into the redefinition of contact interactions employed. Also, the coupling of pions to a pair of heavy mesons involves a derivative, and therefore, this contribution is significantly momentum suppressed near the threshold (at least at tree level). This suppression is lifted at higher energies, which may be one of the reasons that this EFT and lattice data show some tensions when including the data points above the $D\bar D^*$ threshold, as will be discussed.

\item Regarding the contact interactions between $J/\psi\pi$ and $D\bar D^*$, the general covariant operators featuring the fewest derivatives and being invariant under $C$, $P$, chiral, and isospin symmetry transformations are described by
\begin{equation}
\begin{aligned}
\mathcal{L}_{D \bar{D}^* J / \psi \pi}&=  \hat{\lambda}_2 \psi_\mu\left(\nabla^\mu D^{\dagger} u_\nu \bar{D}^{* \nu \dagger}+\bar{D}^{* \nu} u_\nu \nabla^\mu D\right)\\
&+\hat{\lambda}_3 \psi_\mu\left(\nabla^\nu D^{\dagger} u^\mu \bar{D}_\nu^{* \dagger}+\bar{D}_\nu^* u^\mu \nabla^\nu D\right) \\
& +\hat{\lambda}_4 \psi_\mu\left(\nabla^\nu D^{\dagger} u_\nu \bar{D}^{* \mu \dagger}+\bar{D}^{* \mu} u_\nu \nabla^\nu D\right)\\
&+\hat{\lambda}_5 \psi_\mu\left(D^{\dagger} \nabla^\mu u^\nu \bar{D}_\nu^{* \dagger}+\bar{D}_\nu^* \nabla^\mu u^\nu D\right)\>.
\label{ELDDJpsipi}
\end{aligned}
\end{equation}
Here, $\psi_{\mu}$ is the $J/\psi$ field, whereas $u_{\mu}$ refers to the chiral pseudoscalar meson fields as defined in Refs. ~\cite{Gong:2016hlt,Yan:2023bwt}. This Lagrangian was used in Ref. \cite{Gong:2016hlt} and initially in a nonrelativistic form in \cite{Mehen:2011yh}. 

\item The $\mathcal{L}_{J / \psi \pi J / \psi \pi}$ is neglected since the interaction between a pion and a heavy quarkonium is suppressed. This agrees with the small scattering length of the $J/\psi \pi$ interaction (see Sec.~\ref{sec:jpsipi} and Refs. \cite{Yokokawa:2006td,Liu:2008rza,Liu:2009af,Liu:2012dv}) and is discussed also in \cite{Hanhart:2015cua,Guo:2016bjq}.

\end{itemize}

The Lagrangians  \eqref{ELDDDD} and \eqref{ELDDJpsipi} render the transition amplitudes for $\bar D^*(a)D(b)\to\bar D^*(c)D(d)$ and $J/\psi(a)\pi(b)\to \bar{D}^*(c)D(d)$,
\begin{equation}
V_{\bar{D}^* D \rightarrow \bar{D}^* D}=\lambda_1 \varepsilon_c^{\dagger} \cdot \varepsilon_a
\label{EVDDDD}
\end{equation}
and
\begin{equation}
\begin{aligned}
V_{J / \psi \pi \rightarrow \bar{D}^* D}=\frac{1}{F_\pi}    \Big(   & \lambda_2 \left(\varepsilon_a \cdot p_d\right) \left(\varepsilon_c^{\dagger} \cdot p_b\right)\\
+&\lambda_3 \left(\varepsilon_a \cdot p_b\right) \left(\varepsilon_c^{\dagger} \cdot p_d\right)\\
+&\lambda_4 \left(\varepsilon_c^{\dagger} \cdot \varepsilon_a\right) \left(p_b \cdot p_d\right)\\
+&\lambda_5 \left(\varepsilon_a \cdot p_b\right)\left( \varepsilon_c^{\dagger} \cdot p_b\right) \Big)\>,
\label{EVDDJpsipi}
\end{aligned}
\end{equation}
where parameters $\lambda_j$ are proportional to $\hat \lambda_j$ as detailed in \cite{Yan:2023bwt}.\footnote{Notice there is a typo for Eqs.~(6) and (7) in Ref.~\cite{Yan:2023bwt}: the factor of $\sqrt{2}$ should be one.}   

It is then convenient to proceed with partial-wave amplitudes. In scattering processes with spinful particles, both the $\ell S$ and helicity bases can be employed for the partial-wave projections. While these two methods are generally equivalent, the $\ell S$ basis is better suited for our study. The reason is that in the molecular picture of $Z_c(3900)$, the s-wave interaction of $D\bar D^*$ should dominate, with the d-wave part significantly suppressed, at least when focusing on the energy region near the $D\bar D^*$ threshold. We adopt the approach from Ref. \cite{Gulmez:2016scm} to perform partial-wave projections in a covariant manner. This naturally introduces specific energy-dependent terms to the amplitudes from polarization vectors, which are of higher order in the nonrelativistic expansion, without adding extra free parameters. Refs.~\cite{ALBALADEJO2016337,Du:2022jjv} emphasize the crucial role of energy dependence in the interaction kernel ($V(s)$) for generating resonance poles of $Z_c(3900)$ in the complex energy plane near the physical region. Therefore, the covariant scattering amplitudes approach should enable us to gain important insights into the properties of $Z_c(3900)$.

\begin{figure}[h!]
	\centering
	\includegraphics[width=0.5\textwidth,angle=-0]{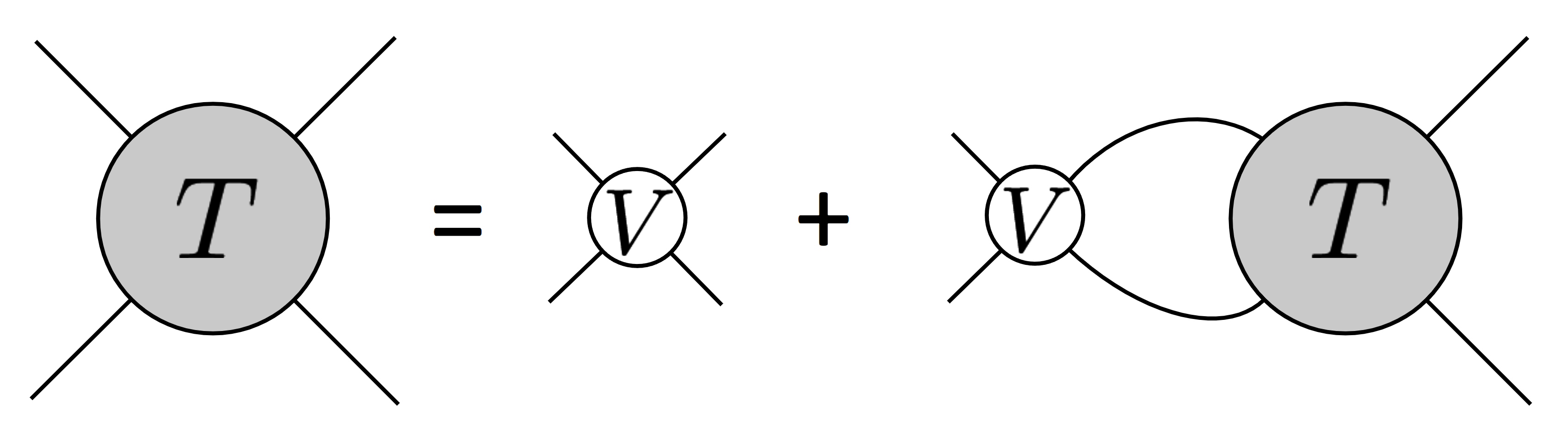}
	\caption{Diagrammatic representation of Eq.~\eqref{ET}, which can be rewritten as $T=V+VGT$, where $G$ denotes the loop.}
	\label{fig:diagramT}
\end{figure}

The $J/\psi \pi$ and $D\bar D^*$ channels are labeled as channels 1 and 2, respectively. The s-wave amplitudes $V_{11}(s)$,\footnote{In the last bullet point in this section, we have already mentioned that the perturbative $J/\psi \pi\to J/\psi \pi$ transition amplitude is insignificantly small. Consequently, the corresponding matrix element vanishes, $V_{11} = 0$.} $V_{12}(s)$,\footnote{The $\hat{\lambda}_3$ term in Eq.~\eqref{ELDDJpsipi} has no impact on the s-wave amplitude.} $V_{22}(s)$ build up a symmetric matrix spanned in the scattering-channel space,
\begin{equation}
V(s)=\left[\begin{array}{ll}
V_{11}(s) & V_{12}(s) \\
V_{12}(s) & V_{22}(s)
\end{array}\right]\>,
\label{ENV}
\end{equation}
where $\sqrt{s}=E_{\textrm{cm}}$ and the explicit expressions of the partial-wave amplitudes $V_{ij}$ can be found in Ref.~\cite{Yan:2023bwt}. 
$V(s)$ is further used to derive the on-shell unitary partial-wave two-body scattering amplitude 
\begin{equation}
T(s)=[1-V(s) \cdot G(s)]^{-1} \cdot V(s)\>,
\label{ET}
\end{equation}
where $G(s)$ is the well-known diagonal loop function matrix that also contains the two free parameters $a_{\textrm{SC,}i}$, $i=1,2$, called subtraction constants. This equation is diagrammatically shown in Fig.~\ref{fig:diagramT}.

\begin{figure}[h!]
	\centering
	\includegraphics[width=0.48\textwidth,angle=-0]{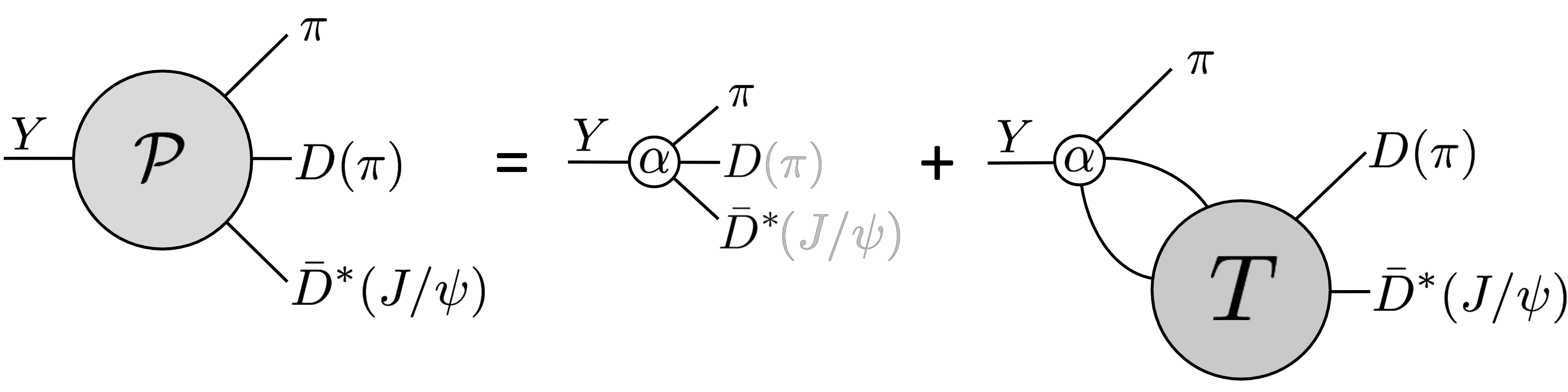}
	\caption{Diagrammatic representation of the production amplitude in Eq.~\eqref{EP}, which can be rewritten as $\mathcal{P}=\alpha+TG\alpha$, where $G$ represents the loop. The negligible $\alpha_1$ is represented with the faded text of the direct product channel $J/\psi \pi$.}
	\label{fig:diagramP}
\end{figure}

The experimental $J/\psi \pi$ and $D\bar D^*$ invariant-mass distributions
\begin{equation}
\frac{d N_i}{d \sqrt{s}}=A_i(P_i(s))+B_i(s), \quad i=1,2\>
\label{EdNdsqrts}
\end{equation}
are projected from the three-body decays $Y\to J/\psi \pi\pi$ and $Y\to D\bar D^*\pi$, respectively. $A_i$ and $B_i$ represent the signal and background part, respectively. The former is a function of the two-body production amplitudes (see Fig.~\ref{fig:diagramP})
\begin{equation}
\footnotesize{
\mathcal{P}(s)=\left[\begin{array}{l}
P_1(s) \\
P_2(s)
\end{array}\right]=[1-V(s) \cdot G(s)]^{-1} \cdot \alpha\>,\quad
\alpha=\left[\begin{array}{c}
\alpha_1 \\
\alpha_2
\end{array}\right]
\approx\left[\begin{array}{c}
0 \\
\alpha_2
\end{array}\right]
\,
}
\label{EP}
\end{equation}
where $\alpha$ are constant production vertices. Given that the cross section of $e^+e^-\to\pi^+D^0D^{*-}$ between \SI{4.2}{} and \SI{4.3}{GeV} is much larger (200--\SI{300}{pb} \cite{BESIII:2018iea}) than that of $e^+e^- \to \pi^+ \pi^-J/\psi$ (50--\SI{80}{pb} \cite{BESIII:2016bnd}), we set the direct production vertex $\alpha_1$ to zero. This is also consistent with the choice of setting $V_{11}=0$ in Eq.\eqref{ENV}.

We will perform a global fit to the experimental and lattice data. Once we establish the values of the unknown parameters in the production amplitudes \eqref{EP}, the unitarized scattering amplitudes \eqref{ET} become unambiguously determined. Consequently, we can extract information about the $Z_c(3900)$ resonance, including its pole positions and coupling strengths in the relevant channels.

To fit also the lattice data, a finite box of length $L$ with periodic boundary conditions and inertial frame with total momentum $|{\bi P}|=0$ or $2\pi/L$ is considered. The potential $V(s)$ and the finite volume energies $E_{\textrm{cm}}=\sqrt{s} $ observed on the lattice are related via the L\"uscher's quantization condition, which was conveniently expressed in Refs.~\cite{multi-channel_scattering,Doring:2012eu,Guo:2018tjx,Gockeler:2012yj} as
\begin{equation}
\det \left[ 1-V(s)\cdot\left(G(s)+\Delta G(s)\right)\right]=0\>~.
\label{Edet}
\end{equation}
Here, $\Delta G(s)$ is the finite volume correction of the loop function.
We remark that correcting our lattice energies with \eqref{EdeltaEn} and \eqref{EEn} enables our lattice energies to be jointly fitted with energies from Refs. \cite{Cheung2017,CLQCD:2019npr}, where a relativistic dispersion relation is employed. This is a notable improvement over the approach in Ref.~\cite{Zc_Sasa_2}. Comparing with Ref.~\cite{Yan:2023bwt}, where only lattice data in the rest frame were included, the present work also considers the EFT fits to the lattice energy levels in the moving frame. We refer to Ref.~\cite{Guo:2018tjx}, especially Eqs.~(31)--(35) of this reference, for the details about calculating the moving-frame spectra in the unitarized EFT approach. 

\subsection{Global fits to the experimental and lattice data}
\label{subsec:fits}
 
The unknown parameters are determined by simultaneously fitting experimental and lattice data, as in Ref. \cite{Yan:2023bwt}. Two types of relevant experimental data are included: first, the $J/\psi\pi^{\pm}$ event distributions from the $e^+e^-\to J/\psi\pi^{+}\pi^-$ process at center-of-momentum energies $\SI{4.23}{GeV}$ and $\SI{4.26}{GeV}$ \cite{PhysRevLett.119.072001}, and second, the $D^0D^{*-}$ and $D^-D^{*0}$ event distributions from the $e^+e^-\to \pi^{\pm}(D\bar{D}^*)^{\mp}$ processes at the same $e^+e^-$ energies \cite{PhysRevD.92.092006}. For lattice data, we take the finite-volume $J/\psi \pi$ and $D\bar D^*$ energies from the left-hand sides of Figs.~\ref{fig:T1pm} and \ref{fig:A2m} and data from Refs. \cite{Cheung2017,CLQCD:2019npr}. 
Fits are uncorrelated since all the lattice data (including ours) is treated on the same footing as the experimental data that are uncorrelated.
The fits employ the masses of the scattered hadrons that feature in the lattice simulations (see Appendix \ref{sec:app:masses} and Ref.~\cite{Yan:2023bwt}).

We assume that the fitted parameters do not depend on masses of scattering hadrons and then provide the scattering amplitudes and poles for the case of physical masses. Our parameters are the dimensionless coupling constants $\lambda_1$ and $\tilde{\lambda}_{j=2, 4, 5}\equiv\lambda_{j=2, 4, 5} \,m_D^{\rm phys}$~\footnote{There is a typo for Eq.~(24) in Ref.~\cite{Yan:2023bwt}: it should be $\tilde{\lambda}_{j=2,3,4,5}=\lambda_{j=2,3,4,5} \,m_D^{\rm phys}$.} ($m_D^{\rm phys}$ being the physical $D$-meson mass), two background parameters $b_1$ from the $Y\to J/\psi \pi \pi$ processes at both $\SI{4.23}{GeV}$ and $\SI{4.26}{GeV}$ center-of-momentum energies, the four production vertices $\alpha_2$ for both $e^+e^-\to J/\psi\pi^{+}\pi^-$ and $e^+e^-\to \pi^{\pm}(D\bar{D}^*)^{\mp}$ processes at both energies and the subtraction constants $a_{\textrm{SC,}1}$, $a_{\textrm{SC,}2}$. 
The background events of the $D\bar D^*$ event distribution are subtracted from the analysis of Ref.~\cite{PhysRevD.92.092006}. Consequently, our fitting approach excludes the background term $B_2(s)$ from \eqref{EdNdsqrts} but retains $B_1(s)$, which includes the parameter $b_1$. The subtraction constants, introduced as part of the unitarization process, are left to vary freely.

\begin{figure*}[ht!]
	\centering
	\includegraphics[width=1.0\textwidth,angle=-0]{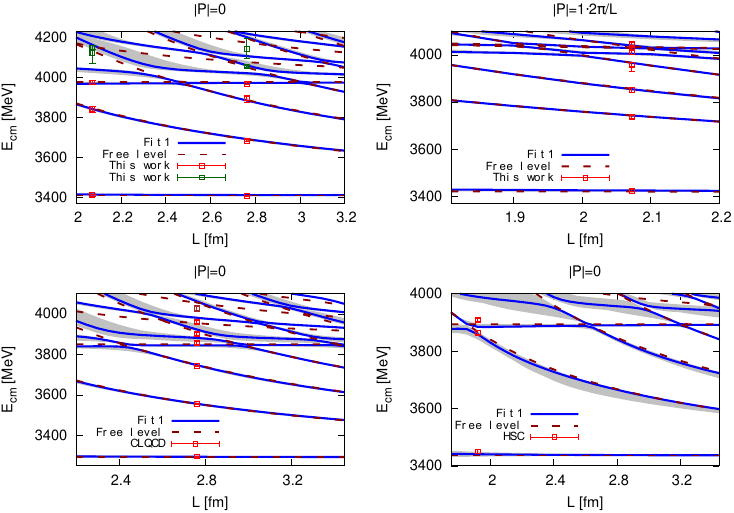}
	\caption{
 Energy levels $E_{\textrm{cm}}$ (red, green and brown squares) from our main lattice results (top), results from HSC (bottom right) \cite{Cheung2017} and CLQCD (bottom left) \cite{CLQCD:2019npr} together with the prediction of the preferred Fit1 (blue solid lines) as a function of $L$. This fit incorporates data points from experimental measurements \cite{PhysRevLett.119.072001,PhysRevD.92.092006}, from lattice studies \cite{Cheung2017,CLQCD:2019npr} and 18 of our lattice data points. Total momentum is $\left|{\bi P}\right|=0$ (top left and bottom) and $\left|{\bi P}\right|=1\cdot2\pi/L$ (top right). The highest brown CLQCD energy point was not used in any of the fits. The dark red dashed lines denote the free energy levels. The gray shadowed areas represent the error bands from Fit1. Only the finite-volume energy levels that are dominantly coupled to s-wave $J/\psi \pi$ and $D\bar D^*$ interpolators are shown and used in the fit.}
	\label{fig:our_P0-P1}
\end{figure*}

Four distinct fits---all including experimental and lattice data---are conducted, and the reason for this will become evident later. The fits incorporate the lattice levels relevant for the coupled scattering of $D\bar D^*$ and $J/\psi \pi$ in s wave. The common feature of all available lattice studies is that all energies are close to noninteracting energies. The distinction between the fits lies in the handling of four of our energy levels with $E_{\textrm{cm}}>\SI{4050}{MeV}$: 
\begin{itemize}
    \item Fit1 is the conventional and preferred fit that includes all the data: the experimental data, all 18 relevant lattice energy levels from our simulation\footnote{Note that no fit incorporates states with significant overlap with $D(1) \bar D^*(2)$ and $D(2) \bar D^*(1)$ operators. Their energies lie around \SI{200}{MeV} above the lattice $D \bar D^*$ threshold, which is the edge of the energy region accessed by the experiment. This and the fact that such high energy levels have not been reliably determined represent the arguments for not including the aforementioned energies.} (red and green squares in Fig.~\ref{fig:our_P0-P1} that correspond to black outlined symbols in top right panes of Figs. \ref{fig:omit_1+-} and \ref{fig:omit_A2-}), and levels from simulations  \cite{Cheung2017,CLQCD:2019npr}.\footnote{Fits here and in \cite{Yan:2023bwt} incorporate the lowest six levels from \cite{CLQCD:2019npr}.}
    \item Fit2: same as Fit1, but the four of our levels at highest energies (green in Fig.~\ref{fig:our_P0-P1}) are not included.
    \item Fit3: same as Fit1, but our four highest levels are included with errors that are artificially reduced by a factor of four; at the same time, we assume that their central values are correct (variations of this fit render similar conclusions as discussed in Appendix \ref{sec:app:more-on-fits}).
    \item Fit4: the original joint fit from Ref. \cite{Yan:2023bwt}, which does not incorporate our lattice levels. 
\end{itemize} 
So, Fit2 excludes four of our highest energy levels, while Fit3 gives these levels more weight in the fit.

\begin{table*}[ht!]
	\centering
	\caption{Fitted parameters and the corresponding $\chi^2/N_{\textrm{dof}}$. All four fits include data from experiment \cite{PhysRevLett.119.072001,PhysRevD.92.092006} and previous lattice calculations \cite{Cheung2017,CLQCD:2019npr}. Additionally, Fit2 includes 14 data points from our new lattice simulation below \SI{4050}{MeV}, while Fit1 and Fit3 include 18 of our data points below \SI{4150}{MeV}. For Fit3, we have artificially reduced the uncertainties of our four lattice levels between \SI{4050}{MeV} and \SI{4150}{MeV} by a factor of 4. The left/right values for the $b_1$ and $\alpha_2$ represent results for the data sets at center-of-momentum energies $\SI{4.23}{GeV}$ and $\SI{4.26}{GeV}$, respectively. \label{tab:param}}
\bgroup
\def\arraystretch{1.3}
		\begin{tabularx}
{1.0\textwidth}
{m{0.22\linewidth}m{0.19\linewidth}m{0.19\linewidth}m{0.19\linewidth}m{0.19\linewidth}}
			\hline\hline
			Parameter  & Fit1 (preferred) & Fit2 & Fit3 & Fit4 \cite{Yan:2023bwt}\\
			\hline
			$a_{\textrm{SC},1}$  &$-6.63^{+1.87}_{-3.69}$&$-4.44^{+0.43}_{-0.52}$& $-9.27^{+2.87}_{-4.23}$   &    $-4.02^{+0.32}_{-0.49}$    \\
			$a_{\textrm{SC},2}$  &$-2.80^{+0.06}_{-0.17}$&$-2.81^{+0.03}_{-0.03}$& $-2.05^{+0.07}_{-0.11}$   &    $-2.80^{+0.02}_{-0.02}$    \\
			$\lambda_1$          &$-102^{+36}_{-13}$     &$-88^{+11}_{-12}$      & $-63^{+71}_{-56}$     &         $-86^{+12}_{-11}$     \\
			$\tilde{\lambda}_2$  &$219^{+151}_{-172}$    &$999^{+115}_{-157}$    & $-2208^{+602}_{-489}$   &        $1082^{+93}_{-116}$    \\
			$\tilde{\lambda}_4$  &$36^{+9}_{-13}$        &$-44^{+17}_{-11}$      & $340^{+71}_{-89}$     &            $-41^{+14}_{-11}$  \\
			$\tilde{\lambda}_5$  &$160^{+119}_{-133}$    &$-734^{+185}_{-121}$   & $3399^{+710}_{-883}$    &        $-751^{+137}_{-111}$   \\
			$b_1\left[10^{-4}\times\rm{MeV^{-3}}\right]$ &$9.0^{+0.2}_{-0.2}/4.9^{+0.1}_{-0.1}$    &$8.8^{+0.2}_{-0.2}/4.8^{+0.1}_{-0.1}$
                                                                            & $8.8^{+0.2}_{-0.2}/4.8^{+0.1}_{-0.1} $    & $8.7^{+0.2}_{-0.2}/4.8^{+0.1}_{-0.2}$ \\
			$\left| \alpha^{J/\psi \pi}_2\right|^2$      &$30.8^{+6.3}_{-10.0}/14.1^{+2.9}_{-4.7}$   &$20.2^{+3.4}_{-2.5}/9.3^{+1.8}_{-1.2}$     
                                                                            & $316.4^{+127.2}_{-117.7}/147.3^{+56.6}_{-55.5}$ & $19.4^{+3.4}_{-2.1}/9.0^{+1.6}_{-1.1}$    \\
			$\left| \alpha^{D \bar D^* }_2\right|^2$     &$1.3^{+0.5}_{-0.4}/0.6^{+0.2}_{-0.2}$      &$2.2^{+0.5}_{-0.4}/1.1^{+0.2}_{-0.2}$
			                                                            & $6.7^{+1.4}_{-1.6}/2.8^{+0.6}_{-0.7}$      &   $2.7^{+0.6}_{-0.6}/1.3^{+0.3}_{-0.3}$   \\
			$\chi^2/N_{\textrm{dof}}$                    &$\frac{484.1}{368-12}=1.36$                &$\frac{412.8}{364-12}=1.17$
			                                                            & $\frac{645.8}{368-12}$=1.81 &   $\frac{398.7}{350-12}=1.18$    \\
			\hline\hline
			\end{tabularx}
			\egroup
\end{table*}

The fitted parameters of all four fits are represented in Table~\ref{tab:param}.
The parameters of Fit2 agree very well with Fit4. On the other hand, certain parameters of Fit1 and Fit3 are not consistent with Fit2 and Fit4. This is more pronounced for Fit3. The $\chi^2/N_{\mathrm{dof}}$ for Fit1 and Fit3 is larger than for the other two. However, the resulting $\chi^2/N_{\mathrm{dof}}$ for Fit1 is still relatively small. 

\begin{figure}[h!]
	\centering
    \includegraphics[width=0.49\textwidth,angle=-0]{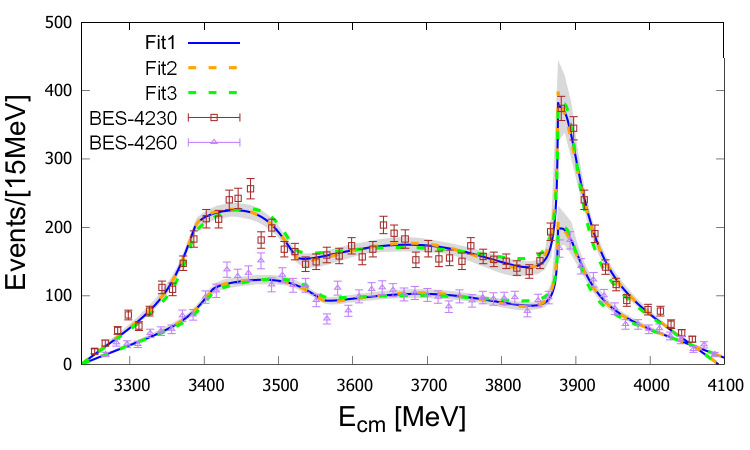}
	\includegraphics[width=0.49\textwidth,angle=-0]{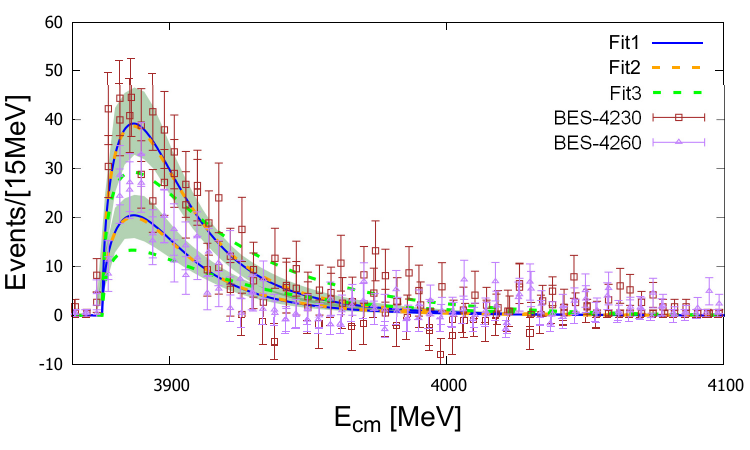}
	\caption{
$J/\psi\pi$ (top) and $D \bar D^*$ (bottom) event distributions from experiment (symbols) and from the EFT approach (lines). The top pane shows $J/\psi\pi^{\pm}$ invariant mass distributions of the process $e^+e^-\to\pi^+\pi^-J/\psi$ at $E_{e^+e^-}=\SI{4.23}{GeV}$ and $\SI{4.26}{GeV}$ from BESIII \cite{PhysRevLett.119.072001}. The bottom pane shows 
 the $D^0 D^{*-}$ and $D^- D^{*0}$ event distributions of $e^+e^-\to\pi^{\pm}(D\bar D^*)^{\mp}$ from BESIII \cite{PhysRevD.92.092006} at the same $E_{e^+e^-}$. Background events from the experimental analysis were subtracted. Blue, orange and green lines represent Fit1, Fit2 and Fit3, respectively. The shaded areas are the uncertainties of Fit1. }
\label{fig:eventsExp}
\end{figure}

Predictions from the fits are compared to the experimental invariant mass distributions for $D\bar D^*$ and $J/\psi \pi$ in Fig.~\ref{fig:eventsExp}. They show good agreement for both $e^+e^-$ energies. In particular, peaks related to the $Z_c$ feature for all fits. Only Fit3 somewhat underestimates the height of the peak in the $D\bar D^*$ distribution.

\begin{figure*}[ht!]
\begin{minipage}{0.33\linewidth} 
		\vspace{3pt}
		\centerline{\includegraphics[width=\textwidth]{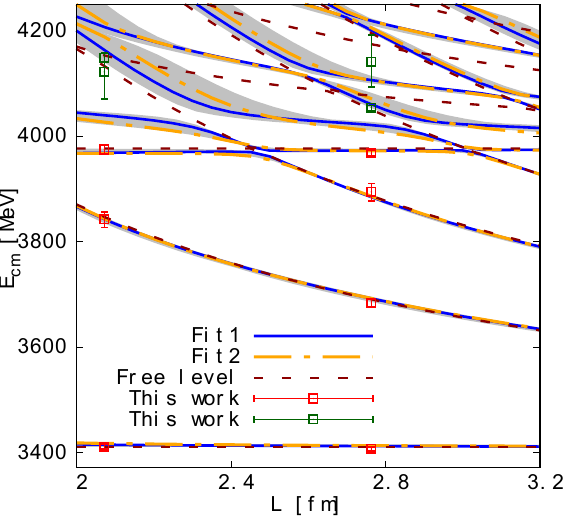}}
	\end{minipage}
	\begin{minipage}{0.308\linewidth}
		\vspace{3pt}
		\centerline{\includegraphics[width=\textwidth]{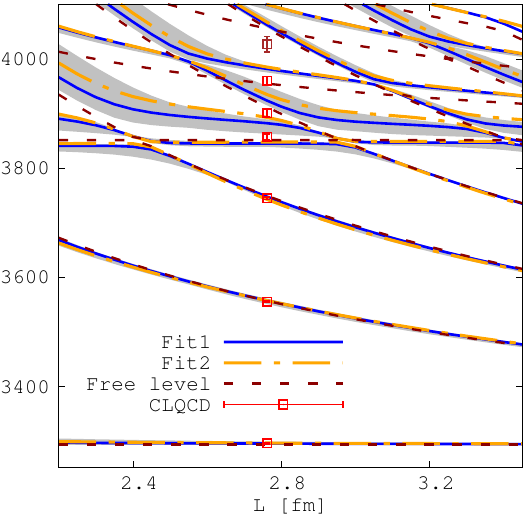}}
	\end{minipage}
	\begin{minipage}{0.308\linewidth}
		\vspace{3pt}
		\centerline{\includegraphics[width=\textwidth]{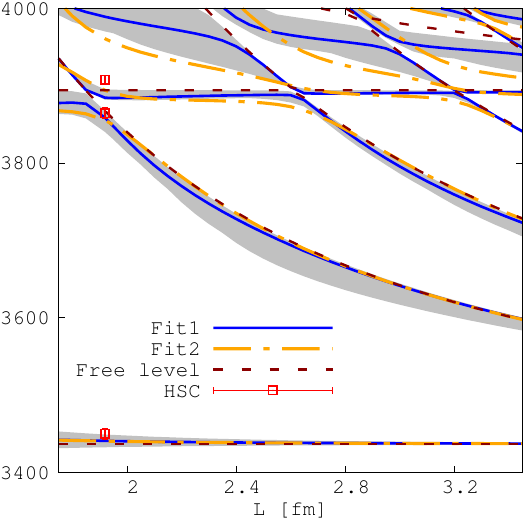}}
	\end{minipage}
\hspace{1cm}
	\caption{
	Lattice energy levels $E_{\textrm{cm}}$ (red, green and brown squares) for zero total momentum $\left|{\bi P}\right|=0$ from our main lattice results (left), results from HSC (right) \cite{Cheung2017} and CLQCD (in the middle) \cite{CLQCD:2019npr}. They are shown together with two different types of performed fits, Fit2 (orange dotted-dashed lines) and the preferred Fit1 (blue solid line). Both fits incorporate data points from experimental measurements \cite{PhysRevLett.119.072001,PhysRevD.92.092006} and from lattice studies \cite{Cheung2017,CLQCD:2019npr}. Fit2 includes 14 of our lattice data points (red squares), while Fit1 includes 18 (red and green squares). The highest brown CLQCD energy point was not used in any of the fits. The dark red dashed lines denote the free energy levels. The gray shadowed areas represent the error bands from Fit1. We omit the error bands from Fit2 to avoid overloading the plots. Only the finite-volume energy levels that are dominantly coupled to s-wave $J/\psi \pi$ and $D\bar D^*$ interpolators are shown and used in the fit.}
\label{fig:our-CLQCD-HSC}
\end{figure*}

\begin{figure*}[ht!]
\begin{minipage}{0.33\linewidth}
		\vspace{3pt}
		\centerline{\includegraphics[width=\textwidth]{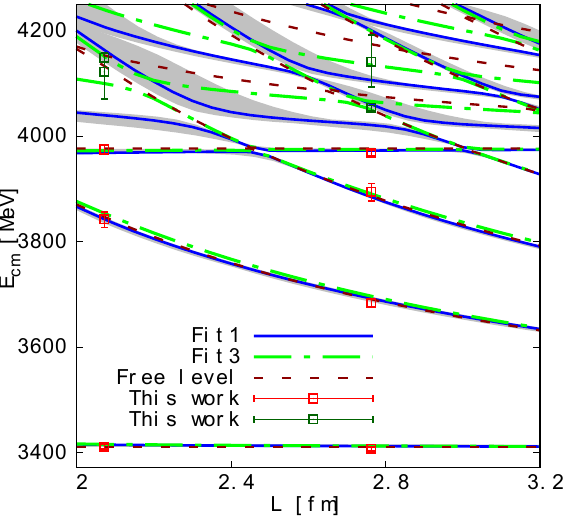}}
	\end{minipage}
	\begin{minipage}{0.308\linewidth}
		\vspace{3pt}
		\centerline{\includegraphics[width=\textwidth]{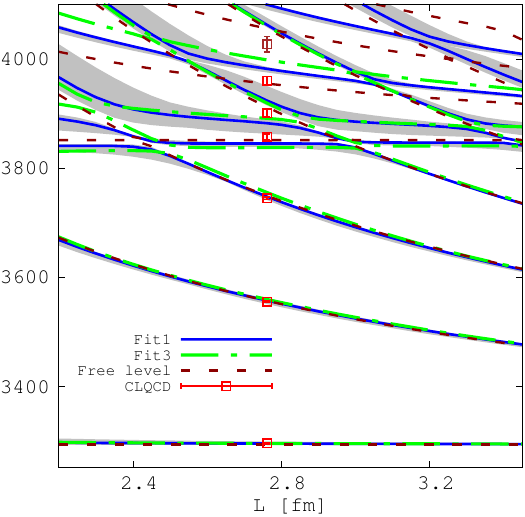}}
	\end{minipage}
	\begin{minipage}{0.308\linewidth}
		\vspace{3pt}
		\centerline{\includegraphics[width=\textwidth]{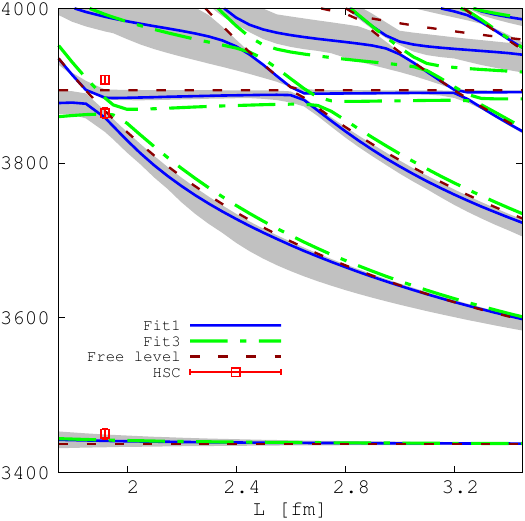}}
	\end{minipage}
\hspace{1cm}
	\caption{
 As Fig.~\ref{fig:our-CLQCD-HSC}, but Fit3 (green dotted-dashed lines) is shown instead of Fit2. The uncertainties of four high-lying levels (shown by green squares) 
 are artificially reduced by a factor of four in Fit3.}
\label{fig:our-CLQCD-HSC+Fit3}
\end{figure*}

Let us confront predictions of the fits with lattice energies from the present and previous simulations: the HadSpec collaboration provides energy levels up to the $D\bar{D}^*$ threshold \cite{Cheung2017}, while CLQCD
\cite{CLQCD:2019npr} and the present study also provide several levels above the threshold. The common feature is that all energies are close to the noninteracting energies.  
Our lattice energies are compared with the preferred Fit1 in Fig.~\ref{fig:our_P0-P1}. All fits are compared to our and other lattice data in Figs.~\ref{fig:our-CLQCD-HSC} and \ref{fig:our-CLQCD-HSC+Fit3}. Note that one should not directly compare the energies from different simulations since they employ different quark masses and volumes.

The predicted energies of the preferred conventional Fit1 manifest some disagreement with our lattice levels above $E_{\textrm{cm}}=\SI{4050}{MeV}$ and we elaborate on possible effects that cause this in Sec.~\ref{subsec:interp}. 
This discrepancy is also the motivation for performing several fits. The high-lying energy levels are not included in Fit2, which achieves a smaller $\chi^2/N_{\mathrm{dof}}$ than Fit1. On the other hand, Fit3 nearly reproduces our high-lying energy levels,\footnote{Our highest level on the large volume is about 1$\sigma$ above the prediction of Fit3.} although they lie close to the noninteracting energies. 
 In summary, the three different fits produce different results since they treat the four decisive lattice data points differently. Fit3 artificially emphasizes these data points, Fit2 excludes them, while the conventional Fit1 includes them with correct weights. This is the reason why we have chosen Fit1 as the preferred fit.

The coupling constants $\lambda_1$ and $\tilde{\lambda}_{j=2, 4, 5}$ as well as two subtraction constants characterize the nonperturbative interactions of $J/\psi\pi$ and $D\bar D^*$. These parameters are crucial in the coupled-channel scattering amplitudes, impacting event distributions across all energy points and influencing the lattice finite-volume spectra. These six parameters simultaneously influence both the experimental and lattice data under consideration. In contrast, the background parameters $b_1$ and the production vertex parameters $\alpha_2$, which serve as normalization constants, are anticipated to vary with the different BESIII datasets at $\SI{4.23}{GeV}$ and $\SI{4.26}{GeV}$.

\subsection{Insights into the $Z_c(3900)$ resonance poles}
\label{subsec:poles}
 
Kinematical singularities, e.g., two-body cusps and triangle and box singularities, offer a potential explanation for specific experimentally observed structures \cite{Guo:2019twa}. These singularities exhibit high sensitivity to the specific kinematics of the processes. In contrast, resonance poles in the complex energy plane possess a universality that extends across all production amplitudes involving the same particles.
The agreement of the employed model with experimental invariant mass distributions at both $e^+e^-$ energies suggests that the experimental $Z_c$ peak does not decisively arise from the triangular singularity but rather from the pole(s) \cite{Wang:2013cya,ALBALADEJO2016337,Du:2022jjv,Gong:2016jzb,Guo:2014iya}. Therefore, we next search for relevant poles in the system under investigation.

 This section focuses only on the poles around the $D\bar D^*$ threshold. In the vicinity of a pole singularity at $s=s_{\textrm{R}}$, the elements of the scattering amplitude follow a behavior described by:
\begin{equation}
T_{ij}(s)=-\frac{\gamma_i\gamma_j}{s-s_{\textrm{R}}}\>,
\label{Epole}
\end{equation}
where indexes $i$ and $j$ stand for the (in this case, two) coupled channels. The real and imaginary parts of the pole position are conventionally identified as the mass and half-width of a resonance, represented as $\sqrt{s_{\textrm{R}}}=m_{\textrm{R}}\pm\frac{i}{2}\Gamma_{\textrm{R}}$. The complex-valued residues $\gamma_i$ offer insights into the strength of the resonance's coupling to scattering channels.

The complex energy plane exhibits branch cuts associated with the opening of new channels. For single-channel scattering, the physical and unphysical Riemann sheets (RS) correspond to $\textrm{Im}(k)>0$ and $\textrm{Im}(k)<0$, respectively. Here $k$ is the magnitude of the three-momentum in the cm-frame.
In our two-channel case ($J/\psi \pi$, $D \bar D^*$), the following labeling of sheets is adopted:
\begin{equation}
\begin{aligned}
\textrm{RS I:}&\qquad\textrm{Im}(k_{J/\psi \pi})>0\quad \textrm{Im}(k_{D \bar D^*})>0\>,\\
\textrm{RS II:}&\qquad\textrm{Im}(k_{J/\psi \pi})<0\quad \textrm{Im}(k_{D \bar D^*})>0\>,\\
\textrm{RS III:}&\qquad\textrm{Im}(k_{J/\psi \pi})<0\quad \textrm{Im}(k_{D \bar D^*})<0\>,\\
\textrm{RS IV:}&\qquad\textrm{Im}(k_{J/\psi \pi})>0\quad \textrm{Im}(k_{D \bar D^*})<0\>.
\label{ERS}
\end{aligned}
\end{equation}
The physical measurements are performed along the real energy axes above threshold on the physical Riemann sheet I.

\begin{figure}[ht!] \includegraphics[width=0.25\textwidth]{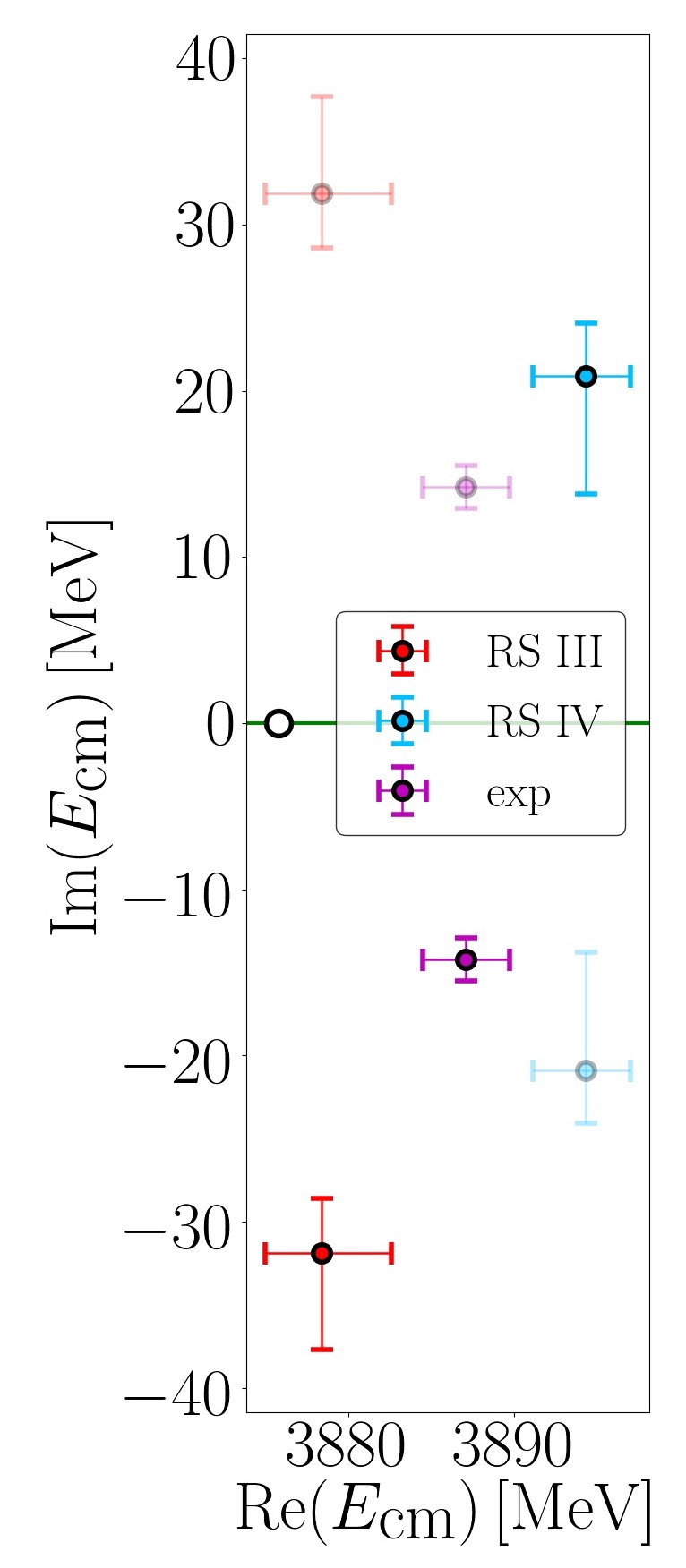}
\includegraphics[width=0.22\textwidth]{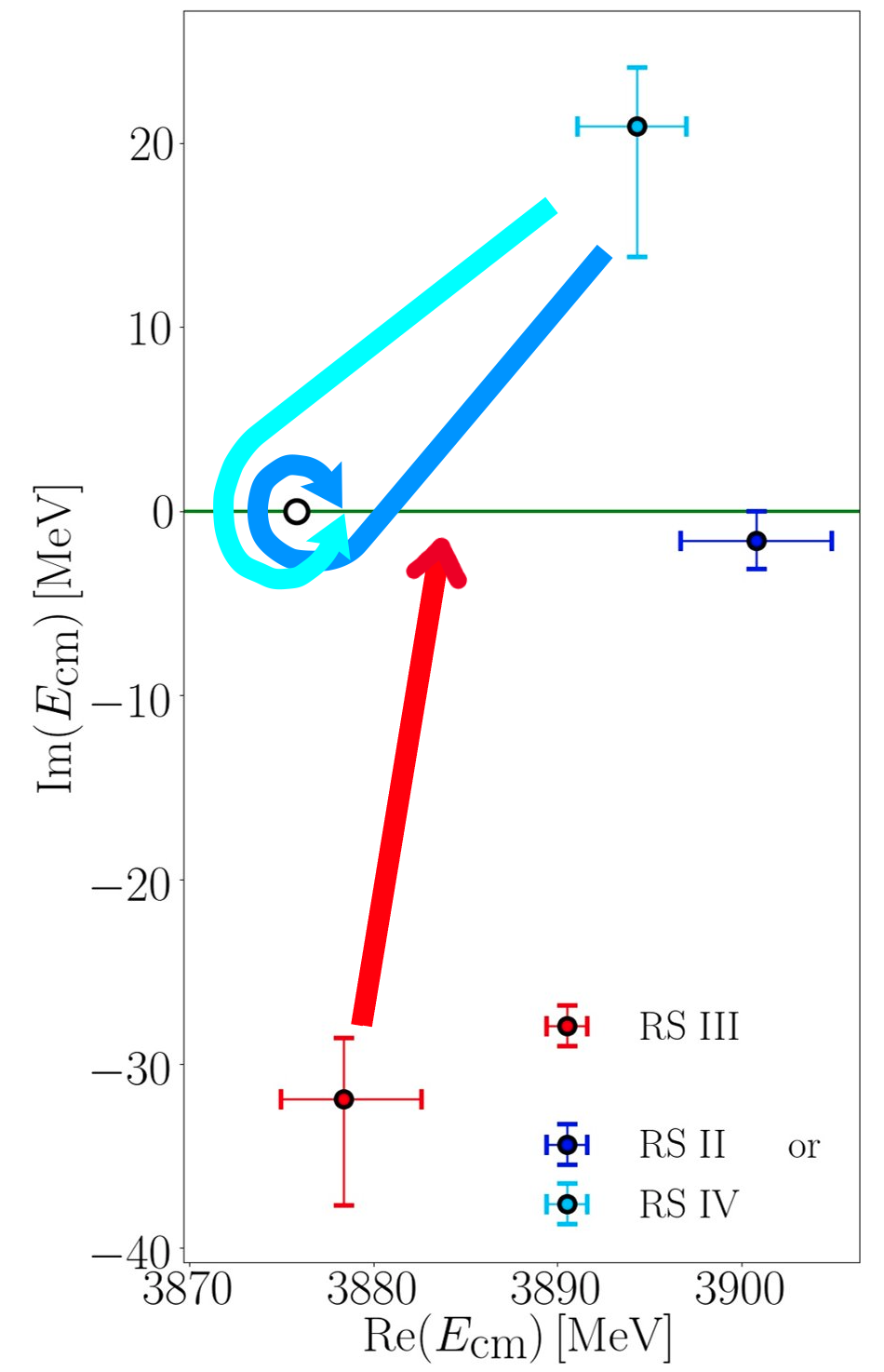} 
\caption{Left panel: The poles corresponding to the central values of Fit1 (RS III and RS IV) and the $Z_c(3900)$ pole from the PDG average $(3887.1\pm2.6)\SI{}{MeV}-i(14.2\pm1.3)\SI{}{MeV}$ \cite{Workman:2022ynf} with their conjugate poles. Right panel: Different paths in the complex energy plane that smoothly connect poles on RS III and RS IV (from Fit1) with the region where the physical $D\bar D^*$ scattering occurs. 
The green lines represent the real energy axis. The white circles represent the physical $D\bar D^*$ threshold, while the $J/\psi \pi$ threshold is not shown since it is too distant.}
\label{fig:poles:Fit1_exp_and_sketch}
\end{figure}

\begin{figure*}[ht!]
\centering
\begin{minipage}{0.338\linewidth}
\centerline{\includegraphics[width=1.\textwidth]{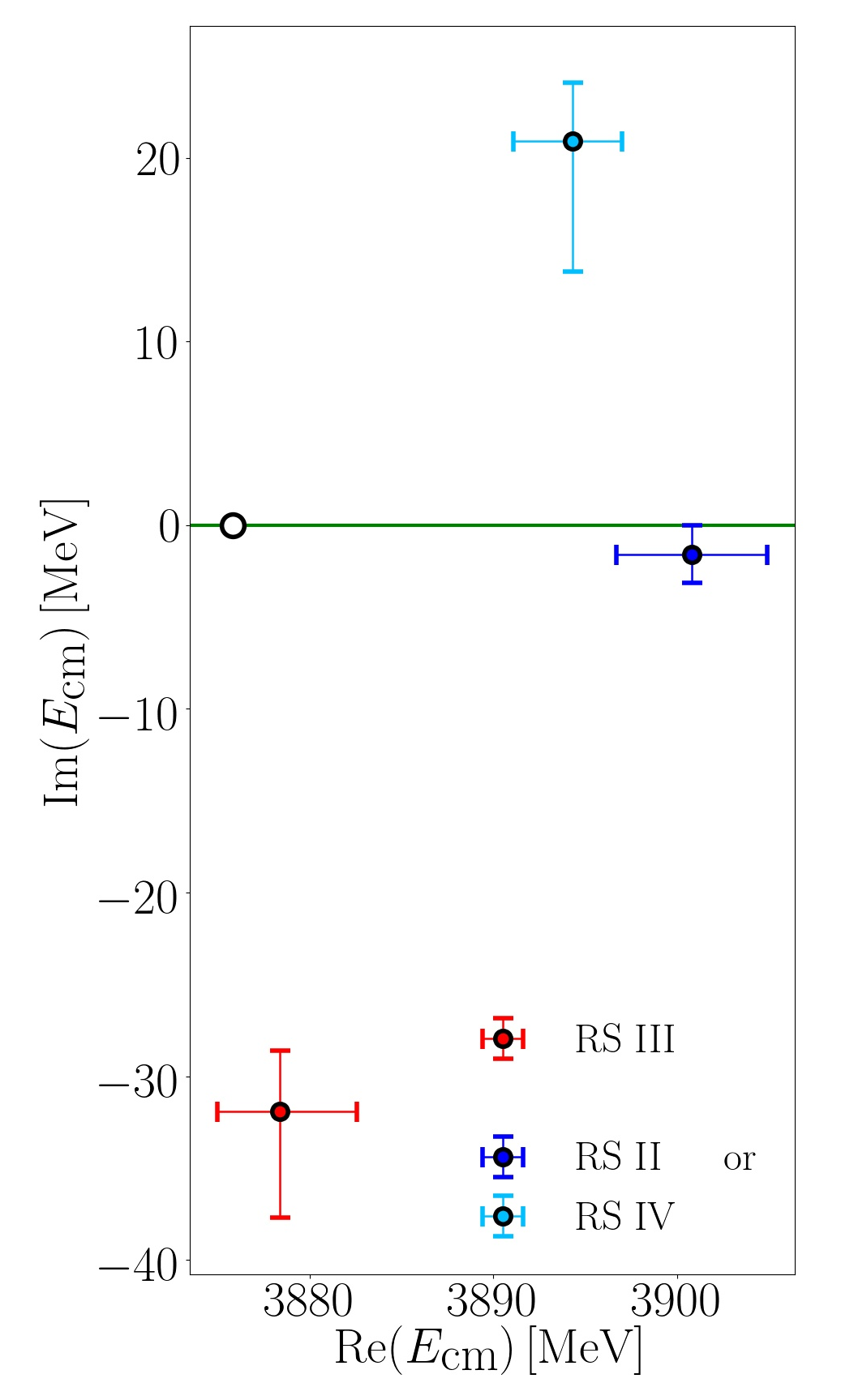}}
    \vspace{-0.9cm}
	\end{minipage}
\begin{minipage}{0.652\linewidth}
\centerline{\includegraphics[width=1.\textwidth]{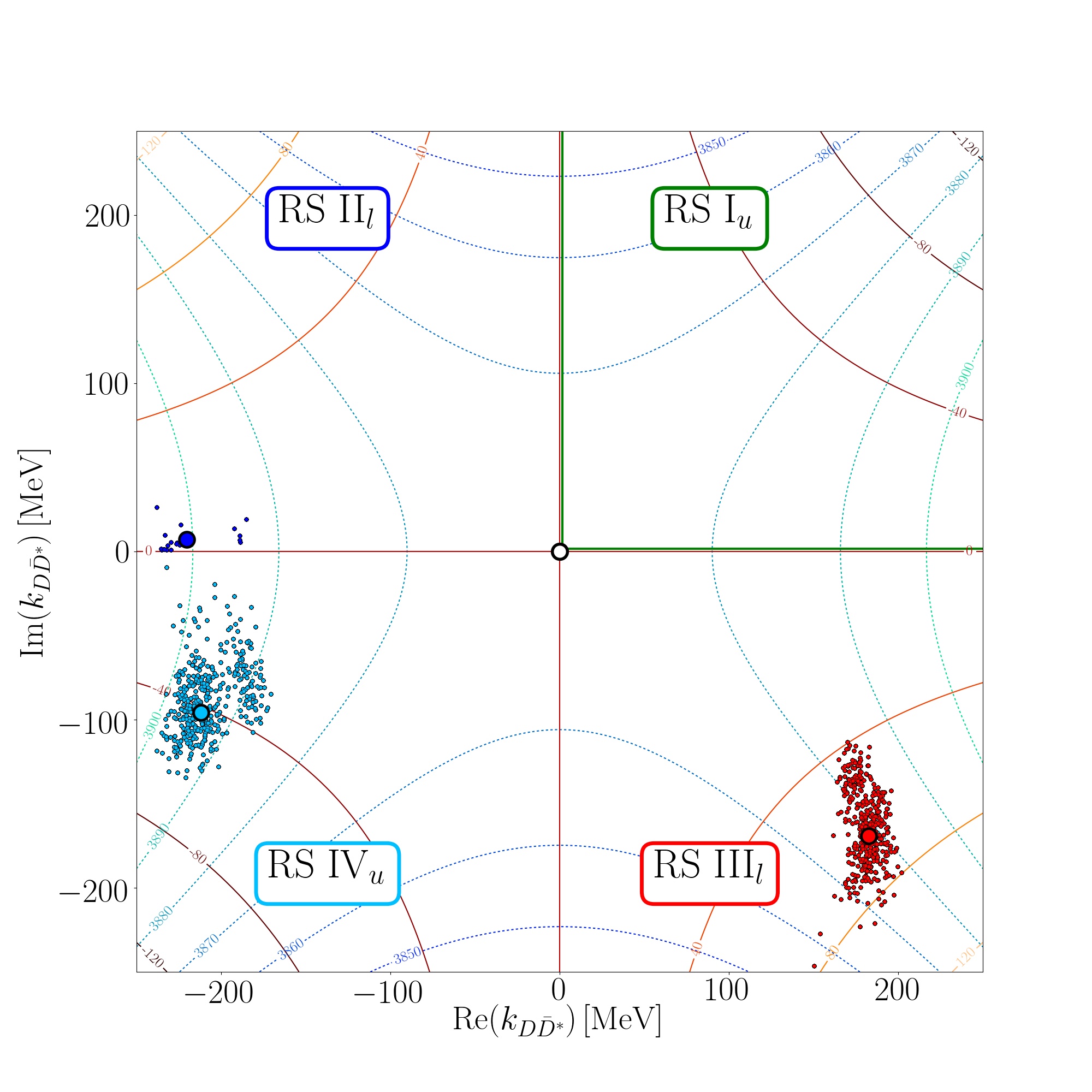}}
	\end{minipage}
\caption{Locations of the resonance poles of the scattering amplitude based on EFT for the physical masses of the scattered mesons. Results from the preferred conventional Fit1 are shown. There are two poles, one on RS III and another one on either RS II or RS IV. 
Left panel: the complex energy plane. The green line represents the real energy axis.
Right panel: the complex plane of the $D\bar D^*$-momentum in the cm frame (complex-$k_{D\bar D^*}$ plane) divided into four quadrants for four RSs. The subscripts $l/u$ represent the lower/upper complex energy half-plane. The dashed and solid contours represent constant $\textrm{Re}\left(E_{\textrm{cm}}\right)$ and $-2 \textrm{Im}\left(E_{\textrm{cm}}\right)$ in MeV, respectively. They aid visualization of the proximity of poles to physical scattering, which occurs along the positive imaginary axis (below the $D\bar D^*$) and the positive real axis (above the $D\bar D^*$ threshold). Small circles represent the spread of poles according to all bootstrap samples. The physical $D\bar D^*$ threshold is represented by the circle (in the origin), while the $J/\psi \pi$ threshold (on the imaginary axis) is not shown since it is too distant.
}
\label{fig:poles:FitA}
\end{figure*}

Two poles relatively close to $D\bar D^*$ threshold and to the physical region are found for all fits. They are located on RS III and RS IV for the central values of the preferred Fit1. Both poles seem to have comparable influence according to their closest connections to the physical axis. This can be seen in Figs.~\ref{fig:poles:Fit1_exp_and_sketch} and \ref{fig:poles:FitA} containing pole singularities of the preferred Fit1 in the complex energy and momentum
planes. The resonance poles' positions from our fits and the corresponding residues are presented in Table~\ref{tab:poles}.  
The left panel of Fig.~\ref{fig:poles:Fit1_exp_and_sketch} emphasizes that the poles on RS III and IV are close to the  $Z_c(3900)$ pole from the Particle Data Group (PDG) average. It is generally known that resonance poles from a single physical state may appear on different Riemann sheets at somewhat different positions. It is possible that the poles on RS III and IV are a manifestation of a single $Z_c$ physical state.
Every bootstrap sample contains a resonance pole on RS III and another either on RS II or RS IV, as in \cite{Yan:2023bwt}.

\begin{table}[h!]
	\caption{Relevant poles of the $Z_c(3900)$ and their residues in different fits. These values have been obtained using the physical masses of the scattering particles. $\gamma_1$ and $\gamma_2$ are residues of the $J/\psi \pi$ and $D\bar D^*$ channels, respectively. All fits have two near-threshold poles. Since the error determination procedure involves bootstrap sampling, different samples can give poles on different RSs. A sample contains either poles on RS III and II or poles on RS III and IV (see Fig.~\ref{fig:poles:all}). \label{tab:poles} }
	\centering
		\begin{tabular}{ l c c c c }
			\hline\hline
			RS & $\quad m_{\textrm{R}}\,$[MeV]$\quad$ & $\Gamma_{\textrm{R}}/2\,$[MeV]& $\left | \gamma_1 \right |\,$[GeV] &  $ \left| \gamma_2 \right|\,$[GeV]\\
			\hline
			\textbf{Fit1}\\
			III 
			& $3878.4^{+4.2}_{-3.4}$& $31.9^{+3.3}_{-5.8}$&$3.2^{+0.5}_{-0.5}$&$10.6^{+0.5}_{-1.6}$ \\
			\hline
			IV$_{\mathrm{(central\ v.)}}$ 
			& $3894.3^{+2.7}_{-3.2}$&$ 20.9^{+3.2}_{-7.1}$&$3.2^{+0.4}_{-0.6}$&$10.1^{+0.4}_{-1.1}$\\
			{ or} & & & &
			\\
                II 
			&$3900.8^{+4.1}_{-4.1}$&$ 1.58^{+1.56}_{-1.56}$&$4.7^{+0.6}_{-0.6}$&$8.2^{+0.9}_{-0.9}$\\
			\hline\hline
			Fit2\\
			III 
			& $3876.0^{+3.5}_{-4.6}$& $32.0^{+2.5}_{-1.6}$&$4.1^{+0.3}_{-0.3}$&$9.1^{+1.0}_{-0.7}$ \\
			\hline
			IV$_{\mathrm{(central\ v.)}}$ 
			& $3900.8^{+1.9}_{-1.9}$&$ 7.7^{+5.4}_{-4.1}$&$4.3^{+0.3}_{-0.4}$&$8.9^{+0.6}_{-0.4}$\\
			{ or} & & & &
			\\
                II 
			& $3902.2^{+1.2}_{-1.2}$  & $1.44 ^{+1.42}_{-1.42}$&$4.8 ^{+0.3}_{-0.3} $&$8.4^{+0.3}_{-0.3}$ \\
   			\hline\hline
			Fit3\\
			III 
			& $3820.9^{+28.9}_{-52.6}$& $41.5^{+23.0}_{-17.3}$&$4.7^{+0.6}_{-0.8}$&$16.2^{+4.5}_{-3.8}$\\
			\hline
			IV
			& $3858.6^{+11.9}_{-5.9}$&$ 29.4^{+4.6}_{-11.8}$&$4.1^{+0.3}_{-0.7}$&$15.4^{+1.7}_{-2.6}$\\
			
                \hline \hline
                Fit4 \\
			III 
			& $3874.8^{+3.7}_{-4.0}$& $32.7 ^{+1.6}_{-1.9}$&$4.3^{+0.3}_{-0.3}$&$8.7 ^{+0.8}_{-0.7}$ \\
			\hline
			IV$_{\mathrm{(central\ v.)}}$ 
			& $3902.4 ^{+1.1}_{-2.3}$&$ 3.3^{+6.3}_{-1.8}$&$4.6^{+0.2}_{-0.4}$&$8.6 ^{+0.6}_{-0.2}$
			\\
			{ or} & & & &\\
                II 
			& $3902.7^{+1.3}_{-1.3}$  & $3.0 ^{+2.4}_{-2.4}$&$4.9 ^{+0.2}_{-0.2} $&$8.3^{+0.3}_{-0.3}$ \\
			\hline\hline
		\end{tabular}
\end{table}

\begin{figure*}[ht!]
\centering
\begin{minipage}{0.495\linewidth}
		\centerline{\includegraphics[width=1.\textwidth]{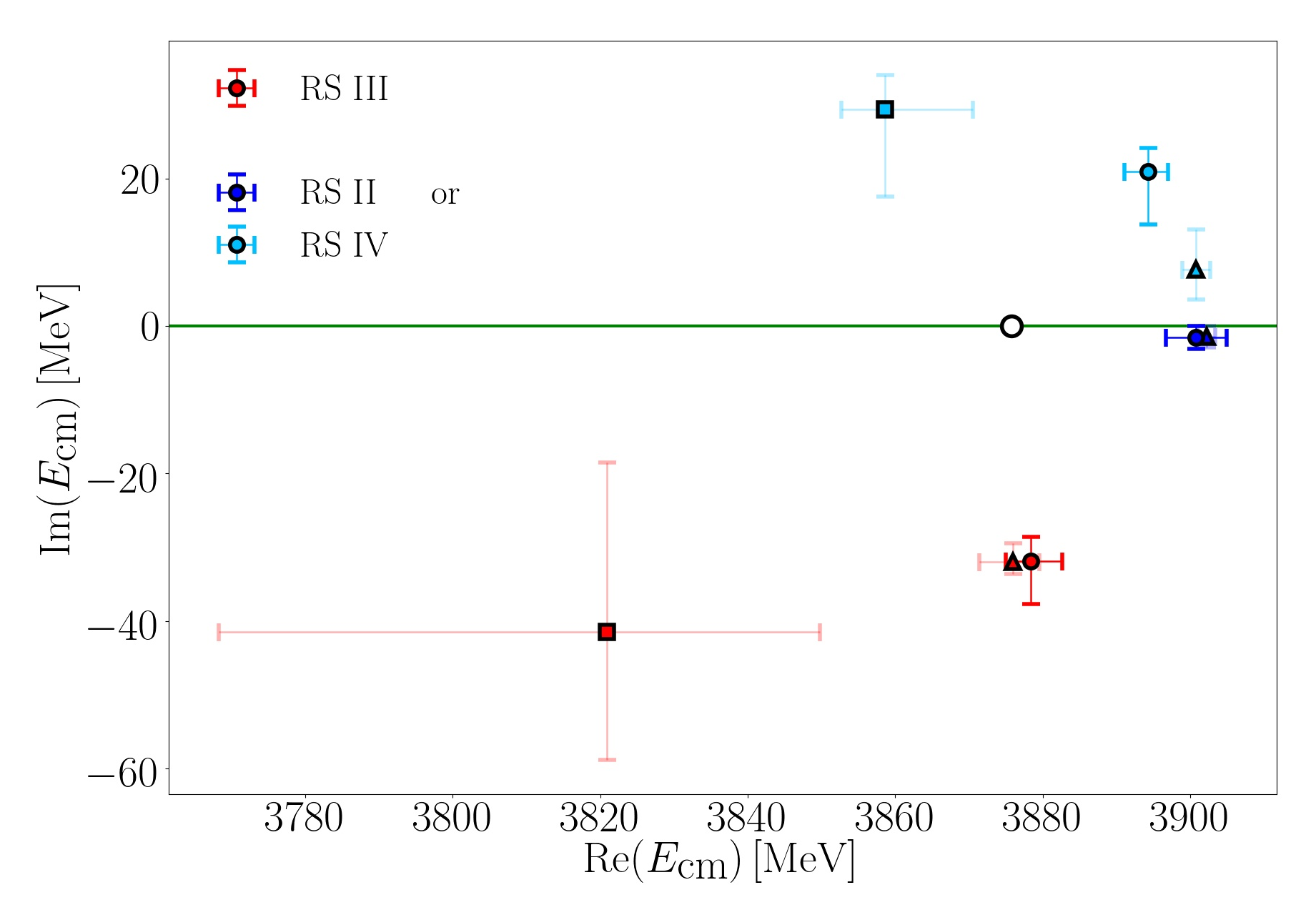}}
    \vspace{-0.3cm}
	\end{minipage}
\begin{minipage}{0.495\linewidth}
		\centerline{\includegraphics[width=1.\textwidth]{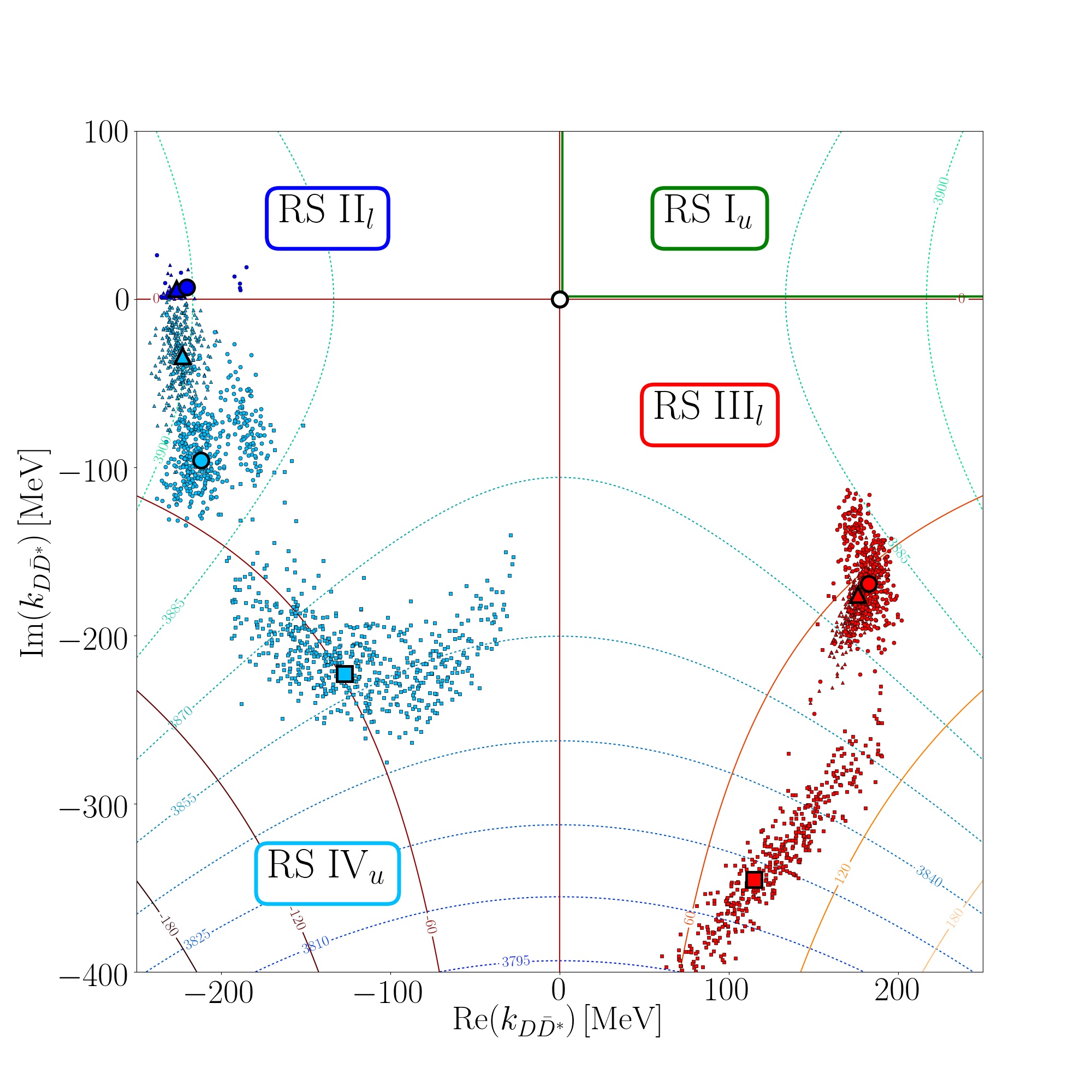}}
	\end{minipage}
\caption{Locations of the resonance poles of the scattering amplitude from Fit1 (circles), Fit2 (triangles) and Fit3 (squares). For more details, see Fig.~\ref{fig:poles:FitA}.}
\label{fig:poles:all}
\end{figure*}

Poles of the scattering amplitude from Fit1, Fit2 and Fit3 are presented in Fig.~\ref{fig:poles:all}. Let us now focus on those from Fit1 and Fit2. RS II/IV poles lie close to each other, which suggests that they are qualitatively the same. On the other hand, the pole on RS II/IV and one on RS III are not so close. Given that the mass of the experimentally observed $Z_c(3900)$ ($\approx \SI{3887}{MeV}$) \cite{Workman:2022ynf} is in between the real parts of both poles' complex energies, it is plausible that the $Z_c(3900)$ peak is a manifestation caused by both poles.

One can see that the poles from Fit3 are further away from the physical region. Their position is still close enough to produce the peaks in the line shapes (see Fig.~\ref{fig:eventsExp}), however, with somewhat smaller peaks in the $D\bar D^*$ distributions (see green lines).

\begin{figure}[h!]
	\centering
	\includegraphics[width=0.52\textwidth,angle=-0]{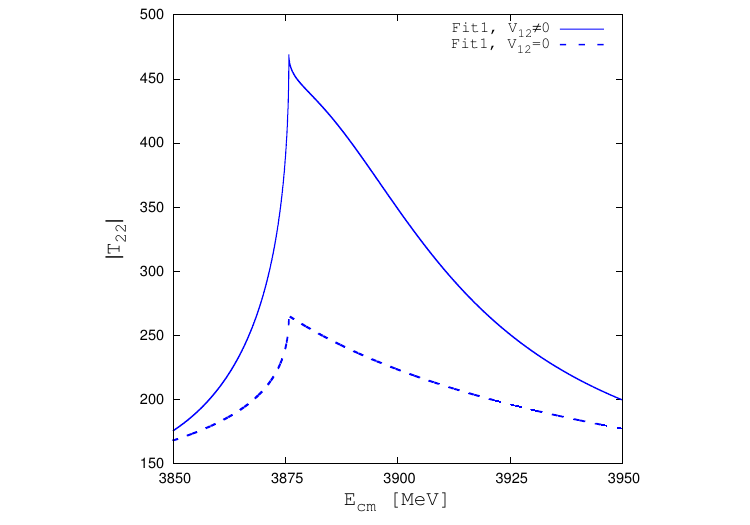}
	\caption{Magnitude of the diagonal $D\bar D^*\to D\bar D^*$ scattering amplitude, with and without $V_{12}$. This suggests that the coupling between both channels is important for the existence of the near-threshold peak and poles.}
	\label{fig:T22}
\end{figure}

We have checked that all the mentioned poles disappear by removing the off-diagonal matrix elements in the interaction kernel ($V_{12}$). The comparison of the magnitude of the $D\bar D^*\to D\bar D^*$ scattering amplitude by including/excluding the $J/\psi\pi\to D\bar D^*$ contribution is given in Fig.~\ref{fig:T22}. This indicates that the coupling between both channels is likely important for the existence of the $Z_c$ peak (see also Sec.~\ref{subsec:pheno1+-}).

\begin{figure*}[ht!]
	\centering
	\includegraphics[width=1.0\textwidth,angle=-0]{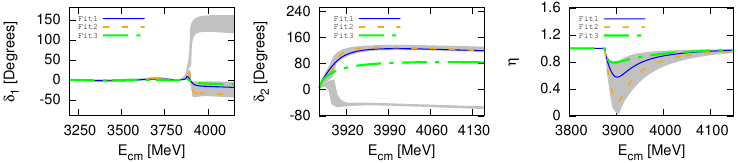}
	\caption{Phase shifts of $J/\psi\pi \to J/\psi\pi$ (left), $D\bar D^*\to D\bar D^*$ (center) and the inelasticity (right) as defined in \eqref{ENSmat}. Shown are the results for Fit1 (blue solid line), Fit2 (orange dashed line) and Fit3 (green dotted-dashed line) with the error band from Fit1.}
 	\label{fig:phase}
\end{figure*}

The pole singularities on the unphysical RSs in the complex energy plane have noticeable effects on the physical amplitudes, which can be characterized by the phase shifts ($\delta_1$, $\delta_2$) and the inelasticity ($\eta$) in the coupled-channel scattering matrix\footnote{We omit the labels $\ell=0$ for the phase shifts denoting the s-wave $J/\psi \pi$ and $D\bar D^*$ scattering.}
\begin{equation}
S=\left[\begin{array}{cc}
\eta e^{2i\delta_1} & i\sqrt{1-\eta^2}e^{i(\delta_1+\delta_2)} \\
i\sqrt{1-\eta^2}e^{i(\delta_1+\delta_2)}  & \eta e^{2i\delta_2}
\end{array}\right]\>.
\label{ENSmat}
\end{equation}
The predicted phase shifts and inelasticity for Fit1 are presented
in Fig.~\ref{fig:phase}. Within the defined uncertainties, two branches of phase shifts emerge above the $D\bar D^*$ threshold.
Our analysis affirms that the upper branch of $\delta_2$ and the lower branch of $\delta_1$ correspond to parameter samples with a pole on the RS IV. The other two branches correspond to parameter samples with a pole on the RS II. Therefore, the rise of $\delta_2$ through $90^\circ$ is likely caused by the pole on RS IV and not the one on RS II. We can not rule out that the pole on RS III also influences the $Z_c$ peak. Therefore, it is plausible that poles on RS IV and III can both influence the physical peak.

The $Z_c(3900)$ (and its strange partner $Z_{cs}(3985)$) is addressed in \cite{Du:2022jjv}, yielding a RS III pole with a position comparable to our results from Fits 1, 2 and 4, as shown in Table~\ref{tab:poles}. A similar pole position on RS III is reported\footnote{In fact, thanks to the communications of Meng-Lin Du and Yun-Hua Chen, another pole on RS IV is present in an almost identical location to the one on RS III in both mentioned studies \cite{Du:2022jjv,Chen:2023def}. However, such a pole is not highlighted since it is considered to have a minor influence on the physical region compared to the RS III pole.} in a study considering the $\pi\pi$ event distributions alongside the $J/\psi\pi$ and $D\bar{D}^*$ ones \cite{Chen:2023def}. To our knowledge, only one study reports two poles at different locations and being close to the threshold \cite{Zhou:2015jta}. Further details on the pole positions identified by other studies are provided in Sec.~\ref{subsec:pheno1+-}.

\subsection{Ratio of partial decay widths}
\label{subsec:br}

Based on the unitarized decay amplitudes, we estimate the ratio of $Z_c(3900)$ partial decay widths into $D\bar D^*$ and $J/\psi\pi$ channels, $R \equiv \Gamma_{Z_c\to D \bar D^*}/\Gamma_{Z_c\to J/\psi\pi}$. This involves performing phase space integrals within the energy range of $E_{\textrm{cm}} =(3900\pm35)\,$MeV,\footnote{We verified that the change of the energy range changes the final values by much less than their $1\sigma$ uncertainties.} following the same energy range and methodology proposed in Ref. \cite{ALBALADEJO2016337}. When evaluating the ratio of partial decay widths for $Z_c(3900)$, we exclude background contributions and set the $\alpha^{J/\psi \pi}_2$ and $\alpha^{D \bar D^*}_2$ factors---accounting for the normalizations of experimental event distributions—to unity. The following are the resulting ratios from various fits:
\begin{equation}
\begin{aligned}
	R&= 6.1^{+2.0}_{-2.3}~ \quad({\rm Fit1})\>,\\
 	R&= 2.4^{+1.0}_{-0.7}~ \quad({\rm Fit2})\>,\\
 	R&= 12.2^{+3.2}_{-3.0}~ \quad({\rm Fit3})\>,\\
  	R&= 1.9^{+0.9}_{-0.6}~ \quad({\rm Fit4})\>.
\label{ERat}
\end{aligned}
\end{equation}
As for other quantities, the agreement between Fit2 and Fit4 also manifests for $R$. Their values are somewhat lower than the experimentally determined $R_{\textrm{exp}} = \SI{6.2}{} \pm \SI{1.1}{} \pm \SI{2.7}{} $ \cite{Zc(3900)_exp_5}.
An interesting finding is that Fit1, with a somewhat larger $\chi^2$, is in closer agreement with experiment. On the other hand, the result from Fit3 is even larger. However, one has to be careful with any conclusions concerning this ratio since it was measured only once by BESIII back in 2014 \cite{Zc(3900)_exp_5}.

\subsection{Compositeness}
\label{subsec:compositeness}

We can estimate the composition of $Z_c(3900)$ based on our results. The partial two-hadron compositeness coefficient $X_i$ is calculated following Ref. \cite{Guo:2015daa}. It corresponds to the probability of having the two-body component from channel $i$ in the considered resonance state. Since we focus on the region around the $D\bar D^*$ threshold, we derive only the corresponding $X_2$ for three fits
\begin{equation}
\begin{aligned}
\label{eq.xivalue}
&
X_2^{\textrm{III}} = 0.65^{+0.05}_{-0.15}, \quad
X_2^{\textrm{IV}} =0.59^{+0.04}_{-0.09} \quad~ (\textrm{Fit1}), \\
&
X_2^{\textrm{III}} = 0.48^{+0.10}_{-0.07},  \quad
X_2^{\textrm{IV}} =0.46^{+0.05}_{-0.04} \quad~(\textrm{Fit2}), \\
& 
X_2^{\textrm{III}} = 0.43^{+0.09}_{-0.06},  \quad
X_2^{\textrm{IV}} =0.42^{+0.05}_{-0.02} \quad~ (\textrm{Fit4}).
\end{aligned}
\end{equation}
We do not consider compositeness coefficients for the Fit3 since the corresponding poles are significantly below the threshold energy, while the way to calculate $X$ in Ref.~\cite{Guo:2015daa} is valid only for resonances above the threshold. The resulting coefficients suggest that elementary and $D\bar D^*$ molecular Fock components are almost equally important. Fit1 slightly favors the molecular component, which is also consistent with the $R$ values in \eqref{ERat} since the $R$ for Fit1 is higher than for Fit2 and Fit4.

\subsection{Interpretations}
\label{subsec:interp}

The locations of both poles suggest that they can equally contribute to the physical peak. The ordinary fits (Fit1 and Fit2) of lattice and experimental data yield a satisfactory replication of the invariant mass distributions. The lattice eigenenergies lie close to noninteracting two-meson energies. The fits nicely reproduce lattice energies below and at the $D\bar D^*$ threshold, while reproduction above the threshold is not optimal. This does not significantly worsen the fit (in terms of the $\chi^2$) since most of the fitted points are experimental, not from the lattice. Still, we tried to put more emphasis on lattice data and performed fits involving the four decisive lattice energies with artificially reduced uncertainties (see Fit3 as the example). As expected, the prediction gets closer to those lattice energy points. One might expect that this fit would not reproduce the $Z_c(3900)$ peaks, but it does: the $J/\psi \pi$ invariant-mass distributions are nicely replicated. The peak in the $D\bar{D}^*$ distributions is still present but somewhat smaller.   
We conclude that this procedure of performing multiple joint fits can reasonably reconcile lattice data and the main features of the experimental data. In particular, the almost noninteracting lattice eigenenergies can be reasonably reconciled with the experimental peaks within the employed effective theory approach.  
All employed fits contain two poles relatively close to the threshold (within $\simeq \SI{50}{MeV}$), and the coupling between $D\bar D^*$ and $J/\psi \pi$ channels plays an important role. 

Let us address the possibilities one could consider when trying to explain the mentioned discrepancy between the global fit (Fit1 and Fit2) and our finite-volume energies slightly above the $D\bar D^*$ threshold.

The EFT omits the $D^*\bar D^*$ channel. This, nevertheless, is not expected to be the primary source of disagreement. It is unlikely that including the $D^*\bar D^*$ channel into our Lagrangians would render almost noninteracting energies, as observed in our simulation. In addition, we extracted the lattice spectrum without $D^*\bar D^*$ and $\eta_c\rho$ interpolators, and it is almost unmodified. Furthermore, considering the decays of $Z_c(3900)$ (near $D\bar D^*$ threshold) and $X(4020)$ (near $D^*\bar D^*$ threshold) into charmed mesons, experiments do not observe $X(4020)$ decaying into $D\bar D^*$, which is on the other hand, true for $Z_c(3900)$. This suggests that the $D\bar D^*$-$D^*\bar D^*$ channel coupling may be insignificant.

A possible reason for the inconsistency between finite-volume energies and the fit at higher energies could be the omission of pion-exchange contributions in EFT, which become significant as we move to higher energies where derivative/momentum suppression does not apply. Any effects from possible left-hand cuts are also omitted.

The origin of the discrepancy may also be inaccurate lattice energies, which do not correspond to the continuum limit. Furthermore, the basis of interpolating operators does not contain local diquark-antidiquark interpolators in our simulation and in \cite{CLQCD:2019npr}. This may be the limiting factor, although previous works \cite{Cheung2017,PRD_2015,Zc_Sasa_2} conclude that the addition of diquark-antidiquark interpolators does not modify the eigenenergies significantly. Finite-volume spectra on more lattice volumes would also be beneficial for more robust conclusions.

\section{Brief resume of phenomenological studies related to $1^{+-}$ and $1^{++}$ channels}\label{sec:pheno}

\subsection{Channel with $J^{PC}=1^{+-}$ and $I=1$}\label{subsec:pheno1+-}

In recent years, the theoretical exploration of the fascinating $Z_c$ states has gained considerable attention. This includes studies on hadronic molecules \cite{Wang:2013cya,Dong:2013iqa,Wilbring:2013cha,PhysRevD.90.016003,Gong:2016hlt,Chen:2023def,Liu:2024ziu,Liu:2024nac}, compact tetraquark states \cite{Maiani:2013nmn} and kinematic effects from a threshold cusp   \cite{Liu:2013vfa,Swanson:2014tra,Swanson:2015bsa} or a triangle singularity \cite{Liu:2015taa,Szczepaniak:2015eza,Chen:2023def,Liu:2024ziu}.

The JPAC collaboration performed a coupled-channel study of the $e^+e^-\to J/\psi \pi \pi$ and $\to D\bar{D}^*\pi$ data from BESIII \cite{Pilloni:2016obd}. They considered several amplitude parametrizations to constrain different phenomenological models for $Z_c(3900)$: compact tetraquark, hadronic molecule, virtual state and a pure kinematical enhancement. All the models fit the data reasonably well. 
The best fit is obtained for a compact tetraquark state; however, the other scenarios have not been ruled out.

An up-to-date analysis of the experimental data related to $Z_c(3900)$ suggests that both the molecular and nonmolecular components are comparable \cite{Chen:2023def}. This conclusion is drawn using a generalized expression for compositeness, which is not model independent \cite{Matuschek:2020gqe}.

The internal structure of the $Z_c(3900)$ is also investigated employing functional methods \cite{Hoffer:2024alv}. Three potential configurations are explored: a tight $\bar{c}c$ core surrounded by the light $\bar{q}q$ pair ($J/\psi\pi$), heavy-light meson clusters ($D\bar{D}^*$) and diquark-antidiquark. 
Norm contributions are computed using the on-shell Bethe-Salpeter amplitude. 
The results suggest that the dominant and subdominant components are $J/\psi\pi$ and $D\bar{D}^*$, respectively, with a substantial $J/\psi\pi$-$D\bar{D}^*$ mixing.

The vicinity of $Z_c$ peak to the threshold led several studies to suggest that this peak might just be a result of a kinematical effect \cite{Swanson:2014tra}. However, if one wants to fit the $Z_c$ within a perturbative approach, the presence of a pronounced peak calls for such a large coupling constant that using a perturbative approach is not justified \cite{Guo:2014iya}. The resummation of the whole series leads to a near-threshold pole and qualifies for a state \cite{Guo:2014iya}.

Our findings emphasize the crucial role of the cross-channel interaction $J/\psi\pi$-$D\bar{D}^*$, as illustrated in Fig.~\ref{fig:T22}. The lattice studies from the HAL QCD collaboration \cite{PhysRevLett.117.242001,Ikeda_2018} also support that this cross-channel interaction is crucial for the existence of $Z_c$.  
A similar conclusion is drawn in Ref. \cite{Ortega2019}, where the constituent quark model with coupled channels ($J/\psi\pi$, $\eta_c\rho$, $D\bar D^*$, $D^*\bar D^*$) is utilized. The study reveals that the $Z_c(3900)$ signal is due to a pole just below the $D\bar{D}^*$ threshold and finds that the nondiagonal $J/\psi\pi$-$D^{(*)}\bar D^*$ and $\eta_c\rho$-$D^{(*)}\bar D^*$ couplings are crucial for the $Z_c$ peak.  

There are many works that report poles near $D\bar D^*$ threshold, either above \cite{PhysRevD.90.016003,Pilloni:2016obd,ALBALADEJO2016337,Du:2022jjv,Chen:2023def} or below \cite{PhysRevD.88.054007,ALBALADEJO2016337,He2018,Ortega2019} the threshold. 
Two poles near the threshold are explicitly reported in Ref.~\cite{Zhou:2015jta}. It utilizes a unitarized coupled-channel model and simultaneously analyses the three invariant mass distributions of $J/\psi\pi^{\pm}$, $(D\bar D^*)^{\pm}$ and $(D^*\bar D^*)^{\pm}$ in the $e^+e^-$ production processes around $\SI{4.26}{GeV}$ \cite{Zc(3900)_exp_1,Zc(3900)_exp_5,Zc(4020)_exp_2}. Their results suggest that the $Z_c$ signal is caused by the two poles at $m=\SI{3846}{MeV}$, $\Gamma/2=\SI{19}{MeV}$ (RS II) and $m=\SI{3875}{MeV}$, $\Gamma/2=\SI{16}{MeV}$ (RS III), where the one on RS III contributes dominantly to the $Z_c$ peak.

\subsection{Channel with $J^{PC}=1^{++}$ and $I=1$}\label{subsec:pheno1++}

There are not many studies on the $1^{++}$ state with isospin 1 that would be an isospin partner of the $\chi_{c1}(3872)$. The isospin-1 state is anticipated in the diquark-antidiquark picture \cite{Maiani:2004vq,Anwar:2018sol}. The leading order of nonrelativistic EFT also predicts such a state within the molecular scenario \cite{Ji:2022uie}. A chiral EFT has recently advanced this understanding by incorporating one-pion exchange and three-body dynamics \cite{Zhang:2024fxy}. They predict the existence of a molecular isospin-1 state related to a virtual state pole and argue that it has not been observed in experiment since it would show up only as either a mild cusp or dip at the threshold. Their results are similar to ours, where a virtual state pole is found in the one-channel $D\bar{D}^*$ scattering (see Sec.~\ref{sec:scattering}).

\section{Elastic s-wave $J/\psi \pi$ scattering in the $1^{+-}$ channel}\label{sec:jpsipi}

\begin{figure}[h!]
	\centering
	\includegraphics[width=0.38\textwidth,angle=-0]{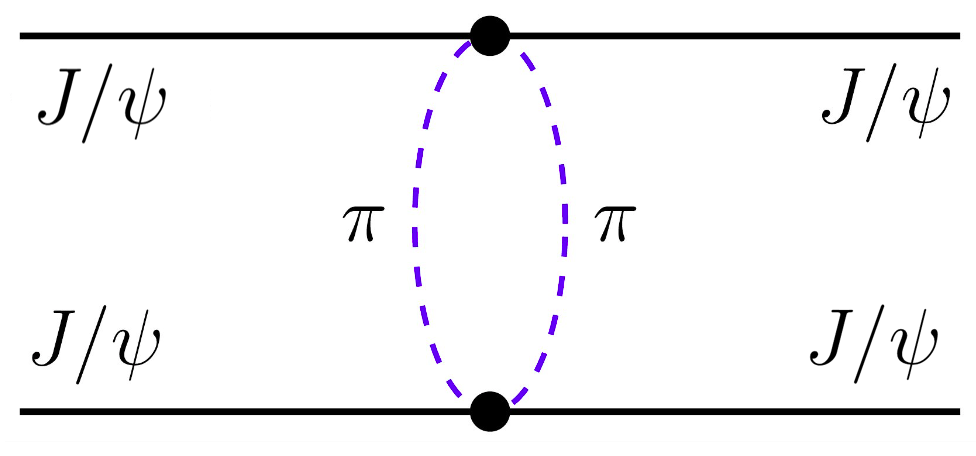}
	\caption{Contribution to the scattering of two $J/\psi$ mesons.}
	\label{fig:JpsiJpsi}
\end{figure}

In this section, we examine elastic s-wave $J/\psi \pi$ scattering for $J^{PC}=1^{+-}$. We investigate this process for two reasons. First, the lowest threshold is $m_{J/\psi}+m_{\pi}$, and the lowest few finite-volume states correspond to the $J/\psi \pi$ channel. Second, the strength of the $J/\psi \pi$ interaction provides insight into whether $\pi\pi$ exchange contributions are sufficiently attractive to render a $\bar cc\bar cc$ state near the $J/\psi J/\psi$ threshold \cite{Dong:2021lkh}, as detailed at the end of this section. 
 This $\pi\pi$ exchange in $J/\psi J/\psi$ scattering can be seen in Fig.~\ref{fig:JpsiJpsi} where each vertex represents the $J/\psi \pi\to J/\psi \pi$ coupling. 

\begin{figure}[h!]
\includegraphics[width=0.55\textwidth,angle=-0,left]{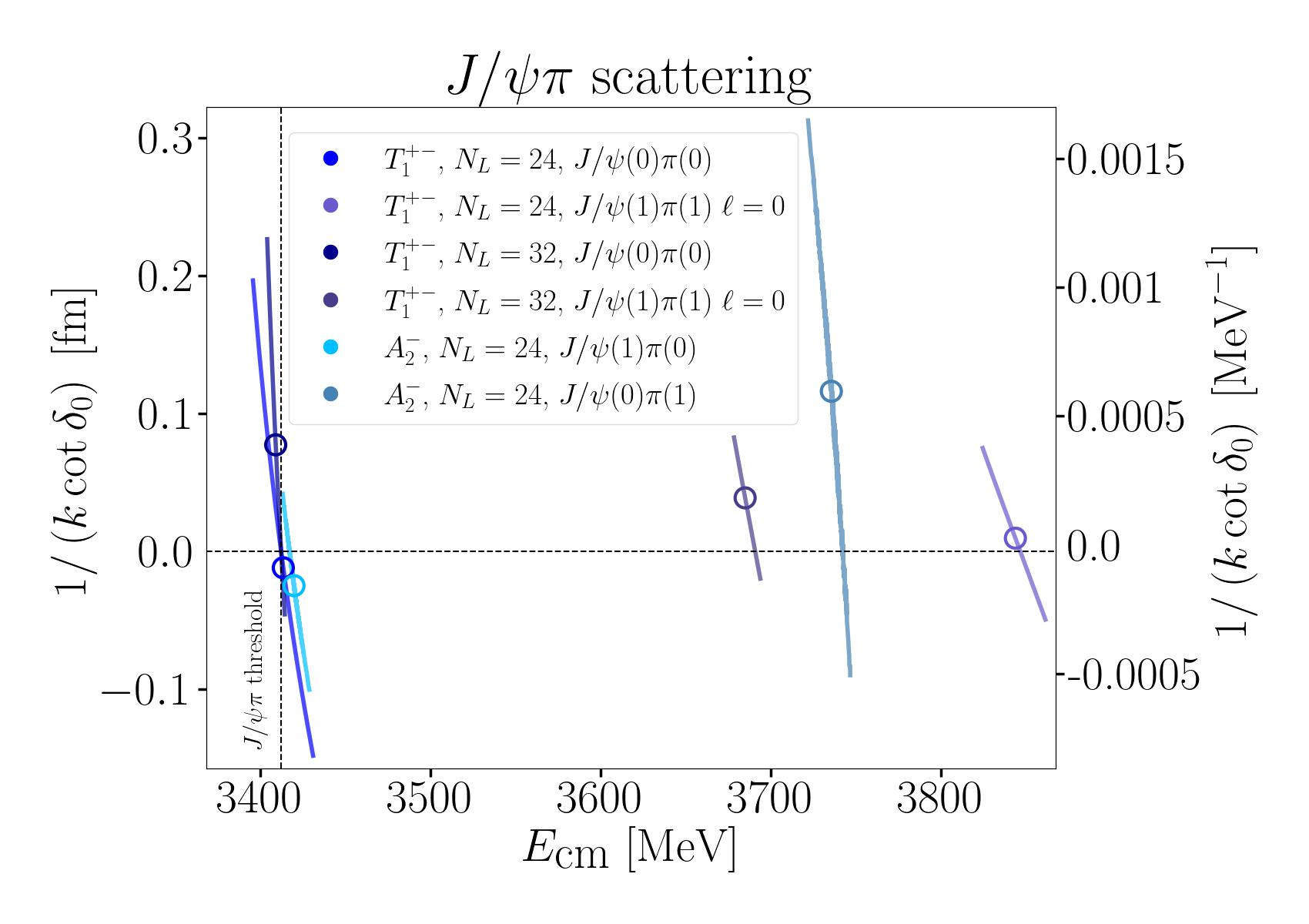}
	\caption{Inverse $k\cot{\delta_{0}}$ from s-wave $J/\psi \pi$ scattering derived from our lattice finite-volume energies. Values around zero represent no interaction. The energy range represents the elastic scattering region.}
	\label{fig:Jpsipi}
\end{figure}

Our lattice energy levels dominated by $J/\psi \pi$ interpolators reveal no substantial interaction between $J/\psi$ and $\pi$. This is expected, as $J/\psi$ and $\pi$ mesons do not contain any valence quarks of the same flavor.
When checking individual energy levels (see Figs. \ref{fig:T1pm} and \ref{fig:A2m}), one encounters exceptions only for the few highest levels. The rest closely align with the free energies, deviating by no more than 1$\sigma$.
To make the connection to the scattering amplitude, values for inverse $k\cot{\delta_{0}}$ are computed from our energy levels via \eqref{ELuscher} and shown in Fig.~\ref{fig:Jpsipi}. Only those below the next threshold, $\eta_c\rho$, are shown. They are all zero within $1\sigma$ uncertainties, corresponding to the noninteracting scenario. 
This gives us upper bounds on the scattering length $|a_{0}|$ near the $J/\psi\pi$ threshold (within the approximation $k\cot{\delta_{0}}\approx1/a_0$)
\begin{equation}
\left|a_{0}\right|\lesssim\SI{0.1}{fm}
\label{Eabound}
\end{equation}
and bound on $\left |1/(k\cot{\delta_{0}})\right |$ elsewhere\hfill
\resizebox{.9\linewidth}{!}{
  \begin{minipage}{\linewidth}
\begin{equation}
\begin{aligned}
\left |1/(k\cot{\delta_{0}(k)})\right |& \lesssim \SI{0.0005}{MeV^{-1}}, \ E_{\textrm{cm}}-m_{J/\psi}-m_{\pi}\approx\SI{300}{MeV}\>,\\
	\left |1/(k\cot{\delta_{0}(k)})\right |&\lesssim \SI{0.0015}{MeV^{-1}}, \  E_{\textrm{cm}}-m_{J/\psi}-m_{\pi}\approx\SI{350}{MeV}\>,
\label{Ekcotbound}
\end{aligned}
\end{equation}
  \end{minipage}
} 
which are all determined at our unphysical pion mass of $m_\pi\simeq \SI{280}{MeV}$. This also constrains the $J/\psi J/\psi$ interaction, as discussed below.

 We also compute the $J/\psi\pi$ scattering length at physical meson masses from the coupled-channel amplitude~\eqref{ET} obtained from fitting the experimental line shapes and lattice data: the results from the different fits in Table~\ref{tab:param} are in the range $|a_0|\simeq 0.01$--$\SI{0.06}{fm}$, which is consistent with the bound in Eq.~\eqref{Eabound} at $m_\pi\simeq\SI{280}{MeV}$.

To our knowledge, only two previous lattice determinations of the $J/\psi \pi$ scattering length are available. The quenched simulation of \cite{Yokokawa:2006td} renders $a_0=\SI{0.0119(39)}{fm}$, while the dynamical simulation of  \cite{Liu:2008rza,Liu:2009af} gives the upper bound $|a_0|\lesssim\SI{0.01}{fm}$. These values are extracted at the physical point from chiral extrapolations of three or four different unphysical values of $m_{\pi}$. These results are significantly more precise than ours. However, they are not directly comparable as our bound \eqref{Eabound} applies at $m_{\pi}\simeq \SI{280}{MeV}$. A phenomenological evaluation from Ref.~\cite{Liu:2012dv} suggests $|a_0|\lesssim\SI{0.02}{fm}$. Another phenomenological estimate that uses the central value of a quark model calculation of the charmonium chromopolarizability \cite{Brambilla:2015rqa,Dong:2022rwr} and relations from \cite{Dong:2021lkh} predicts $|a_0|\approx\SI{0.04}{fm}$.  

Let us now discuss the possible implications of $J/\psi \pi$ interaction for the $\bar cc\bar cc$ tetraquarks. The spectroscopy of fully charm tetraquarks became more active in 2020 when LHCb detected the $X(6900)$ in the $J/\psi J/\psi$ final state, supporting the $cc\bar{c}\bar{c}$ quark content \cite{LHCb:2020bwg}. Subsequently, additional significant peaks were observed, representing further states with the same proposed quark content. There are interpretations that these states are compact tetraquarks \cite{Zhang:2022qtp,Wang:2022yes,An:2022qpt,Lu:2020cns,liu:2020eha,Yu:2022lak,Deng:2020iqw,Dong:2022sef,Maiani:2022psl,Ortega:2023pmr,Wang:2023jqs,Anwar:2023fbp,Nefediev:2021pww} or hadronic molecules \cite{Dong:2020nwy,Dong:2021lkh,Baru:2022vmi}. Latter studies suggest that the exchange of correlated light mesons, such as two pions, contributes significantly to the attraction within a double-charmonium system. The absence of any interaction between $J/\psi$ and $\pi$ would disfavor this scenario.

The $J/\psi \pi$ scattering length and the long-range interaction potential between $J/\psi J/\psi$ are connected through the contact interaction vertex $J/\psi J/\psi \pi \pi$ of the chiral effective theory (see Fig.~\ref{fig:JpsiJpsi}). These two different channels are related in Ref.~\cite{Dong:2021lkh}, where the authors conclude that a relatively small s-wave $J/\psi \pi$ scattering length $\mathcal{O}(\SI{0.01}{fm})$ is significant enough to explain the attraction between two $J/\psi$ mesons via the exchange of correlated light mesons to a near-threshold state. Such small scattering lengths are allowed, but not guaranteed, from the only available dynamical QCD simulations that render only the upper bounds $\left|a_{0}\right|\lesssim\SI{0.1}{fm}$ (present work) and $|a_0|\lesssim\SI{0.01}{fm}$ \cite{Liu:2008rza,Liu:2009af}. Therefore, progress on the role of two-pion exchange calls for a more precise determination of the $J/\psi\pi$ scattering length from lattices with dynamical quarks. This is particularly challenging due to the smallness of this interaction.

\section{Conclusions}\label{sec:conclusions}
 
A charmoniumlike tetraquark $Z_c^+(3900)$ with charged decay channels and $I(J^{PC})=1(1^{+-})$ was discovered in 2013. Its exotic quark content is $\bar{c}c\bar{q}q$, and its peak lies just above the $D\bar{D}^*$ threshold. Our investigation aims to understand whether the interaction between $D\bar{D}^*$ and its coupling to $J/\psi \pi$ contributes to the existence of $Z_c(3900)$ in the $1^{+-}$ sector. Furthermore, we explore whether the $D\bar{D}^*$ interaction provides clues about an undiscovered near-threshold state in the $1(1^{++})$ sector.

This work investigates the channels $\bar{c}c\bar{q}q$ with $I=1$ and $J^{PC}=1^{+\pm}$ on the lattice. Our and other lattice finite-volume energies for $1^{+-}$ are compared to experimental $Z_c(3900)$ data via an EFT. The main challenge is that such states decay to $D\bar{D}^*$ as well as lower-lying final states of a charmonium and a light meson.

In the lattice part of the study, meson-meson interpolator types \eqref{HLMM} are employed, and the finite-volume spectrum is extracted (see Figs.~\ref{fig:T1pm}--\ref{fig:A2pCloseup}). This is the first study considering hadronic states with $1^{+\pm}$, a nonzero total momentum and two different lattice volumes. The operator basis omits $\psi(2S)\pi$, $\psi(3770)\pi$ and three-particle  operators. The determined eigenenergies are compared to the noninteracting energies of the corresponding two-hadron systems $\bar{c}c+\bar{q}q$ and $\bar{c}q+\bar{q}c$. 
The number of extracted energy levels equals the number of expected noninteracting meson-meson levels. All eigenenergies lie relatively close to the noninteracting meson-meson energies, which agrees with the findings of all previous lattice studies of these channels. Only certain eigenstates, predominantly coupled to $D\bar D^*$, exhibit slightly negative energy shifts, which are more pronounced in the $1^{++}$ compared to the $1^{+-}$ channel.

After extracting the finite-volume energies, our first approach is to investigate whether single-channel $D\bar D^*$ scattering supports the existence of $Z_c(3900)$ and its $1^{++}$ partner. This part of the study employs the standard Lüscher formalism for one-channel s-wave $D\bar D^*$ scattering and the energy dependence is parametrized via the effective range expansion. The resulting scattering amplitudes exhibit a virtual state pole slightly below the threshold in both sectors, $1^{+-}$ and $1^{++}$. However, we acknowledge that this approach neglects the impact of the left-hand cut and assumes that $D\bar{D}^*$ does not couple to other channels, which may not be valid assumptions. Due to these limitations, our emphasis lies in the results discussed below.

The second procedure applies to $1^{+-}$ and involves $J/\psi \pi$-$D\bar{D}^*$ coupled-channel scattering based on a contact EFT. Considering contact interactions $D\bar{D}^*$-$D\bar{D}^*$ and $D\bar{D}^*$-$J/\psi \pi$, we jointly fit the experimental $J/\psi \pi$ and $D\bar{D}^*$ invariant-mass distributions from \mbox{BESIII} \cite{PhysRevLett.119.072001,PhysRevD.92.092006} and lattice finite-volume energies (from our work and others \cite{Cheung2017,CLQCD:2019npr}) with an effective field theory approach. We perform four different fits, which differ in treating higher-lying energy levels from our lattice study. All fits reasonably reproduce $Z_c$ peak in the experimental $J/\psi \pi$ and $D\bar D^*$ line shapes. They all reproduce also our and other lattice energy levels up to slightly above the $D\bar D^*$ threshold. The reproduction of our higher-lying energy levels is more favorable in fits where these levels are incorporated with artificially reduced uncertainties. This implies that the employed EFT model can reconcile the peaks in the experimental line shapes and the lattice energies that lie close to noninteracting energies. 
 All the fits yield two poles within $\simeq\SI{50}{MeV}$ of the $D\bar D^*$ threshold: both (RS III and RS IV) poles are possibly equally responsible for the $Z_c$ peak. The central values of the two pole locations for the preferred conventional fit lie relatively close to the experimental $Z_c$ pole from PDG, as shown in Fig.~\ref{fig:poles:Fit1_exp_and_sketch}. Both poles may be manifestations of the same $Z_c(3900)$ state. Another common feature of all fits is that the coupling between the $D\bar D^*$ and $J/\psi\pi$ channels is significant and plays a crucial role, in agreement with the findings of the HAL QCD study \cite{PhysRevLett.117.242001}.

Although the lattice determination of finite-volume spectrum in $Z_c$ channel represents a considerable challenge, we expect that the energies employed in the scattering analysis are roughly within the rather conservative errors provided. One conclusion from this work is that even lattice energy shifts consistent with zero or small (observed in this and previous lattice simulations) can be reasonably reconciled with the experimental line shapes and the presence of a pole. We believe this conclusion would not be modified even if some of our energy levels were slightly different. 

Lastly, elastic s-wave $J/\psi \pi$ scattering with $1^{+-}$ is considered and the resulting $1/(k\cot{\delta_{0}})$ is presented in Fig.~\ref{fig:Jpsipi}. It is zero within the uncertainties, which implies nearly noninteracting $J/\psi$ and $\pi$. We, therefore, set an upper bound on the $J/\psi\pi$ scattering length and compare it to other studies. It would be important to determine this scattering length in the future more precisely since one could use it to explain the binding of $J/\psi$-$J/\psi$ mesons via the exchange of correlated light mesons to a near-threshold bound state.

The shortcomings of the present work could be improved by future lattice studies. Improved actions and smaller lattice spacings would reduce possible discretization effects. The method of optimized distillation profiles introduced in Ref. \cite{Knechtli:2022bji} and used in \cite{Urrea-Nino:2023ysv,Neuendorf:2024ekv} could be applied to define operators with large overlaps onto energy eigenstates of interest, significantly improving on standard distillation \cite{HadronSpectrum:2009krc}.
Our large (statistical) uncertainties call for ensembles with higher statistics.
Since $\rho$ and $a_0$ resonances are involved, one should perform three-particle scattering. To tackle this, one has to incorporate the scattering formalism outlined, for example, in \cite{Hammer:2017kms,Mai:2017bge,Hansen:2019nir}, where the presence of the left-hand cut is not discussed. It is advisable to incorporate techniques that address the left-hand cut \cite{Du:2023hlu,Raposo:2023oru,Meng:2023bmz,Hansen:2024ffk} even before attempting to reduce the pion mass. Simulating at the physical pion mass will introduce challenges related to strongly decaying $D^*\to D\pi$ (consequently $Z_c(3900)\to D\bar{D}\pi$). To incorporate $D\bar{D}\pi$ effects, the method from Ref. \cite{Hansen:2024ffk} can be used.
Figures \ref{fig:our_P0-P1}--\ref{fig:our-CLQCD-HSC+Fit3} with predictions of discrete lattice spectra in a wide range of volume sizes can provide helpful guidelines for future lattice calculations.

\section*{Acknowledgments}
We thank Michael D\"oring, David R. Entem, Feng-Kun Guo, Christoph Hanhart, Luka Leskovec, Marko Maležič, Mikhail Mikhasenko, Daniel Mohler, Raquel Molina, Alexey V. Nefediev, Pablo G. Ortega, and Eulogio Oset for valuable discussions. 
We are grateful to the members of CLS for the effort in generating the gauge field ensembles, which form a basis for our computation.
We thank Christian B. Lang and Stefano Piemonte for contributions to the computing codes we used and the authors of the {\footnotesize\textsc{TwoHadronsInBox}} package Ref. \cite{MORNINGSTAR2017477} for making the {\footnotesize\textsc{TwoHadronsInBox}} package public.
Our code implementing the distillation method was written within the framework of the Chroma software package \cite{EDWARDS2005832}.
The correlators were computed on two HPC systems: we gratefully acknowledge the HPC RIVR consortium (\href{https://www.hpc-rivr.si/}{www.hpc-rivr.si}) with EuroHPC JU (\href{https://eurohpc-ju.europa.eu/}{eurohpc-ju.europa.eu}) for one part and the University of Regensburg for funding this research by providing computing resources of the HPC system Vega at the Institute of Information Science (\href{https://www.izum.si/en/home/}{www.izum.si}) and the HPC-Cluster Athene, respectively.
M. S. acknowledges the financial support by the Slovenian Research and Innovation Agency ARIS (Grant No. 53647). S. C. acknowledges the support from the DFG Grant SFB/TRR 55. The work of S. P. is supported by the Slovenian Research and Innovation Agency ARIS (research core funding No. P1-0035 and No. J1-8137) and (at the beginning of the project) DFG Grant No. SFB/TRR 55. M.P. gratefully acknowledges support from the Department of Science and Technology, India, SERB Start-up Research Grant No. SRG/2023/001235 and Department of Atomic Energy, India. Z. H. G. is partially supported by the Natural Science Foundation of China (NSFC) under Contracts No.~12475078, No.~12150013 and No.~11975090, and the Science Foundation of Hebei Normal University with Contract No.~L2023B09. 

\appendix

\section{Masses of mesons on our lattices}\label{sec:app:masses}

The masses of single hadrons from lattice and experiment are listed in Table~\ref{tab:m}. The masses from lattice reveal a slight dependence on the lattice size $N_L=24,~32$. Consequently, the curves representing the noninteracting energies of a meson pair with zero momentum are not precisely horizontal, as previously elaborated upon in Sec.~\ref{sec:results}. The difference of $m_1+m_2$ (for all channels studied) from the two volumes is smaller than the statistical uncertainties of eigenenergies for $\bar cc\bar du$ system, and the results alter by less than one sigma depending on which volume is chosen as an input for the stable meson masses.

\begin{table*}[ht!]
\caption{Masses of relevant mesons measured on both lattices. The experimental masses are also given, where (if applicable) the masses of both the charged and neutral particles are shown.}
\label{tab:m}
\centering
\begin{tabularx}
{1.\textwidth}
{m{0.09\linewidth}m{0.14\linewidth}m{0.19\linewidth}m{0.19\linewidth}m{0.12\linewidth}m{0.12\linewidth}m{0.12\linewidth}}
\hline\hline
 & $m_{\pi^{\pm}}/m_{\pi^0}$ [MeV] & $m_{D^{\pm}}/m_{D^0}$ [MeV] & $m_{D^{*\pm}}/m_{D^{*0}}$ [MeV] & $m_{J/\psi}$ [MeV] & $m_{\eta_c}$ [MeV] & $M_{\textrm{av}}$ [MeV] \\ \hline
$N_L=24$ & $289(4)$ & $1933(2)$ & $2055(3)$ & $3126(1)$ & $3028(1)$ & $3101(1)$ \\ 
$N_L=32$ & $280(3)$ & $1927(1)$ & $2049(2)$ & $3129(1)$ & $3030.7(3)$ & $3104.3(4)$ \\
exp. \cite{Workman:2022ynf} & $139.57/134.98$ & $1869.66(5)/1864.84(5)$ & $2010.26(5)/2006.85(5)$ & $3096.90(1)$ & $2983.9(4)$ & $3068.7(1)$ \\ \hline\hline
\end{tabularx}
\end{table*}

\section{Lists of implemented two-hadron interpolators}\label{sec:app:interpolators}

All  interpolators incorporated for $1^{+-}$ and $1^{++}$ are schematically shown in Tables~\ref{tab:Cm} and \ref{tab:Cp}, respectively. The multi\-plicity of all linearly independent interpolators is indicated by ``$[\times 2]$'', ``$[\times 3]$''.

Single-meson pseudoscalar and vector operators are realized using $\gamma_5$ and $\gamma_i$, respectively. In addition, specific meson-meson interpolators are duplicated (as indicated by ``$\times 2$'') by using also $\gamma_5\gamma_t$ and $\gamma_i\gamma_t$, for example
\begin{equation}
\begin{aligned}
O^{J/\psi(0)\pi(0)}_{S=1,\ell=0,J=1}&=\bar c\gamma_z c({\bi 0})\, \bar q\gamma_5 q({\bi 0}), \quad \bar c\gamma_z\gamma_t c({\bi 0})\,\bar q\gamma_5\gamma_t q({\bi 0})\>.\nonumber
\label{EJpsipideggamma4}
\end{aligned}
\end{equation}

\begin{table*}[ht!]
\caption{Table of employed meson-meson interpolators \eqref{HLMM} transforming under irreps $\Lambda^{PC}=T_1^{+-}$ and $\Lambda^{C}=A_2^{-}$ which correspond to $J^{PC}=1^{+-}$. All momenta $|{\bi P}|^2$, $|{\bi p}_i|^2$ are in units of $(2\pi/L)^2$. The interpolators below the partition horizontal line are in addition to those above added to the ensemble with $N_L=32$.}
\label{tab:Cm}
\centering
\begin{tabular}{cccc|ccc} 
\hline
\hline
\multicolumn{4}{c|}{$|{\bi P}|^2=0,\ \Lambda^{PC}=T_1^{+-}$}             & \multicolumn{3}{c}{$|{\bi P}|^2=1,\ \Lambda^{C}=A_2^{-}$}  \\ 
\hline
                      &                  &  $J/\psi(0)\pi(0)$    &  $\times 2$ &  $J/\psi(1)\pi(0)$  &  $\times 2$ &                              \\
                      &                  &  $J/\psi(1)\pi(1)$    &  [$\times 2$] &  $J/\psi(0)\pi(1)$  &  $\times 2$ &                              \\
                      &                  &  $J/\psi(2)\pi(2)$    &  [$\times 3$] &  $J/\psi(2)\pi(1)$  &  [$\times 2$] &                              \\
{ $N_L=24$}              &                  &  $\eta_c(0)\rho(0)$   &            &  $J/\psi(1)\pi(2)$  &  [$\times 2$] &                              \\
{ 15 interpolators}      & $N_L=32$         &  $\eta_c(1)\rho(1)$   &  [$\times 2$] &  $J/\psi(4)\pi(1)$  &            &                              \\
                      & 21 interp. &  $\bar{D}^*(0)D(0)$   &  $\times 2$ &  $\eta_c(1)\rho(0)$ &            &  $N_L=24$                     \\
                      &                  &  $\bar{D}^*(1)D(1)$   &  [$\times 2$] &  $\eta_c(0)\rho(1)$ &            &  21 interp.             \\
                      &                  &  $\bar{D}^*(0)D^*(0)$ &            &  $\eta_c(2)\rho(1)$ &  [$\times 2$] &                              \\
\hhline{=~~~~~~} \cline{3-4}
\multicolumn{1}{c}{} &                  & $J/\psi(3)\pi(3)$    & [$\times 2$] &  $\bar{D}^*(0)D(1)$ &  $\times 2$ &                              \\ 
\multicolumn{1}{c}{} &                  & $\eta_c(2)\rho(2)$   & [$\times 3$] &  $\bar{D}^*(1)D(0)$ &  $\times 2$ &                              \\
\multicolumn{1}{c}{} &                  & $h_c(1)\pi(1)$       &            &  $\bar{D}^*(1)D(2)$ &  [$\times 2$] &                              \\
\multicolumn{1}{c}{} &                  &                      &            &  $\bar{D}^*(2)D(1)$ &  [$\times 2$] &                              \\
\hhline{~======}
\end{tabular}
\end{table*}

\begin{table*}[ht!]
\caption{Table of interpolators \eqref{HLMM} transforming under irreps $\Lambda^{PC}=T_1^{++}$ and $\Lambda^{C}=A_2^{+}$ which correspond to $J^{PC}=1^{++}$. All momenta $|{\bi P}|^2$, $|{\bi p}_i|^2$ are in units of $(2\pi/L)^2$. The interpolators below the partition horizontal lines are in addition to those above added to the ensemble with $N_L=32$.}
\label{tab:Cp}
\centering
\begin{tabular}{cccc|cccc} 
\hline
\hline
\multicolumn{4}{c|}{$|{\bi P}|^2=0, \ \Lambda^{PC}=T_1^{++}$}                                                   & \multicolumn{4}{c|}{$|{\bi P}|^2=1, \ \Lambda^{C}=A_2^{+}$}                                  \\ 
\hline
\multicolumn{1}{c}{ $N_L=24$}        & \multicolumn{1}{c}{}                 &  $J/\psi(0)\rho(0)$   &            &  $\eta_c(1)a_0(0)$    &            &                  &                                        \\
\multicolumn{1}{c}{ $\ 5\ $ interpolators} & \multicolumn{1}{c}{}                 &  $\bar{D}^*(0)D(0)$   &  $\times 2$ &  $\chi_{c0}(1)\pi(0)$ &            &                  &                                        \\
\multicolumn{1}{c}{}                & \multicolumn{1}{c}{$N_L=32$}         &  $\bar{D}^*(1)D(1)$   &  [$\times 2$] &  $\chi_{c0}(0)\pi(1)$ &            &                  &                                        \\ 
\hhline{=~~~~~~~}\cline{3-4}
\multirow{10}{*}{}                    & \multicolumn{1}{c}{10 interp.} & $J/\psi(1)\rho(1)$   & [$\times 3$] &  $J/\psi(1)\rho(0)$   &            &                  &  $N_L=24$                               \\
                                      & \multicolumn{1}{c}{}                 & $\chi_{c0}(1)\pi(1)$ &            &  $J/\psi(0)\rho(1)$   &            &                  &  13 interp.                       \\
                                      & \multicolumn{1}{c}{}                 & $\chi_{c1}(1)\pi(1)$ &            &  $\bar{D}^*(0)D(1)$   &  $\times 2$ & $N_L=32$         &                                        \\ 
\hhline{~===~~~~}
                                      & \multirow{7}{*}{}                     &                      &            &  $\bar{D}^*(1)D(0)$   &  $\times 2$ & 17 interp.&                                        \\
                                      &                                       &                      &            &  $\bar{D}^*(1)D(2)$   &  [$\times 2$] &                  &                                        \\
                                      &                                       &                      &            &  $\bar{D}^*(2)D(1)$   &  [$\times 2$] &                  &                                        \\ 
\cline{5-6}\hhline{~~~~~~~=}
                                      &                                       &                      &            & $\eta_c(0)a_0(1)$    &            &                  & \multicolumn{1}{c}{\multirow{4}{*}{}}  \\
                                      &                                       &                      &            & $\chi_{c0}(2)\pi(1)$ &            &                  & \multicolumn{1}{c}{}                   \\
                                      &                                       &                      &            & $\chi_{c0}(4)\pi(1)$ &            &                  & \multicolumn{1}{c}{}                   \\
                                      &                                       &                      &            & $\chi_{c1}(2)\pi(1)$ &            &                  & \multicolumn{1}{c}{}                   \\
\hhline{~~~~===~}
\end{tabular}
\end{table*}

\section{Overlaps between eigenstates and operators}\label{sec:app:overlaps}

The correlation matrix \eqref{E0} is constructed with the interpolating operators $O_i$ in \eqref{HLMM}, with the quantum numbers of the states we are interested in, and can be written as 
\begin{equation} \begin{split}C_{ij}(t) 
&=\langle 0|O_i(t)O_j^{\dagger}(0)|0\rangle\\
&= \sum_{n=1}^{\infty}\langle 0|O_i|n\rangle\langle n|O^{\dagger}_j|0\rangle e^{-E_nt}\\
&\approx\sum_{n=1}^{N}Z_i^{(n)}Z_j^{(n)*} e^{-E_nt}\>.
\label{E0:app}\end{split}\end{equation}
Here, we identify an infinite tower of eigenstates of the Hamiltonian,\footnote{The lattice eigenenergies $E_n^{\textrm{lat}}$ are denoted for simplicity by $E_n$ in this and the following appendix.} denoted as $|n\rangle$, and normalized as $\langle m|n\rangle=\delta_{mn}$. The overlaps between the eigenstates and the operators are expressed as:
\begin{equation}
\begin{split}
Z_i^{(n)}=&\langle 0|O_i|n\rangle\>. \label{Zs:app}
\end{split}
\end{equation}
In the final step of \eqref{E0:app}, the summation over eigenstates is truncated such that the number of states is the same as the number of employed interpolators ($N$).

The overlaps \eqref{Zs:app} are used to identify an eigenstate $|n\rangle$, and, in practice, they are determined by the following equation
\begin{equation}
Z_i^{(n)}(t)= e^{E_nt/2}\frac{\left|C_{ij}(t)u_j^{(n)}(t)\right|}{\left|C(t)^{\frac{1}{2}}{\bi u}^{(n)}(t)\right|}\>,
\label{ZsDerivation:app}
\end{equation}
where ${\bi u}^{(n)}$ are the eigenvectors from Eq.~\eqref{EGEVP}.

Since the overlaps are time dependent, they may not be constant within the region of interest. This challenge is particularly pronounced when dealing with many interpolators. To address this issue, the following procedure is adopted:
The usual GEVP from Eq.~\eqref{EGEVP} is solved only for $t$ up to $t_{\mathrm{fix}}$. The eigenvectors at higher times $t>t_{\textrm{fix}}$ are therefore not calculated. Instead, they are defined as ${\bi u}^{(n)}(t>t_{\textrm{fix}})={\bi u}^{(n)}(t_{\textrm{fix}})$. The eigenvalues $\lambda^{(n)}(t>t_{\textrm{fix}},t_0)$ are computed with $C(t){\bi u}^{(n)}(t_{\textrm{fix}})=\lambda^{(n)}(t,t_0)C(t_0){\bi u}^{(n)}(t_{\textrm{fix}})$ ($t>t_{\textrm{fix}}$) and the overlaps $Z_i^{(n)}(t>t_{\textrm{fix}})$ become constant by definition.
Our specific choices for $t_{\textrm{fix}}$ are documented in Table~\ref{tab:t0tfix}. These choices are partially dictated by constraints from mixing, described in Appendix \ref{sec:app:states}. Ideally, values of $t_{\textrm{fix}}$ should be consistent for all eigenstates of a given correlator. However, in some cases, different values are necessary to disentangle some states high up in the spectrum.

\begin{figure*}[ht!]
\makebox[\textwidth][c]{\includegraphics[width=.97\textwidth]{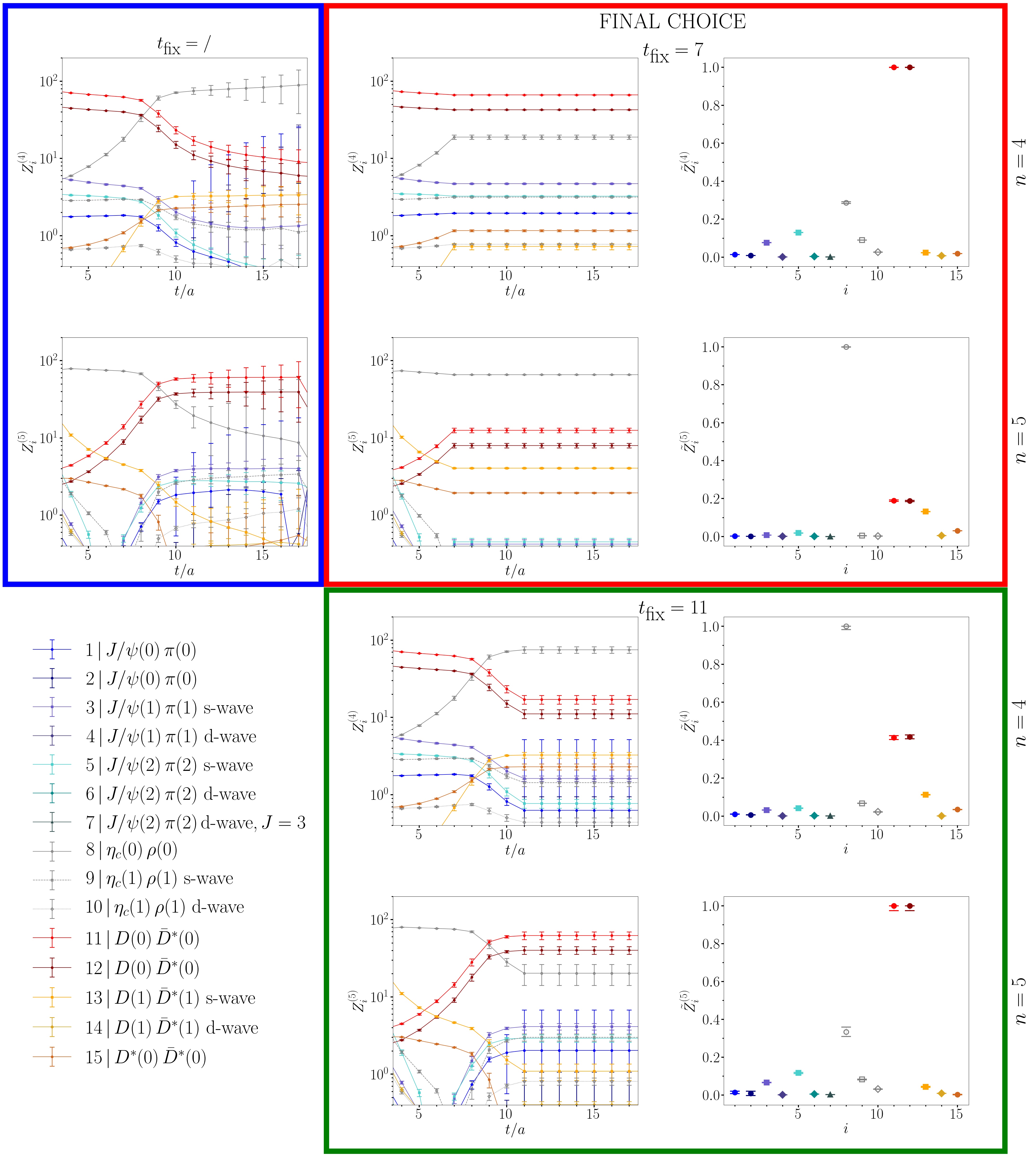}}
\caption{Examples of time-dependent overlaps $Z$ and normalized overlaps $\tilde Z$ for the states $n=4,\ 5$ of the irrep $T_1^{+-}$ and $N_L=24$. These levels highlight an example where a careful choice of $t_{\textrm{fix}}$ is valuable. In the blue box, overlaps are not fixed over time, whereas for the plots in the red (green) box, they are fixed at $t_{\textrm{fix}}=7\,(11)$. The legend lists all the interpolators. Color and symbol schemes are similar to those from other figures. Due to the crossover between interpolators $D(0)\bar{D}^*(0)$ and $\eta_c(0)\rho(0)$ (blue box), a different choice of $t_{\textrm{fix}}$ leads to different normalized overlaps. The state $n=4$ is, e.g., dominated by $D(0)\bar{D}^*(0)$ when $t_{\textrm{fix}}=7$ and by $\eta_c(0)\rho(0)$ when $t_{\textrm{fix}}=11$, where the latter $t_{\textrm{fix}}$ was the final choice. Note that fitted eigenenergies $E_n^{\textrm{lat}}$ mostly do not depend on the choice of $t_{\textrm{fix}}$. This is particularly true for the low-lying states. Overlaps $Z_i^{(4)}(t)$ ($Z_i^{(5)}(t)$) are averaged over time interval $[8,20]$ ($[10,20]$).}
\label{fig:Z}
\end{figure*}

The overlaps $Z_i^{(n)}(t)$ are averaged over a time interval that encompasses the plateau region of effective energies $E_n^{\textrm{eff}}(t)$ defined in Appendix \ref{sec:app:En}. The time-averaged overlaps are denoted as $Z_i^{(n)}$. A valuable measure for overlaps is the normalized overlap  
\begin{equation}
\tilde Z_i^{(n)}\equiv \frac{Z_i^{(n)}}{\max^N_{m=1}  Z_i^{(m)}}
\label{tildeZs:app}
\end{equation}
as it is independent of the normalization of the inter\-polator. An eigenstate $|n\rangle$ is connected with the identity of an interpolator $O_i$ if overlap $\tilde Z_i^{(m)}$ is the largest for eigenstate $m=n$ among all eigenstates $m$. The choice of $t_{\textrm{fix}}$ is irrelevant in many cases, while Fig.~\ref{fig:Z} shows an example of levels where a careful choice of $t_{\textrm{fix}}$ is valuable.

\section{Fitting finite-volume energy levels}\label{sec:app:En}

\begin{figure}[h!]
\begin{center}
\includegraphics[width=.51\textwidth]{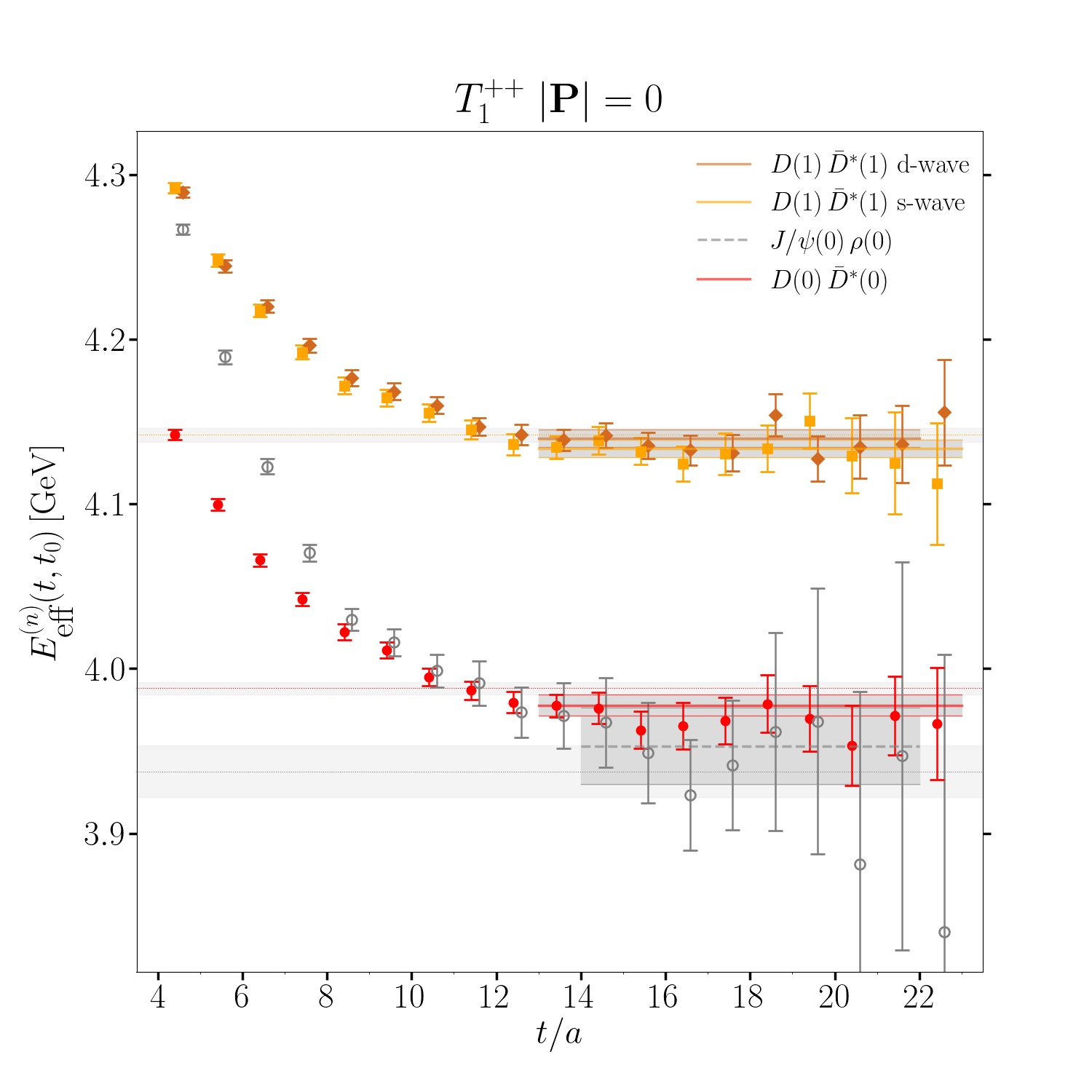}
\caption{Effective energies $E^{(n)}_{\textrm{eff}}(t,t_0=3)$ for the irrep $T_1^{++}$ on the smaller lattice ($N_L=24$) and all eigenstates $n=1,\ldots,4$. They render eigenenergies ($E_n^{\textrm{lat}}$) in the plateau region shown by horizontal lines with gray bands representing their uncertainties. States dominantly couple to the interpolators shown in the legend. Dotted lines represent the noninteracting energies $E_{\textrm{ni}}^{\textrm{lat}}$.}
\label{fig:EeffEn}
\end{center}
\end{figure}

The eigenvalues, $\lambda^{(n)}(t,t_0)\propto e^{-E_n^{\textrm{lat}}(t-t_0)}$, from \eqref{EGEVP} have an asymptotically exponential behaviour at large $t$ and give the effective energies, $E^{(n)}_{\textrm{eff}}(t,t_0)\equiv\frac{1}{a}\ln[\lambda^{(n)}(t,t_0)/\lambda^{(n)}(t+a,t_0)]$, which equal the finite-volume eigenenergies $E_{n}=\lim_{t\to \infty}E^{(n)}_{\textrm{eff}}(t,t_0)$ in the plateau region. An illustrative example of effective energies, as well as the $E_n$, extracted asymptotically through single-exponential fits to $\lambda^{(n)}(t,t_0)$, is plotted in Fig.~\ref{fig:EeffEn}. Our choice of $t_0$ is shown in Table~\ref{tab:t0tfix}. No temporal boundary effects are visible in our variational analysis.

\begin{table}[ht!]
\caption{Table with $t_0$ and $t_{\textrm{fix}}$ used when extracting the eigenenergies of all states below the highest states coupled to $D\bar D^*$ from the full correlation matrix. Symbol $ ^{\dagger}$ means that those numbers are valid for all states except those listed below. }
\label{tab:t0tfix}
\centering
\begin{tabular}{>{\centering\hspace{0pt}}m{0.16\linewidth}>{\centering\hspace{0pt}}m{0.102\linewidth}>{\centering\hspace{0pt}}m{0.428\linewidth}>{\centering\hspace{0pt}}m{0.092\linewidth}>{\centering\arraybackslash\hspace{0pt}}m{0.145\linewidth}} \hline \hline
Irrep & $N_L$ & States & $t_0$ & $t_{\textrm{fix}}$ \\ \hline 
$T_1^{+-}$ & 24 & all$^{\dagger}$ & 2 & 11 \\
 &  & $J/\psi(2) \pi(2)$~s wave& 3 & 11 \\
 &  & $D(1)\bar D^*(1)$~s wave & 3 & 11 \\
 &  & $D(1)\bar D^*(1)$~d wave & 2 & 14 \\
 &  & $\eta_c(1) \rho(1)$~s wave & 2 & 7 \\ 
 &  & $\eta_c(1) \rho(1)$~d wave & 2 & 16 \\
 &  & $D^*(0)\bar D^*(0)$ & 3 & 7 \\ \hline
$T_1^{+-}$ & 32 & all$^{\dagger}$ & 2 & 10 \\
 &  & $D(1)\bar D^*(1)$~s wave & 2 & 6 \\
 &  & $D(1)\bar D^*(1)$~d wave & 2 & 9 \\
 &  & $\eta_c(1) \rho(1)$~s wave & 2 & 9 \\
 &  & $\eta_c(1) \rho(1)$~d wave & 2 & 8 \\
 &  & $D^*(0)\bar D^*(0)$ & 2 & 9 \\ 
 &  & $J/\psi(3) \pi(3)$~s wave& 2 & 9 \\
 &  & $J/\psi(3) \pi(3)$~d wave& 2 & 9 \\\hline
$T_1^{++}$ & 24 & all & 3 & 16 \\ \hline
$T_1^{++}$ & 32 & all & 2 & 10 \\ \hline
$A_2^{-}$ & 24 & all & 2 & 13 \\ \hline
$A_2^{+}$ & 24 & all$^{\dagger}$ & 2 & 16 \\
 &  & $J/\psi(0) \rho(1)$ & 3 & 8 \\ \hline
$A_2^{+}$ & 32 & all$^{\dagger}$ & 2 & 8 \\
 &  & $\eta_c(1) a_0(0)$ & 3 & 10 \\
 &  & $\chi_{c0}(2) \pi(1)$ & 2 & 10 \\
 &  & $D(1)\bar D^*(2)$ & 3 & 10 \\ \hline \hline
\end{tabular}
\end{table}

\section{Mixing between the states}\label{sec:app:states}

\begin{figure*}[ht!]
\centering
\begin{minipage}{0.515\linewidth}
\centerline{\includegraphics[width=1.\textwidth]{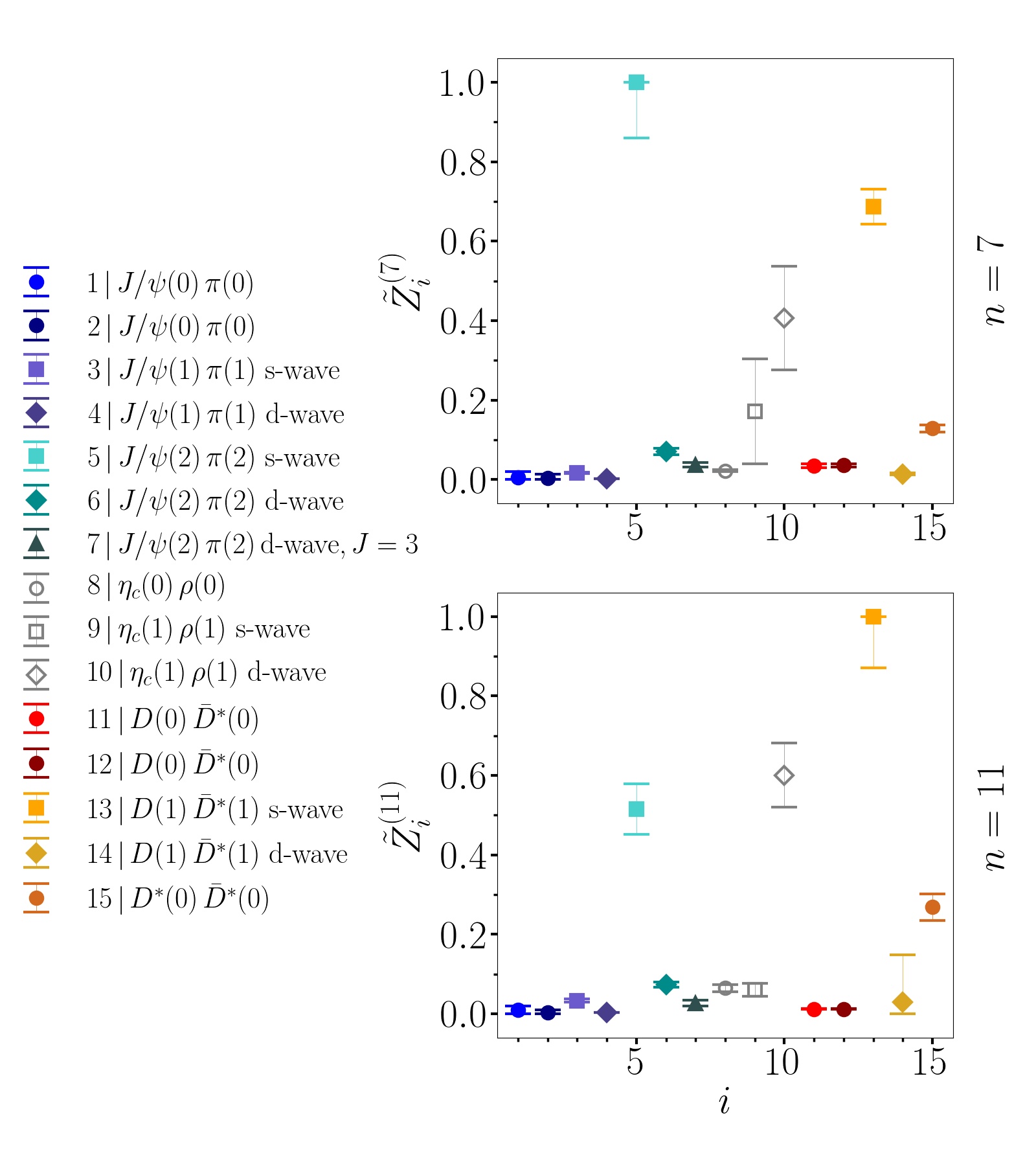}}
  \vspace{-0.8cm}
	\end{minipage}
\hspace*{1cm}
\begin{minipage}{0.415\linewidth}
\centerline{\includegraphics[width=1.\textwidth]{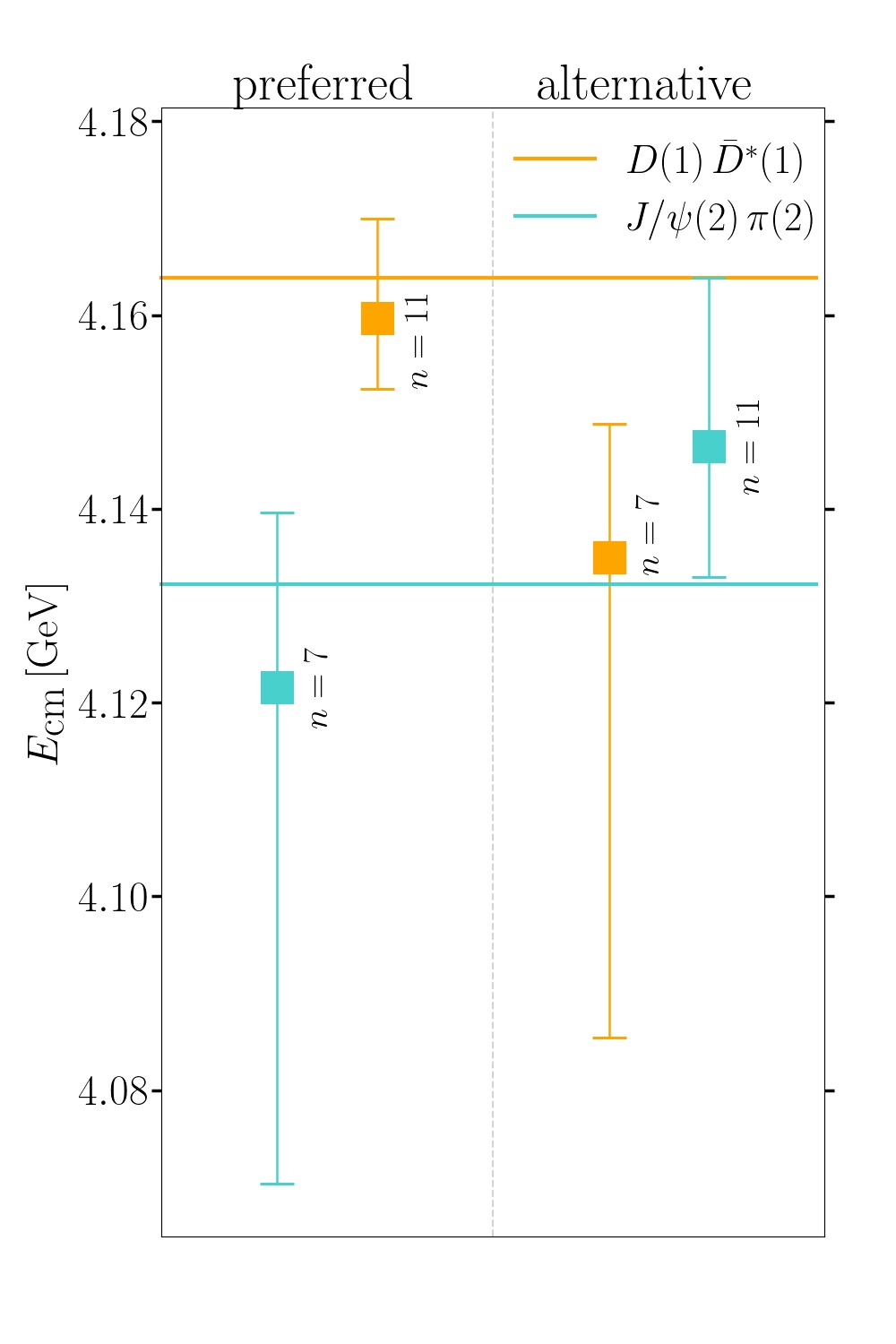}}
	\end{minipage}
\caption{Two states containing significant admixtures of $D(1)\bar{D}^*(1)$ s wave, $J/\psi(2)\pi(2)$ s wave [but also $\eta_c(1)\rho(1)$ d wave] are presented. They come from the irrep $T_1^{+-}$ and $N_L=24$. Both panels' color and symbol schemes are similar to those in the other figures.
Left panel: normalized overlaps for the two states $n=7$ (above) and $n=11$ (below). Right panel: Eigenenergies from two identifications: preferred identifications where $n\!=\!7$ is $J/\psi(2)\pi(2)$ and $n\!=\!11$ is $D(1)\bar{D}^*(1)$, and alternative identification where they are reversed. Both lead to eigenenergies that agree within errors.}
\label{fig:Zmix}
\end{figure*}

In most cases, identifying states via overlaps with specific interpolators is clean. Exceptions do not necessarily appear in cases where both states have similar energies. For example, one would expect to encounter mixing between $D(1)\bar{D}^*(1)$ and $J/\psi(3)\pi(3)$ in $T_1^{+-}$, $N_L=32$, but there is none. A physically important example featuring relatively mixed states is presented in the left-hand side plots in Fig.~\ref{fig:Zmix}, which exhibits an admixture of $D(1)\bar{D}^*(1)$ s wave and $J/\psi(2)\pi(2)$ s wave ($T_1^{+-}$, $N_L=24$). Energies from two different identifications are shown in the right pane, and they agree within the uncertainties. Similarly, the two interpolators $D(1)\bar{D}^*(0)$ and $J/\psi(1)\pi(2)$ mix in two states of the irrep $A_2^{-}$ ($N_L=24$). These examples suggest that the interaction between $J/\psi\pi$ and $D\bar{D}^*$ could be significant in understanding experimental $Z_c(3900)$ peaks.

\section{More fits based on EFT}\label{sec:app:more-on-fits}

The EFT analysis of the $Z_c$ channel in Sec.~\ref{sec:comparison} was based on four fits listed in Sec.~\ref{subsec:fits}.  

In addition, we also performed several variations of Fit3 with the same central value and artificially reduced errors on our energy levels above $E_{\mathrm{cm}}=\SI{4050}{MeV}$. They involve the four fitted lattice points (green squares in Figs.~\ref{fig:our_P0-P1}, \ref{fig:our-CLQCD-HSC} and \ref{fig:our-CLQCD-HSC+Fit3}) with uncertainties artificially lowered by a factor of 2, 7 and 10. Their results show qualitatively similar behavior to Fit3. With more artificially enhanced precision, we can make the following general observations. We obtain
\begin{itemize}[noitemsep,nolistsep]
    \item a slightly larger $\chi^2/N_{\textrm{dof}}$;
    \item better replication of our lattice energy points;
    \item a smaller, but still present, peak of the predicted $D\bar D^*$ line shapes;
    \item a more significant difference in specific fitted parameters compared to Fit1, Fit2 and Fit4.
\end{itemize}
All those fits (together with Fit3) reproduce the $J/\psi\pi$ line shapes very similarly to other fits.

\begin{figure}[h!]
\begin{center}
\includegraphics[width=0.52\textwidth,angle=-0]{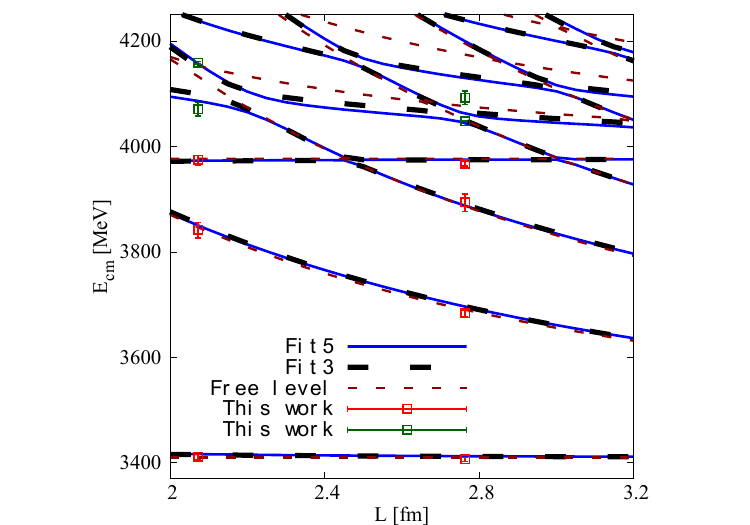} 
\caption{Energy levels from Fit3 and Fit5, which are obtained by artificially reducing the original error bars by a factor of 4. For the former situation, the central values of the energy levels in the fit are fixed to their original determinations, while for the latter case, i.e., Fit5, we vary the central values in the fit within the original uncertainties. To be more specific, the central value of the fifth energy level for $N_L = 24$ takes the upper end of the original boundary, and for the fourth energy level of $N_L = 24$, the fifth and sixth energy levels for $N_L = 32$, we consider the lower ends from their original boundaries as the central values for Fit5. In this plot, the error bars shown have been divided by four for the four green points.}  
\label{fig:Fit5}
\end{center}
\end{figure}

We also performed variations of Fit3, where the central value of four mentioned higher-lying lattice levels vary within their original uncertainties, while the error is artificially reduced by a factor of four. A variation of such a fit (Fit5) represented in Fig.~\ref{fig:Fit5} reproduces such energy levels, and there is no qualitative change in the reproduction of experimental data or location of $Z_c$ poles in Table~\ref{tab:Fit5}.

\begin{table}[ht!]
\caption{ Central values of pole positions and residues  for Fit5.}  
\label{tab:Fit5}
\centering
		\begin{tabular}{ l c c c c }
			\hline\hline
			RS & $\quad m_{\textrm{R}}\,$[MeV]$\quad$ & $\Gamma_{\textrm{R}}/2\,$[MeV]& $\left | \gamma_1 \right |\,$[GeV] &  $ \left| \gamma_2 \right|\,$[GeV]\\
             \hline
			Fit5\\
			III 
			& $3814.8$& $41.3$&$5.2$&$15.9$\\
			\hline
			IV
			& $3861.9$&$ 26.8$&$4.4$&$14.9$\\
			
                \hline \hline
               
		\end{tabular}
\end{table}

\section{Finite-volume spectra with omitted states}
\label{sec:app:omitted_states}

\begin{figure*}[ht!]
\begin{minipage}{0.4\linewidth}\centerline{\includegraphics[width=1.0\textwidth]{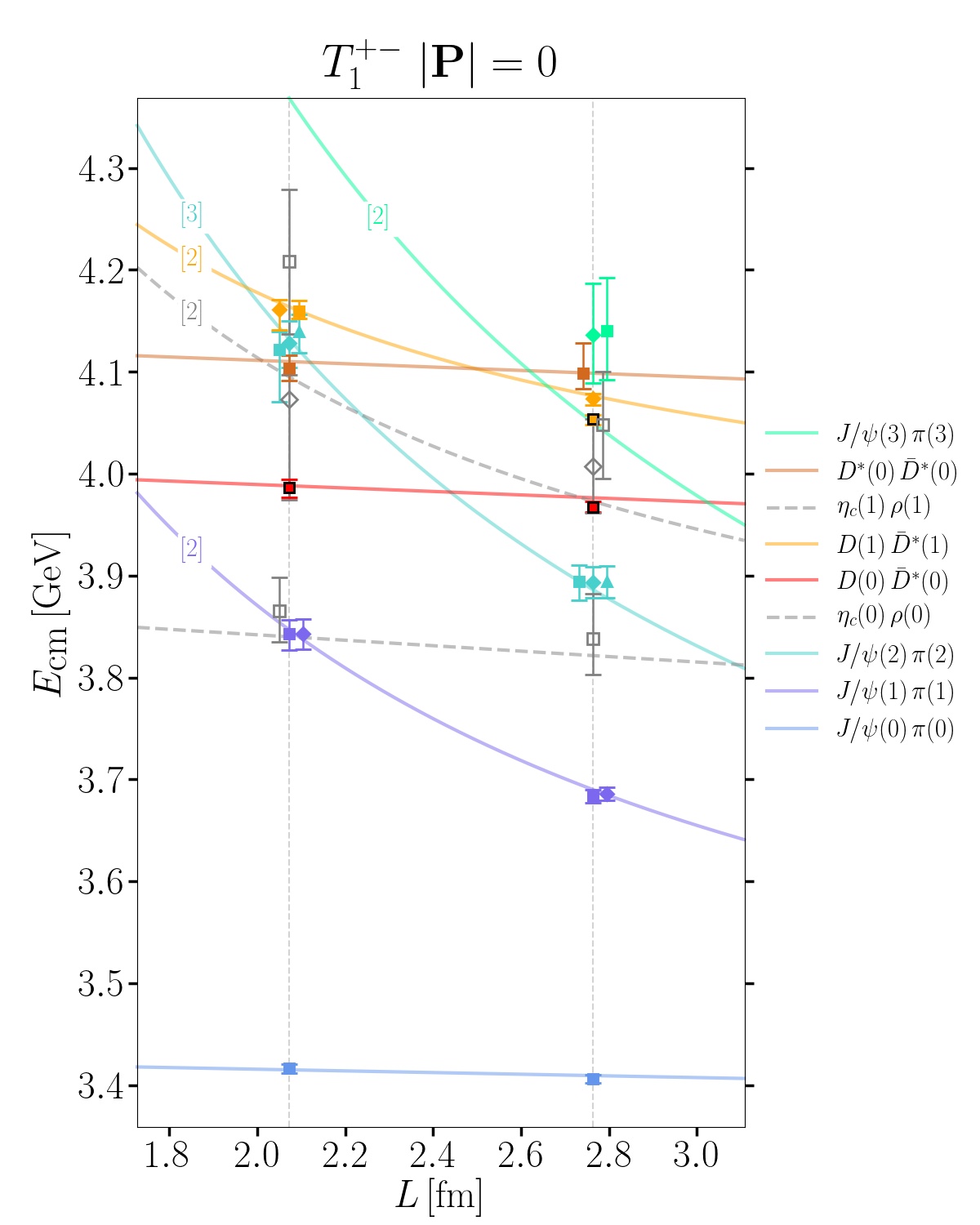}}\end{minipage}
\begin{minipage}{0.4\linewidth}\centerline{\includegraphics[width=1.0\textwidth]{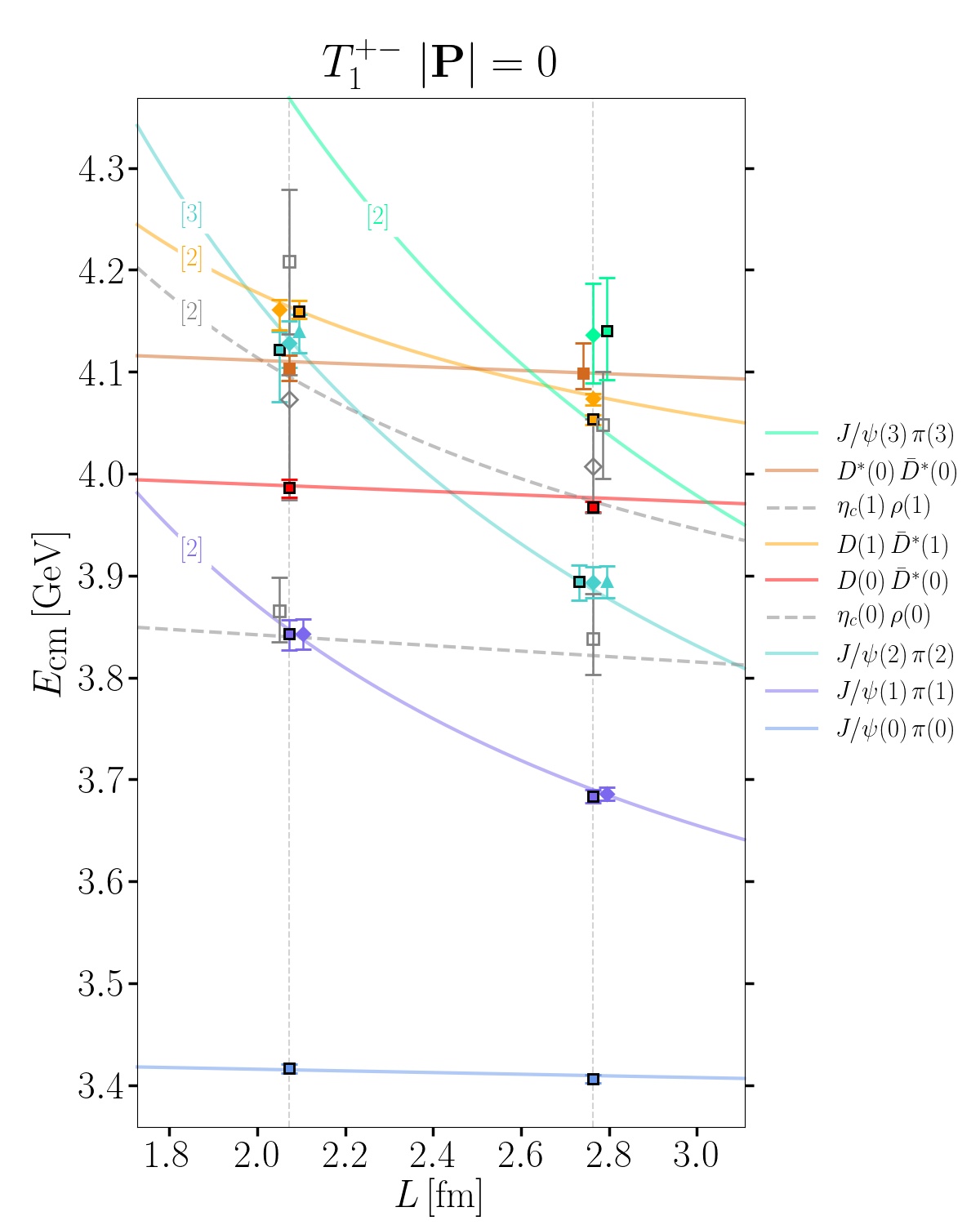}}\end{minipage}
\begin{minipage}{0.4\linewidth}\centerline{\includegraphics[width=1.0\textwidth]{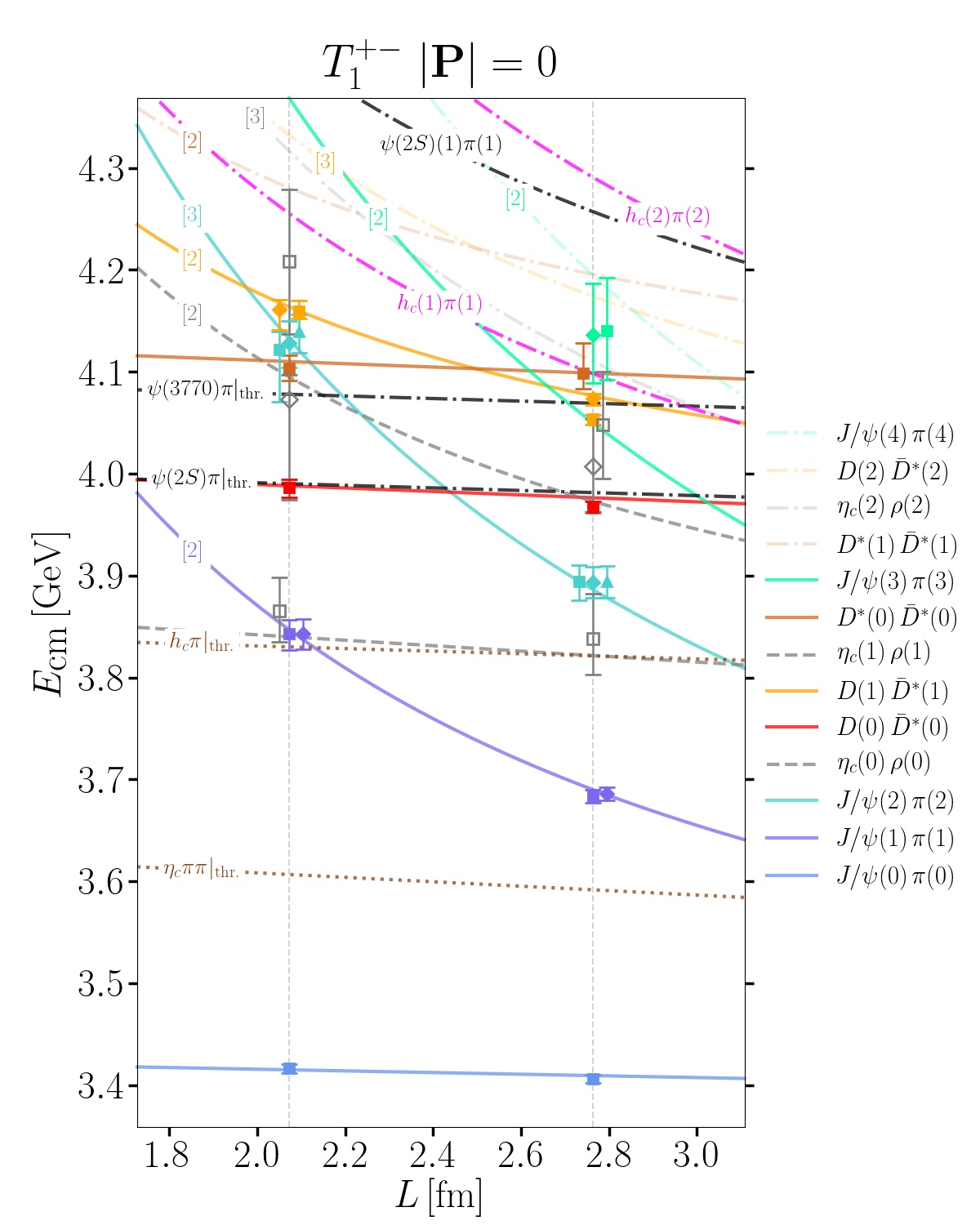}}\end{minipage}
\caption{
Top: finite volume spectra from the left pane of Fig.~\ref{fig:T1pm} with black outlines added to eigenenergies used in elastic (left) or coupled channel (right) scattering analysis. Bottom: same spectrum with additional lines: these lines represent neglected operators with higher internal momenta (transparent dash-dotted lines with the same color as the highest lines for included operators), omitted channels in the operator basis (bold dash-dotted lines) and thresholds with no noninteracting level at the threshold (brown dotted lines). For example, $h_c(0)\pi(0)$ and $\eta_c(0)\pi(0)\pi(0)$ are not present in $T_1^{+-}$ therefore thresholds are indicated by dotted lines. The exception with the given line-style convention is $h_c(1)\pi(1)$ that was in the basis, but this high-lying level was not extracted. The energy scale is the same as in Fig.~\ref{fig:T1pm}.
}
\label{fig:omit_1+-}
\end{figure*}

\begin{figure*}[ht!]
\begin{minipage}{0.4\linewidth}\centerline{\includegraphics[width=1.09\textwidth]{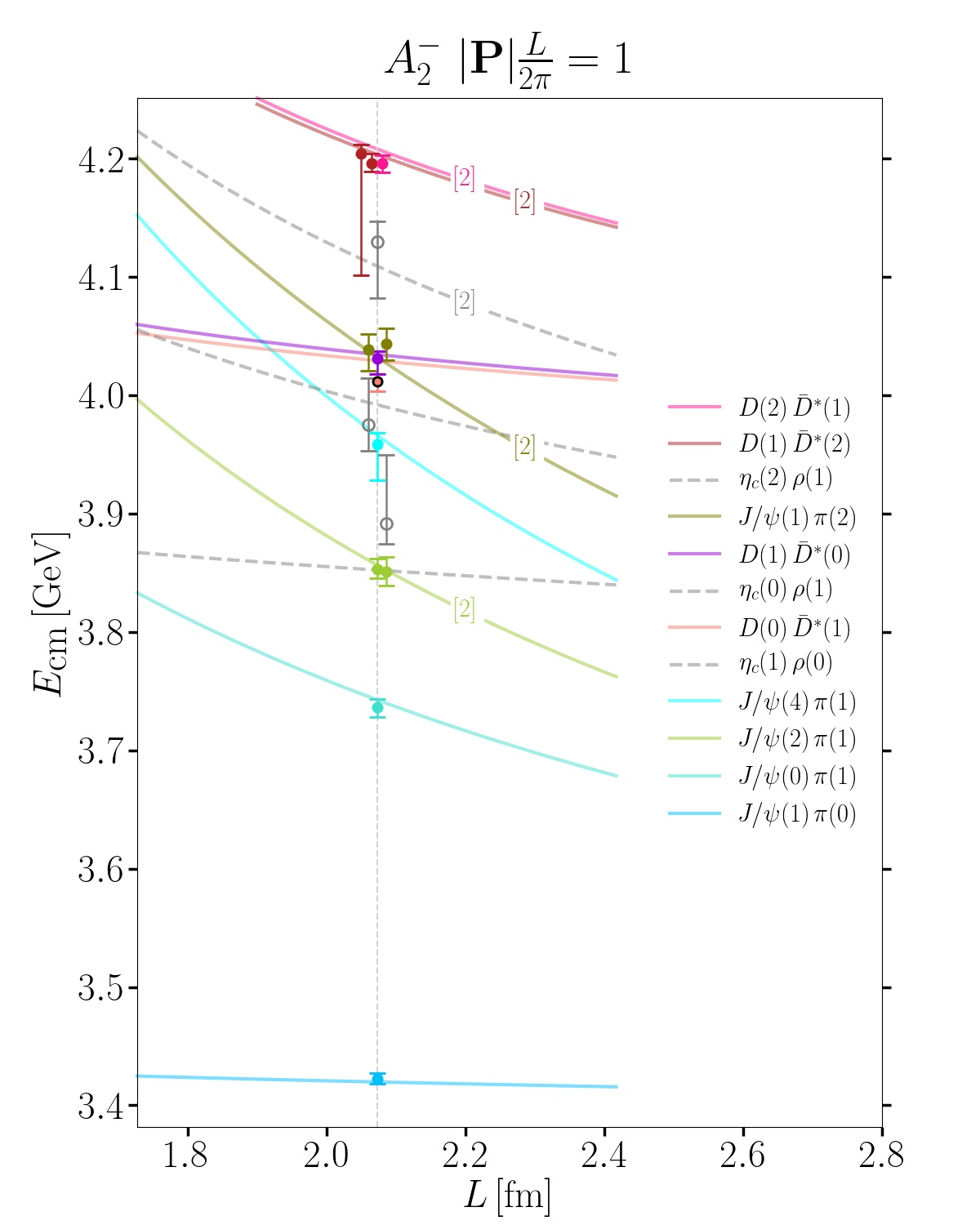}}\end{minipage}
\begin{minipage}{0.4\linewidth}\centerline{\includegraphics[width=1.09\textwidth]{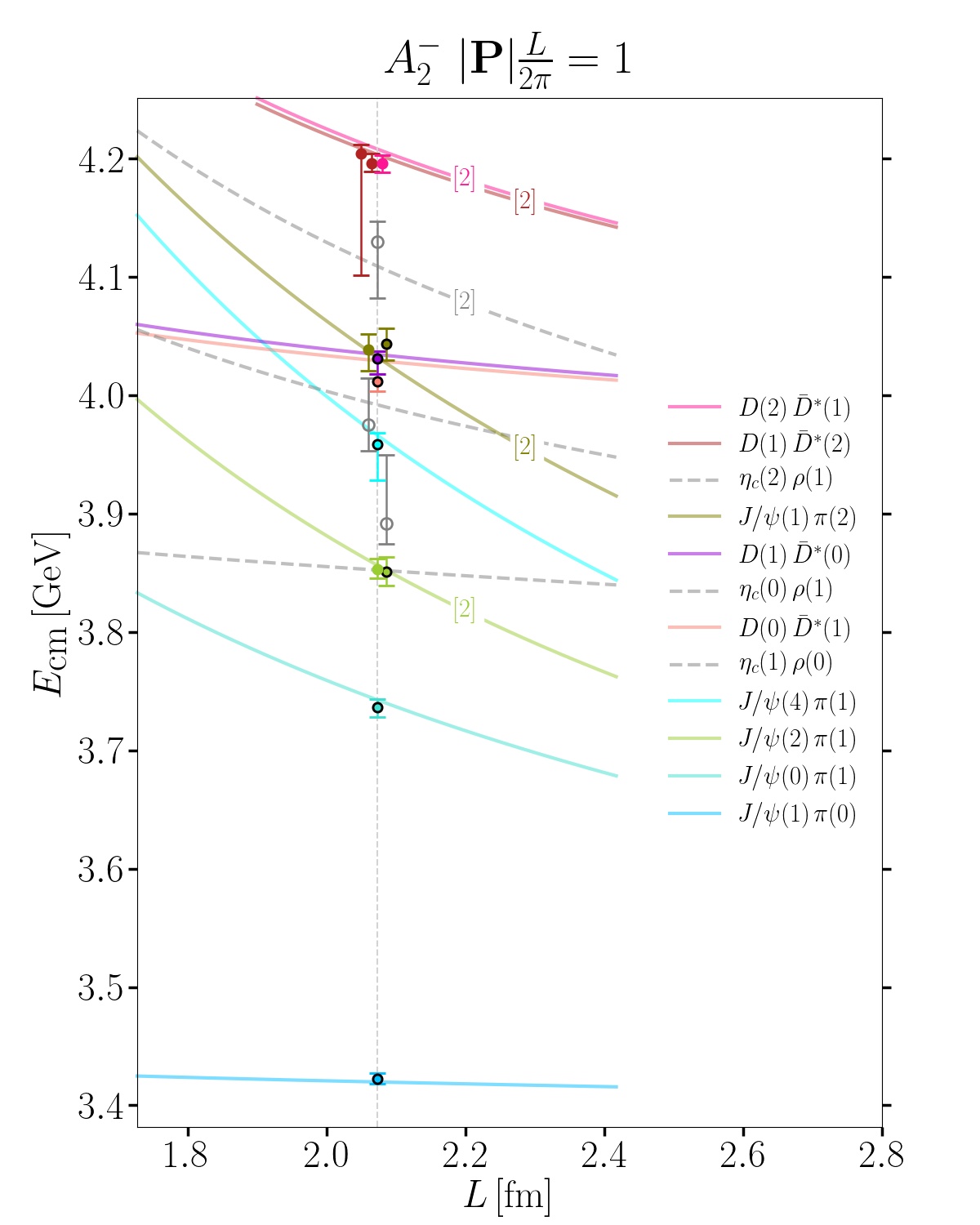}}\end{minipage}
\begin{minipage}{0.4\linewidth}\centerline{\includegraphics[width=1.09\textwidth]{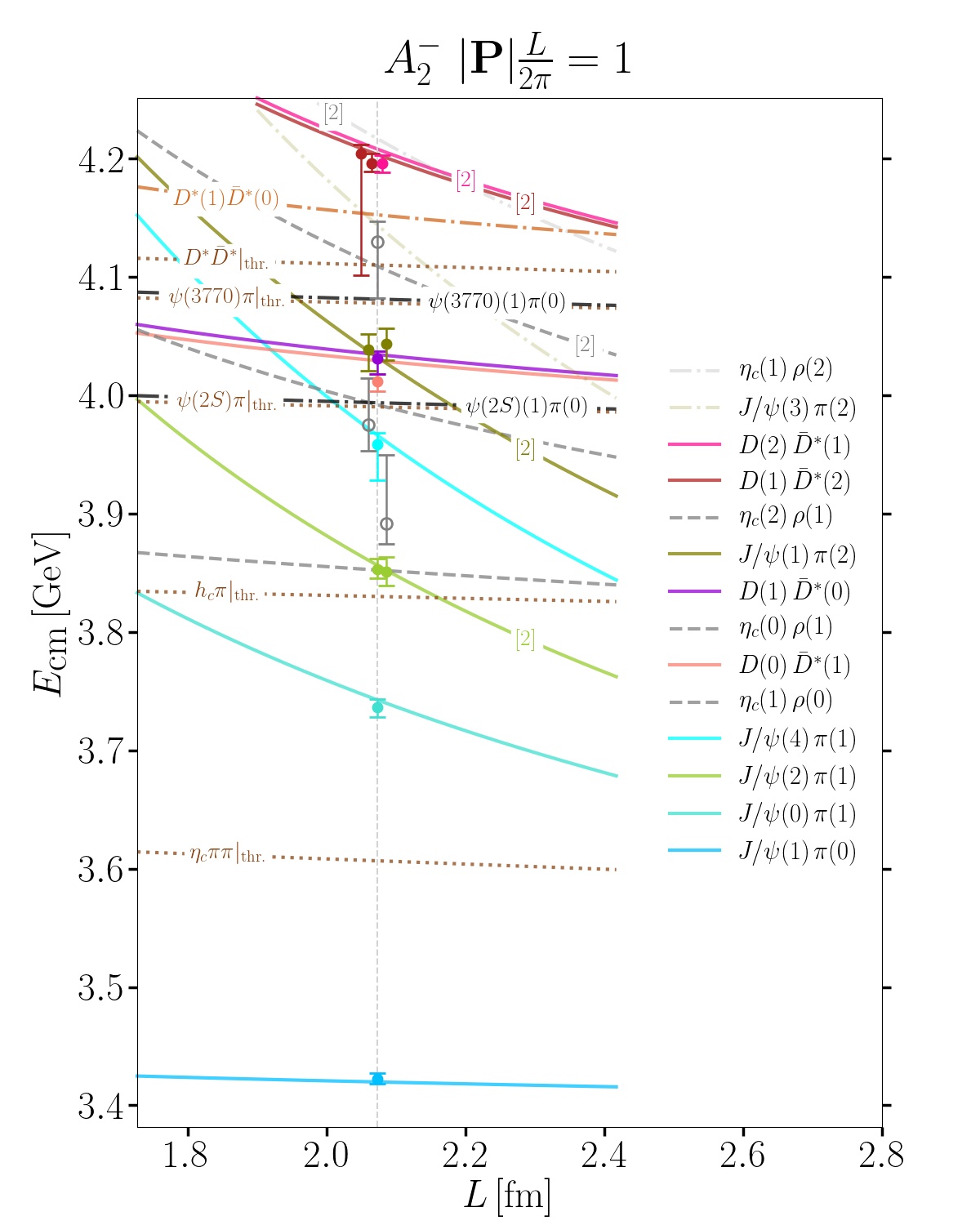}}\end{minipage}
\caption{
Finite volume spectra from the left pane of Fig.~\ref{fig:A2m} using analogical representation as in Fig.~\ref{fig:omit_1+-}. Noninteracting levels $h_c(1)\pi(0)$ and $h_c(0)\pi(1)$ are not present in $A_2^{-}$. The next $h_c\pi$ level $h_c(2)\pi(1)$ is out of our energy range. The energy scale is the same as in Fig.~\ref{fig:A2m}.
}
\label{fig:omit_A2-}
\end{figure*}

\begin{figure*}[ht!]
\begin{minipage}{0.45\linewidth}
\centerline{\includegraphics[width=1.29\textwidth]{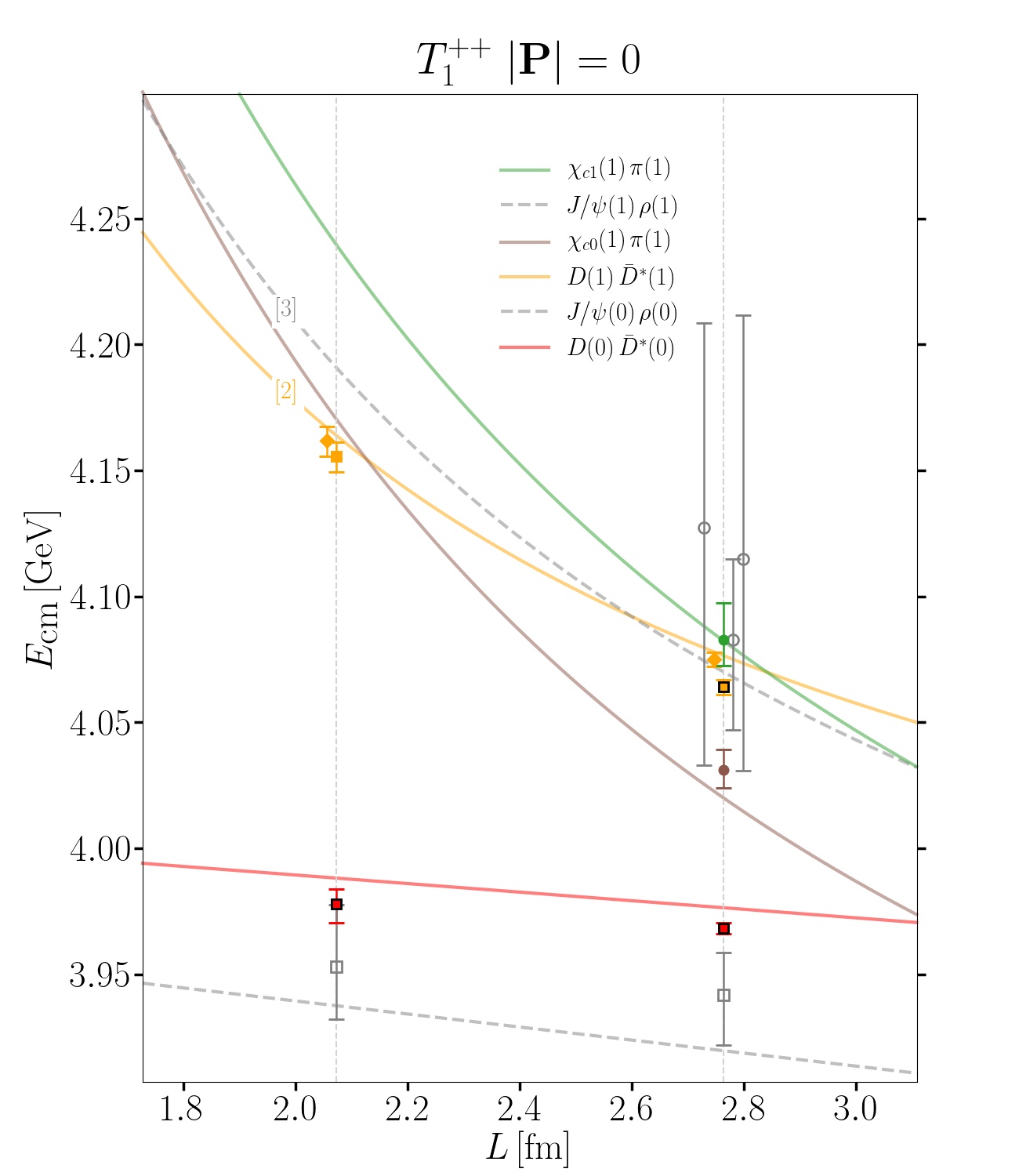}}
	\end{minipage}
\hspace*{1.55cm}
\begin{minipage}{0.45\linewidth}
\centerline{\includegraphics[width=1.29\textwidth]{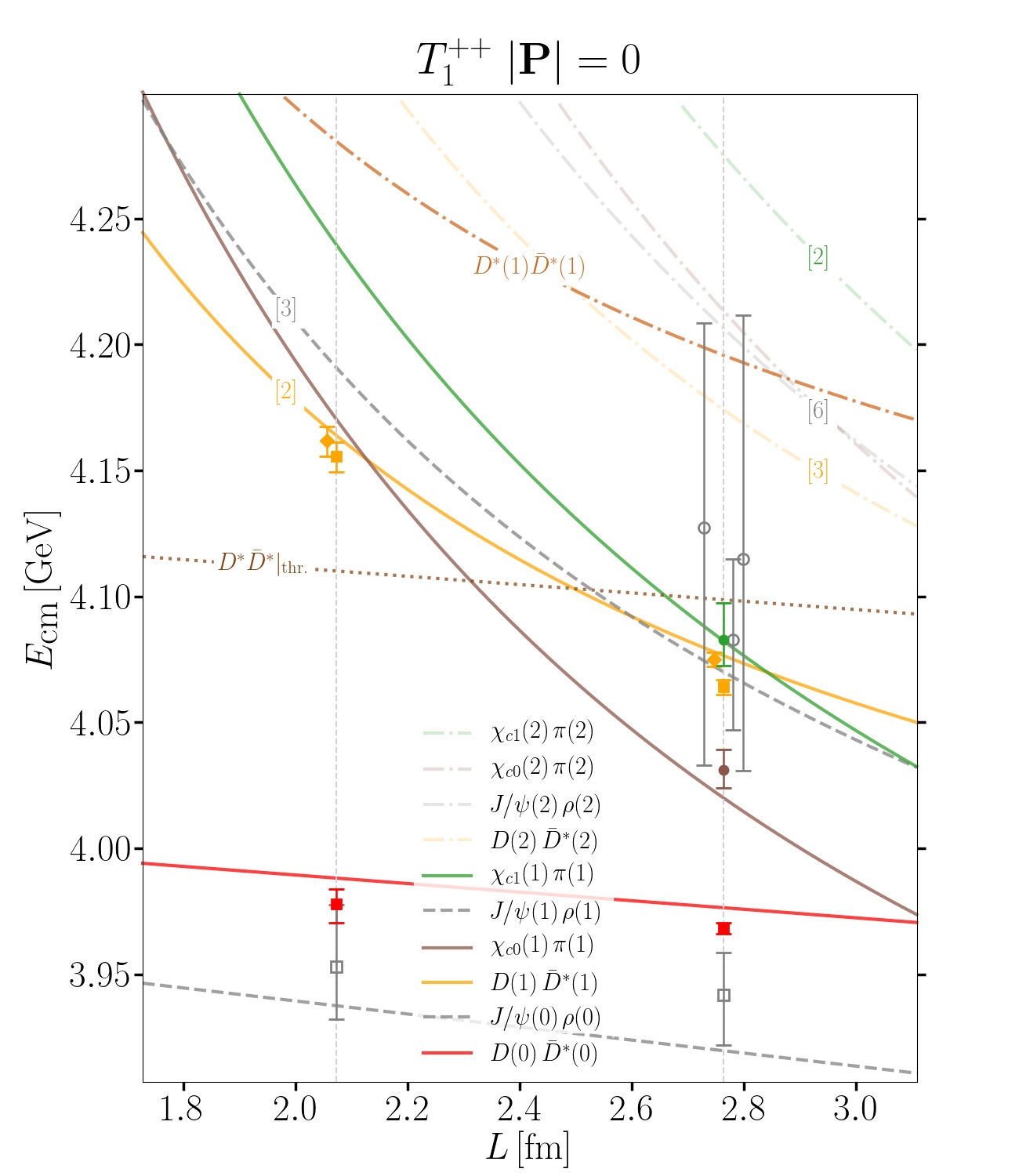}}
	\end{minipage}
\caption{
Finite volume spectra from the left pane of Fig.~\ref{fig:T1pp} with black outlined eigenenergies used in elastic scattering analysis (left pane) and additional lines (right pane). The notion of various lines' styles is provided in Fig. \ref{fig:omit_1+-}.    State $D^*(0)\bar{D}^*(0)$ is not present in $T_1^{++}$ with charge parity $C=+$. The energy scale is the same as in Fig.~\ref{fig:T1pp}, therefore, the right pane does not contain the threshold $J/\psi\pi\pi$, which appears below the lowest extracted states.
}
\label{fig:omit_1++}
\end{figure*}

\begin{figure*}[ht!] 
\begin{minipage}{0.45\linewidth}
\centerline{\includegraphics[width=1.24\textwidth]{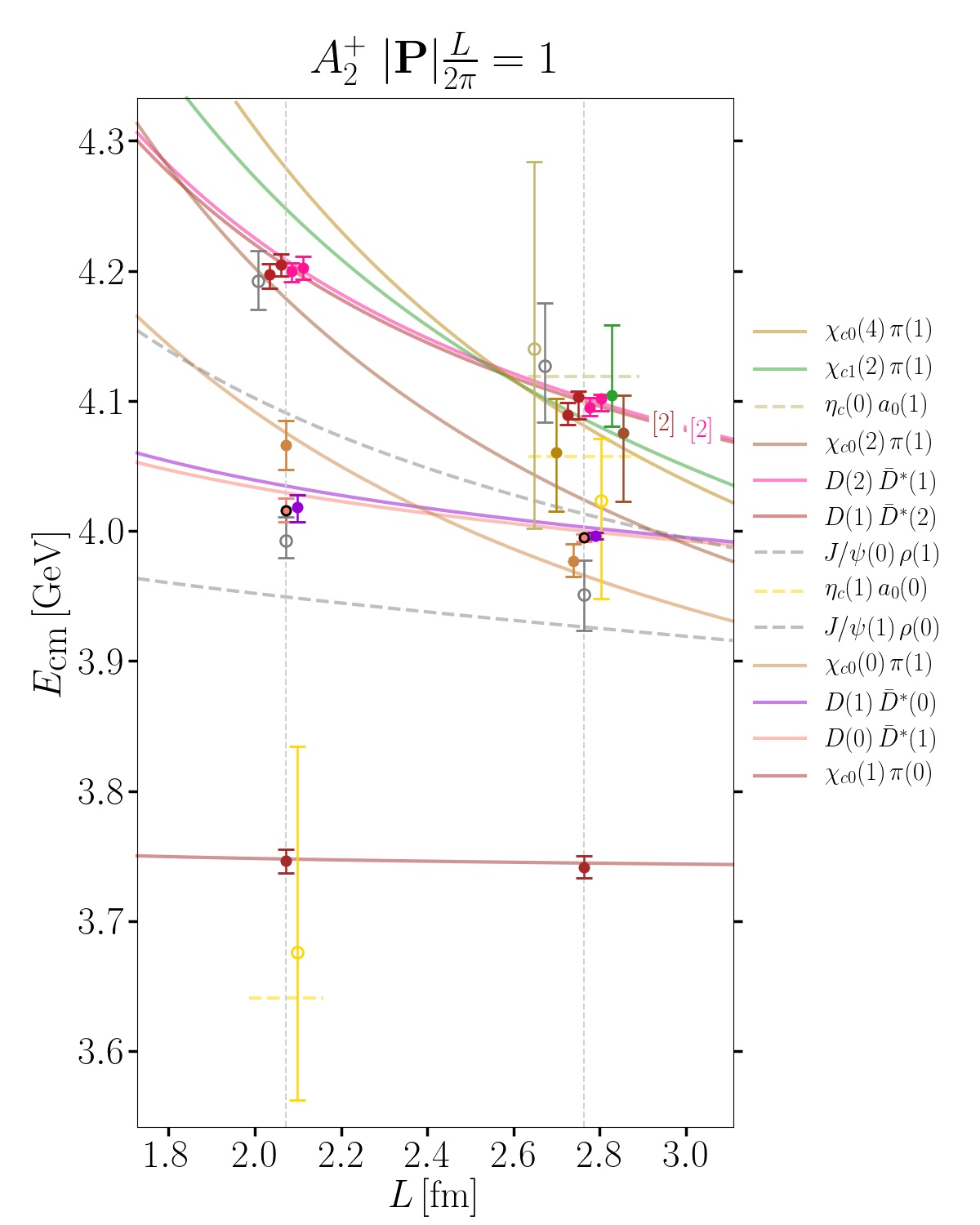}}
	\end{minipage}
\hspace*{1.55cm}
\begin{minipage}{0.45\linewidth}
\centerline{\includegraphics[width=1.24\textwidth]{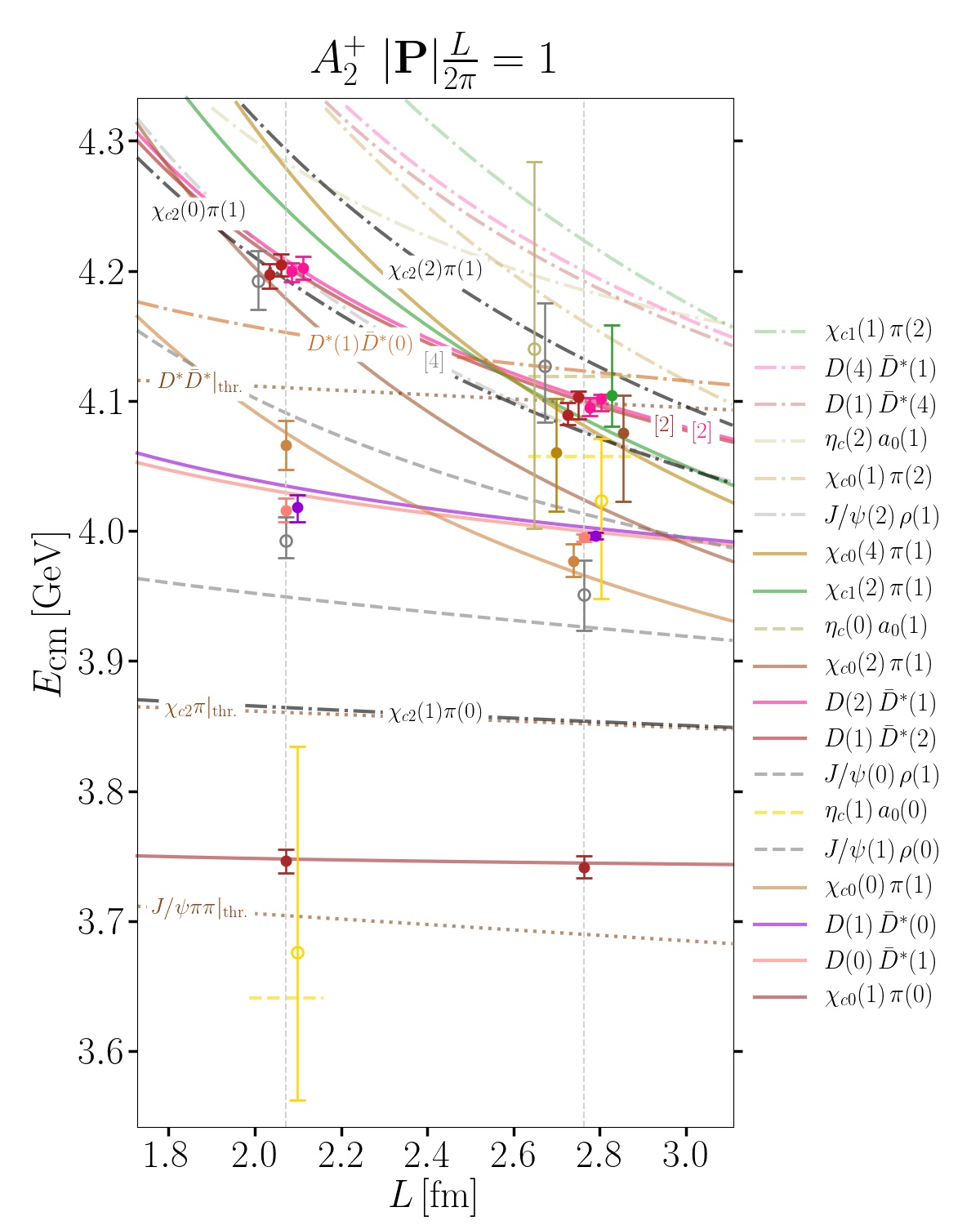}}
	\end{minipage}
\caption{
Finite volume spectra from the left pane of Fig.~\ref{fig:A2p} with black outlined eigenenergies used in elastic scattering analysis (left pane) and additional lines   (right pane).  The noninteracting line $\eta_c(2)a_0(1)$ is continuous and assumes that $m_{a_0}$ does not depend on $L$. The energy scale is the same as in Fig.~\ref{fig:A2p}.
}
\label{fig:omit_A2+}
\end{figure*}

To assess the robustness of the computed finite-volume spectra, we reproduce the left panes of Figs.~\ref{fig:T1pm}--\ref{fig:A2p} and add 
\begin{itemize}
    \item noninteracting energies of the lowest neglected interpolators with higher internal momenta (dash-dotted lines with the same but more transparent colors as the highest lines that correspond to the included interpolators),
    \item noninteracting energies (dash-dotted lines) and thresholds (brown dotted lines) of neglected two-particle channels,
    \item thresholds of neglected three-particle channels (brown dotted lines).
\end{itemize}
These spectra are in bottom panes of Figs.~\ref{fig:omit_1+-} (irrep $T_1^{+-}$) and \ref{fig:omit_A2-} ($A_2^-$), and in right panes of Figs.~\ref{fig:omit_1++} ($T_1^{++}$) and \ref{fig:omit_A2+} ($A_2^+$). These figures additionally contain reproductions of left panes of Figs.~\ref{fig:T1pm}--\ref{fig:A2p}, where the eigenenergies incorporated in the scattering analyses are outlined in black.

\bibliographystyle{apsrev4-2}
\bibliography{bibliography_for_paper_cbarcqbarq}

\end{document}